\documentclass[12pt]{article}
\usepackage{graphicx}
\usepackage{dcolumn} 
\usepackage{bm}      
\usepackage{umlaut}
\usepackage{epsfig}
\usepackage{dcolumn}
\usepackage{amsmath}
\usepackage{cite}
\usepackage{changebar}
\usepackage{latexsym}
\begin{document}
\hbox{}
 \mbox{} \hfill \hspace{1.0cm}
         \today 
 \mbox{} \hfill BI-TP 2006/01r\hfill
~\\
\begin{center}
{\Large{\bf{Lattice QCD Calculations\\
~\\ 
for the\\
~\\ 
Physical Equation of State}}}$^{*}$
\\[1cm]
David E. Miller$^{a,b}$
\\[1cm]
$^{a}$Department of Physics\\
Pennsylvania State University\\
Hazleton Campus\\
Hazleton, PA 18202 USA\\
E-mail:om0@psu.edu
\\[0.2cm]
and
\\[0.2cm]
$^{b}$Fakult\"at f\"ur Physik\\
Universit\"at Bielefeld\\
D-33501 Bielefeld\\
F. R. Germany\\
E-mail:dmiller@physik.uni-bielefeld.de
\\[2cm]
{\underline{\bf{ABSTRACT:}}}
\end{center}
In this report we consider the numerical simulations at finite
temperature using lattice QCD data for the computation of the 
thermodynamical quantities including the pressure, energy density 
and the entropy density. These physical quantities can be related 
to the equation of state for quarks and gluons. We shall apply the 
lattice data to the evaluation of the specific structure of the 
gluon and quark condensates at finite temperature in relation to 
the deconfinement and chiral phase transitions. Finally we mention 
the quantum nature of the phases at lower temperatures.
~\\
$^{*}${\it{Physics Reports}}, to appear.\\
\newpage
\begin{center}
{\Large{\underline{\bf{TABLE OF CONTENTS:}}}}
\\[1.0cm]
\end{center}
I.  Introduction: Thermodynamics of Strong Interactions~~~~~~~~~~~~~~~~~~~~~~~~~~~~~~~~~~~~~~~~~~~~~~~~~~~~~~~~~~~~~~~~~3\\
~\\
II.  Pure Gauge Theory~~~~~~~~~~~~~~~~~~~~~~~~~~~~~~~~~~~~~~~~~~~~~~~~~~~~~~~~~~~~~~~~~~~~~~~~~~~~~~~~~~~~~~~~~~~~~9\\
\indent
II.1  Lattice Thermodynamics~~~~~~~~~~~~~~~~~~~~~~~~~~~~~~~~~~~~~~~~~~~~~~~~~~~~~~~~~~~~~~~~~~~~~~~~~~~~~~~9\\
\indent
II.2  Thermal Field Theoretical Evaluation~~~~~~~~~~~~~~~~~~~~~~~~~~~~~~~~~~~~~~~~~~~~~~~~~~~~~~~~~~~~~12\\
\indent
II.3  Numerical Evaluation of Physical Quantities~~~~~~~~~~~~~~~~~~~~~~~~~~~~~~~~~~~~~~~~~~~~~~~~~~~~13\\
\indent
II.4  Comparison of Physical Quantities~~~~~~~~~~~~~~~~~~~~~~~~~~~~~~~~~~~~~~~~~~~~~~~~~~~~~~~~~~~~~~~~~14\\
\indent
II.5  Discussion of Conformal Symmetry~~~~~~~~~~~~~~~~~~~~~~~~~~~~~~~~~~~~~~~~~~~~~~~~~~~~~~~~~~~~~~~~~~~~15\\
~\\
\noindent
III.  Dynamical Quarks~~~~~~~~~~~~~~~~~~~~~~~~~~~~~~~~~~~~~~~~~~~~~~~~~~~~~~~~~~~~~~~~~~~~~~~~~~~~~~~~~~~~~~~~~~~~17\\
\indent
III.1  Equation of State for two light Quark Flavors~~~~~~~~~~~~~~~~~~~~~~~~~~~~~~~~~~~~~~~~~~~~~~~~~~~~17\\
\indent
III.2  Chiral Symmetry and Dynamical Quarks~~~~~~~~~~~~~~~~~~~~~~~~~~~~~~~~~~~~~~~~~~~~~~~~~~~~~~~~~19\\ 
\indent
III.3  Thermodynamics with Dynamical Quarks~~~~~~~~~~~~~~~~~~~~~~~~~~~~~~~~~~~~~~~~~~~~~~~~~~~~~~~~20\\
\indent
III.4  Equation of State with Different Quark Flavors~~~~~~~~~~~~~~~~~~~~~~~~~~~~~~~~~~~~~~~~~~~~~~~22\\
\indent
III.5  Discussion of Recent Massive Quark Data~~~~~~~~~~~~~~~~~~~~~~~~~~~~~~~~~~~~~~~~~~~~~~~~~~~~~~26\\
~\\
\noindent
IV.  Gluon and Quark Condensates~~~~~~~~~~~~~~~~~~~~~~~~~~~~~~~~~~~~~~~~~~~~~~~~~~~~~~~~~~~~~~~~~~~~~~~~~~~~~~~~~~27\\
\indent
IV.1  Pure Gluon Condensates~~~~~~~~~~~~~~~~~~~~~~~~~~~~~~~~~~~~~~~~~~~~~~~~~~~~~~~~~~~~~~~~~~~~~~~~~~~~~~~~~~27\\
\indent
IV.2  The Effect of Quark Condensates~~~~~~~~~~~~~~~~~~~~~~~~~~~~~~~~~~~~~~~~~~~~~~~~~~~~~~~~~~~~~~~~~~~~~~~~~~~~~~28\\
\indent
IV.3  Gluon and Quark Condensates with two light Flavors~~~~~~~~~~~~~~~~~~~~~~~~~~~~~~~~~~~~~~~~~~~~~~~~~~~~~~29\\
\indent
IV.4  Gluon and Quark Condensates with more massive Flavors~~~~~~~~~~~~~~~~~~~~~~~~~~~~~~~~~~~~~~~~~~~~~~~~~~~31\\
\indent
IV.5  Comparisons of Properties of Gluon Condensates~~~~~~~~~~~~~~~~~~~~~~~~~~~~~~~~~~~~~~~~~~~~~~~~~~~~~~34\\
~\\
\noindent
V. Discussion of Physical Results~~~~~~~~~~~~~~~~~~~~~~~~~~~~~~~~~~~~~~~~~~~~~~~~~~~~~~~~~~~~~~~~~~~~~~~~~~~~~~37\\
\indent
V.1  Phenomenological Models of Quark and Gluon Properties~~~~~~~~~~~~~~~~~~~~~~~~~~~~~~~~~~~~~~~~~~~~~~~~~~~~~~~37\\
\indent
V.2  Theoretical Models for the Quark and Gluon Properties~~~~~~~~~~~~~~~~~~~~~~~~~~~~~~~~~~~~~~~~~~~~~~~~~~~~~~~38\\ 
\indent
V.3 Scaling Properties in Matter~~~~~~~~~~~~~~~~~~~~~~~~~~~~~~~~~~~~~~~~~~~~~~~~~~~~~~~~~~~~~~~~~~~~~~~~~~~~40\\
\indent
V.4 Scalar meson dominance~~~~~~~~~~~~~~~~~~~~~~~~~~~~~~~~~~~~~~~~~~~~~~~~~~~~~~~~~~~~~~~~~~~~~~~~~~~~~~~~~41\\
\indent
V.5 Entropy for the Hadronic Ground State~~~~~~~~~~~~~~~~~~~~~~~~~~~~~~~~~~~~~~~~~~~~~~~~~~~~~~~~~~~~~~~~~~~~~~~~~~~~44\\
~\\
\noindent 
VI. Conclusions, Deductions and Evaluations~~~~~~~~~~~~~~~~~~~~~~~~~~~~~~~~~~~~~~~~~~~~~~~~~~~~~~~~~~~~~~~~~46\\
\indent
VI.1  Summary of Results and Related Ideas~~~~~~~~~~~~~~~~~~~~~~~~~~~~~~~~~~~~~~~~~~~~~~~~~~~~~~~~~~~~~~~~~~~~~~~~~~~46\\
\indent
VI.2 Implications from the Analysis~~~~~~~~~~~~~~~~~~~~~~~~~~~~~~~~~~~~~~~~~~~~~~~~~~~~~~~~~~~~~~~~~~~~~~~~~~~~~~~~~~47\\
\indent
VI.3 Applications to Physical Processes~~~~~~~~~~~~~~~~~~~~~~~~~~~~~~~~~~~~~~~~~~~~~~~~~~~~~~~~~~~~~~~~~~~~~~~~~~~~~~48\\
~\\
\noindent 
Acknowledgements~~~~~~~~~~~~~~~~~~~~~~~~~~~~~~~~~~~~~~~~~~~~~~~~~~~~~~~~~~~~~~~~~~~~~~~~~~~~~~~~~~~~~~~~~~~~~~~~51\\  
~\\
\noindent 
Appendix A: Phenomenological Models for Confinement~~~~~~~~~~~~~~~~~~~~~~~~~~~~~~~~~~~~~~~~~~~~~~~51\\
\noindent 
Appendix B: Mathematical Forms relating to Relevant Physical Currents~~~~~~~~~~~~~~~~~~~~~~~~~52\\
~\\
\noindent 
References~~~~~~~~~~~~~~~~~~~~~~~~~~~~~~~~~~~~~~~~~~~~~~~~~~~~~~~~~~~~~~~~~~~~~~~~~~~~~~~~~~~~~~~~~~~~~~~~~~~~~~~~~~~55
\newpage
\noindent
{\Large{\bf{I. Introduction: Thermodynamics of Strong Interactions}}}\\
~\\
\noindent
We start our considerations after this opening discussion by describing 
the numerical calculations on the lattice using the essential properties 
of quantum chromodynamics (QCD) as the basic theory for the evalution 
of the equation of state for thermodynamical systems of elementary 
particles with strong interactions. The interest in the results discussed 
in this report has grown greatly due to the evaluations of the heavy ion 
experiments performed at CERN and BNL. Nevertheless, the details of our
investigations are not directly dependent upon the outcomes of any given
experiment.\\ 
~\\
\indent
    Any opening discussion of QCD calculations on the lattice begins with the
original work of Kenneth Wilson~\cite{Wil74} which is followed by the numerical 
simulations of Michael Creutz~\cite{Creu83} as the important early developments 
in the field. In their early works the formulation of the lattice gauge theory
and the suitable methods for evaluation were developed. Although we shall not
discuss the actual numerical methods in detail, we shall try to indicate the 
approach. Since QCD is a gauge invariant theory, an important issue is the
advancement of methods which uphold this property on discrete space-time points. 
Thereafter we can continue with a more general discussion of the methods and 
results~\cite{MonM96}, which we today can associate with the very extensive 
numerical evaluations in QCD at finite temperatures and densities. In the 
following sections of this report we shall discuss first the numerical 
evaluations for pure gauge theory at finite temperature with only heavy 
static quark sources, following which we include the dynamical quarks in 
the thermodynamics~\cite{LeBel96,Roth92}. Throughout this work we shall 
concentrate on writing the lattice QCD results in terms of the actual 
physical variables which are not simply the ratios of the measurable 
thermodynamical quantities. In this context a number of 
new quantities or variables have been introduced for the needs 
of these simulations on the lattice. In most cases these 
lattice quantities can be readily related to the corresponding 
quantites appearing in the usual quantum thermal field theory. In the cases
of the thermodynamical quantities or densities it has been a custom to set in
ratios of the densities for the sake of the computation instead of the actual 
physical variables of the continuum field theories. In this context we must 
explain why the defined~\cite{EKSM82,EFRSW89} lattice quantity $\Delta(T)$, 
which is commonly called the {\it{interaction measure}} written as 
$(\varepsilon~-~3p)/T^4$. It replaces the actual physical quantities 
in the equation of state of the form $\varepsilon-3p$, where these 
are defined as the (internal) energy density $\varepsilon$ and the 
pressure $p$. From the lattice point of view the pressure ratio 
$p/T^4$ may be easily and accurately computed by direct integration
techniques~\cite{EFKMW90}. However, the quantity $\Delta(T)$ involves the 
renormalization group beta-function which was somewhat later for $SU(2)$ 
gauge theory well computed~\cite{EKR94}. This work led to a useful general
procedure for the computation of the thermodynamical functions~\cite{Boyd} for
lattice gauge theories with $SU(N_c)$ symmetries with different numbers of
colored states, which of particular physical interest is clearly $N_c = 3$.\\
~\\
\indent
     Above and beyond the pure numerical computations is fact that we must 
declare how these actual physical quantities may be related to the expected phase 
transitions which have been generally believed to take place. This consideration 
has been an active subject over the last quarter century~\cite{Satz} under the title 
of {\it{Quark Matter}} for which the disordered phase is oftentimes called the 
{\it{Quark Gluon Plasma}} (QGP). The relationship to the lattice computations as 
well as to perturbative QCD has been previously discussed~\cite{CleGS}. In the very 
simplest picture an ideal gas of hadrons is converted into an ideal gas of quarks 
and gluons. This picture fails to explain in any way why such a conversion 
should or even could take place. The first obvious improvement to this highly 
oversimplified model is the introduction of the bag constant $\mathcal{B}$, 
which offers a means of holding the quarks and gluons together inside the 
hadrons (see Appendix A: Phenomenological Models for Confinement).
The {\it{MIT bag model}}~\cite{ChJJThW} provides a constant vacuum 
energy density $\varepsilon~=~+{\mathcal{B}}$ and a vacuum pressure of 
$p~=~-{\mathcal{B}}$ for the hadrons in all directions. This special assumption 
makes the ground state of the hadrons more favorable than that of the free 
quark-gluon gas. Out of these two conditions on $\varepsilon$ and $p$ the trace 
condition on the hadronic ground state average of the energy momentum tensor
${\langle{\theta}^{\mu}_{\mu}\rangle}_0~=~4{\mathcal B}$ has been previously 
presented for the bag model~\cite{BJBP}. We recall that the presence of a finite
trace yields a breaking of the scale and conformal symmetry. Thus it is with 
these special properties of the bag model that the statistical theory for the 
hadrons with the inclusion of the strong interactions really starts.\\
~\\
\indent
   For QCD at low energies there arises another important issue which comes out 
of the spontaneous breaking of the chiral symmetry~\cite{Shur,DoGoHo}. It is known 
from atomic physics with many particles that the Nambu-Goldstone modes appear in the 
ground state when there is a spontaneous symmetry breaking\footnote[1]{A very recent 
discussion on this subject by Fran\c{c}ois Englert with the title "Broken Symmetry 
and Yang-Mills Theory" can be found in Gerardus 'tHooft's collection~\cite{GtH}.} 
relating to a conserved current. In many known cases like the ferromagnet and the 
superconductor these modes remain with the quantum state even at finite temperatures. 
From the standard model of elementary particle physics the relation of this 
phenomenon to the lattice data will be discussed more extensively using a model 
in a later section. In this manner these modes are generally understood~\cite{DoGoHo} 
using the sigma model with the $SU(2)$ chiral symmetry for very low energies. 
Within small corrections to isospin symmetry the light quarks have the pion 
as the approximately massless Nambu-Goldstone mode, for which the vacuum 
expectation value $\langle{\bar{q}q}\rangle_0$ has the same value for both 
of the light quarks. If the masses of the light quarks were taken to be zero, 
the pion would be the true Goldstone boson with $M_{\pi}=0$. In the low energy 
limit with quarks of masses $m_u$ and $m_d$ the pion mass can be written as 
$M_{\pi}=(m_u+m_d)b_0$ and $\langle{\bar{q}q}\rangle_0=-F^2_{\pi}b_0$, where
$b_0$ is a positive constant of dimension energy and  $F^2_{\pi}$ is the
square of the pion decay constant~\cite{DoGoHo}.\\
~\\ 
\indent
     Along this same line another important example of this type of
approach was used long ago in the discussion of the evidence for the
scalar meson dominance by Peter G. O. Freund and Yoichiro Nambu~\cite{FreuNam}. 
Out of this discussion we note the possiblilty of mesonic bound states at high 
temperatures. In this approach the trace of the energy momentum tensor is coupled 
through the Klein-Gordon wave equation to a single massive scalar field. 
In this early work they provide an effective Lagrangian for the dominance 
of the scalar meson. We shall develop this topic further in a later section. 
Another model with spontaneous symmetry breaking which relates to the low
energy properties of QCD is the Nambu-Jona-Lasinio Model~\cite{NamJon,Kugo}.\\
~\\
\indent 
     In the course of this article we shall look at some of the work of the more
recent past involving the transitions between the hadronic states and the 
the long speculated QGP. It involves both the restoration of the chiral symmetry
as well as the deconfinement transition~\cite{NoRhoZah,AdHatZah}. We start our 
ideas with an development closely tied to some earlier work~\cite{BJBP} on the
hadron to quark-gluon phase transitions motivated by the relativistic 
heavy-ion collisions. Here we expect the properties of the quark condensate
to have a significant part in the development of any new phase. The rules
for the scaling properties then relate to the surrounding medium as proposed
some time ago by Gerry Brown and Mannque Rho~\cite{BroRho91,BroRho02} for the 
ratios of the decay constants and the masses in the presence of the surrounding 
medium to those of the vacuum. In this context we shall look into the use of the 
effective Lagrangian approach which will be used  primarily in relation to the 
restoration of the chiral symmetry. Furthermore, we note that some more recent 
work on the nature of the chiral restoration transition~\cite{BGLR} has been 
performed. A related discussion arises with the chiral bag 
model~\cite{NoRhoZah,ZahBro}, where the action 
is constructed in such a way that it is invariant 
under global chiral rotation (see Appendix A for more detail). 
This model is an extension of the usual bag model which had ignored 
the properties of chiral symmetry~\cite{NoRhoZah}. \\
~\\
\indent
     Next we note the importance of some work on the use of the QCD sum rules 
at low temperatures~\cite{DoGoHo,AdHatZah}. This work was then related to some 
earlier numerical lattice simulations at finite temperatures~\cite{kochbrown} 
which involved the structures of the the electric and magnetic condensates
separately. Along this line we should mention the important distinction 
between the "hard" and "soft" glue arising from the type of breaking of the 
different symmetries. Now it is quite necessary to note that there are {\it two} 
different types of symmetry breaking involved--that mentioned above as the 
spontaneous breaking involving the chiral symmetry 
and that which appears as the anomalous breaking of the
conformal symmetry. The spontaneous breaking of chiral symmetry already
appears in the hadronic ground state with the destruction of the chiral 
invariance in the axial current. It involves the operator average in the 
hadronic ground state of the form ${\langle \bar{\psi}_q{\psi}_q \rangle}_{0}$
which has lower field dimension than the lagrangian density since the quark-antiquark
pair $\bar{\psi}_q{\psi}_q$ alone has the operator dimension three. However, 
the anomalous breaking of the scale and conformal symmetry arises from 
the square of the gluon field strength tensor, which we write 
symbolically as ${\langle G^2 \rangle}_{0}$ in the hadronic ground state. 
It possesses the field dimension four. In the more general context 
it appears that with the loss of conformal symmetry, which we 
shall later see relates to the gluon condensate itself. It is {\it never} 
really restored even at very high temperatures. In the finite temperature field 
theory~\cite{LeBel96} another type of breaking occurs from the renormalization
group equation at finite temperatures. The effect of the finite temperature
renormalization first takes out the vacuum gluon condensate. Then, as it was 
clearly stated by Heinrich Leutwyler~\cite{Leut}, it then continues to 
decondense with the increasing temperatures. This particular situation 
we shall discuss in the following paragraphs.\\
~\\
\indent 
      Here we discuss further the ideas concerning the two different types 
of symmetry, which have been recently related to the problem of the mass in the 
mesonic bound states~\cite{PaLeBr}, which we shall discuss later in relation 
to the scalar meson dominance~\cite{FreuNam}. The study of the breaking 
of the chiral symmetry in gauge theories has had a very long history 
in quantum field theory. It had already arisen in other models well 
before QCD. The anomalous electromagnetic decay of the neutral pion, 
that is $\pi^0~\rightarrow~2\gamma$, served as a major problem even before 1950. 
Its relationship to the study of anomalies was realized later in 1969 from
spinor electrodynamics by Stephen Adler~\cite{Adler,AdBar} and the work of 
John Bell and Roman Jackiw on the nonlinear sigma model~\cite{BellJack}. 
It was later recognized as the anomalous breaking of chiral symmetry in QCD, 
which is usually known as the chiral or the ABJ anomaly~\cite{TreJaZuWit,Bert,Jack}. 
It represents the anomalous divergence\footnote[2]{A very interesting discussion 
of both the chiral and the trace anomalies has been recently written by Stephen Adler 
entitled "Anomalies to all orders", in reference~\cite{GtH}.}
of the axial current arising out of an axial Ward identity~\cite{Bert}. 
In contrast to the conformal or trace anomaly, which will be very essential 
to all further discussions of the physical equation of state in QCD, 
the chiral or the ABJ anomaly had always a higher rating in its acceptance 
because of the clear advantage from the already well known experimental 
verifications using the decays of the $\pi^0$ as well as that of such 
hadronic processes as the $K^{+}$ going into pions~\cite{Bert}. \\
~\\
\indent
     The discovery of anomalous terms appearing as a finite value of the trace
of the energy momentum tensor was pointed out as a result of nonperturbative
evaluations in low-energy theorems~\cite{nonpert} many years ago. Furthermore,
it was also somewhat later realized how this factor arose with the process
of renormalization in quantum field theory which became known as the trace 
anomaly~\cite{tracanom,ColDunJog,Niel} since it was found in relation 
to an anomalous trace of the energy momentum tensor. The basic idea of 
the relationship between the trace of the energy momentum tensor and 
the gluon condensate has already been studied for finite temperature 
by Leutwyler~\cite{Leut} in relation to the problems of deconfinement and 
chiral symmetry. The starting point of this work begins with a detailed 
discussion of the trace anomaly based on the interaction between the 
Goldstone bosons in chiral perturbation theory. Quite central to his 
investigation is the role of the energy momentum tensor averaged over 
all the states, whose trace is directly related to the averaged gluon 
field strength squared. 
Here it is important to state that the averaged total energy 
momentum tensor $T^{\mu\nu}(T)$ can be separated into the vacuum or   
confined part, $\theta^{\mu\nu}_{0}$, involving only the temperature 
independent states in the average, and the finite temperature 
contribution $\theta^{\mu\nu}(T)$ as follows:
\begin{equation}
  \label{eq:emtensor}
  T^{\mu\nu}(T) = \theta^{\mu\nu}_{0} + \theta^{\mu\nu}(T) .
\end{equation}
\noindent
The temperature independent part, $\theta^{\mu\nu}_{0}$, has the standard 
problems with infinities of any ground state, which has been previously 
discussed~\cite{SVZ1} in relation to the nonperturbative effects in QCD 
and the operator product expansion. In the following analysis we shall 
start our discussion with a bag type of model~\cite{ChJJThW,BJBP} as a means 
of stepping around these difficulties with the QCD vacuum since at this 
time we are primarily interested in the thermal properties\footnote[3]{For 
a discussion of the process of gluon condensation in relation to confinement 
see the contribution of David Pottinger in the collection on the 
{\it{statistical mechanics of quarks and hadrons}}~\cite{Satz}.}. The 
finite temperature part, which clearly vanishes at zero temperature, 
has no such problems with the divergences. We shall discuss in the following 
sections of this report how at finite temperatures the diagonal elements 
of $\theta^{\mu\nu}(T)$ are calculated in a straightforward way on the 
lattice. Furthermore, the trace  $\theta^{\mu}_{\mu}(T)$ is connected 
to the thermodynamical contribution to the internal energy density 
$\varepsilon(T)$ and pressure $p(T)$ for relativistic fields as well as 
for relativistic hydrodynamics in the following simple form:
\begin{equation}
  \theta^{\mu}_{\mu}(T) = \varepsilon(T) - 3p(T) .
  \label{eq:epstemp}
\end{equation}
\noindent
The gluon field strength tensor including the coupling $g$ is denoted by
$G^{\mu\nu}_a$, where $a$ is the color index for $SU(N_c)$. Thus the
basic equation for the relationship between the gluon condensate
and the trace of the energy momentum tensor at finite temperature was
written down by Leutwyler \cite{Leut} using the trace anomaly.
Leutwyler's equation takes the following form:
\begin{equation}
\langle G^2 \rangle_{T} = \langle G^2 \rangle_0~-~
{\theta}^{\mu}_{\mu}(T),
\label{eq:condef}
\end{equation}
\noindent
for which the brackets with the subscript $T$ mean thermal average and
the zero the ground or confined state average.
The renormalized gluon field strength tensor is squared inside of the 
brackets which is then summed over all the colors to yield 
\begin{equation}
G^{2}~=~{{-\beta(g)}\over{2g^3}} G^{{\mu}{\nu}}_{a}G_{{\mu}{\nu}}^{a}.
\label{eq:glucondop} 
\end{equation}
\noindent
The renormalization group beta function $\beta(g)$ in terms of the
coupling may be generally written as
\begin{equation}
\beta(g)~=~\mu{dg \over{d\mu}}
         =~-{1 \over{48\pi^{2}}}(11N_{c}~-~2N_{f})g^{3}~+~O(g^{5}).
\label{eq:betafun}
\end{equation}
\noindent
Out of these relationships
Leutwyler~\cite{Leut} has calculated the trace of the energy momentum 
tensor at finite temperature for two massless quarks using the low 
temperature chiral expansion .\\
~\\
\indent
     The most immediate generalization of Leutwyler's equation~(\ref{eq:condef}) 
has been previously considered in an earlier work~\cite{Mil99}. In the presence 
of {\it{massive}} quarks the averaged trace of the energy-momentum tensor takes 
the following form from the trace anomaly:   
\begin{equation}
 \theta^{\mu}_{m \mu}~=~m_q\langle \bar{\psi}_q{\psi}_q \rangle~
       +~\langle G^2 \rangle,
\label{eq:thetaquark} 
\end{equation}
\noindent
where $m_q$ is the renormalized quark mass and ${\psi}_q$,
$\bar{\psi}_q$ represent the quark and antiquark fields respectively.
As an operator relation this equation (unaveraged) would not make
any sense, since these three operators have different operator 
dimensions and also carry different symmetries therein.
We include with these averages the renormalization group functions
$\beta(g)$ and $\gamma(g,m)$, which appear in this trace from both the 
coupling and mass renormalization processes. This averaged form holds 
for both the confined and temperature dependent structures. We shall
use these properties in relation to the massive dynamical quark
lattice simulations in Section III.\\ 
~\\
\indent
     We start with a discussion of previous evaluations~\cite{BoMi} 
from an earlier collaboration with Graham Boyd. Originally 
the motivation for this work was just to study the pure 
lattice gauge theory data for both the $SU(2)$~\cite{EKR94} and 
the $SU(3)$~\cite{Boyd} simulations, which had at that time been 
recently finished in Bielefeld. Of particular interest at that time 
was the relation of the equation of state to the pure gluon condensate 
above the deconfinement temperature, which had been in both cases 
very accurately computed and denoted as $T_c$. As it had been stated 
previously by Leutwyler~\cite{Leut}, we did, indeed, find that 
with increasing temperatures the pure gluon condensate was unbounded 
from below over its range of negative values . This observation was 
quite contrary to the then commonly accepted ideal gas models which 
were supposed  to appear at high temperatures because 
of "asymptotic freedom" in QCD. Actually in our present understanding
we know that the important property of QCD, asymptotic freedom, is already
in the renormalization group beta function whose stable ultraviolet
fixed point determines the critical behavior at $T_c$.  During the time that 
this earlier work~\cite{BoMi} was being carried out there appeared some 
numerical lattice computations from the MILC collaboration~\cite{BKT94,MILC96}
which included {\it two light} dynamical quarks. At the same time
in Bielefeld~\cite{Eng5} there were similar computations for four 
flavored not so light dynamical quarks. These results we included at
the end of our work~\cite{BoMi}(see our Figure 4 where these results
are compared to a rescaled $SU(3)$ curve). We noted, but could
not explain why, that the heavier Bielefeld data followed the
rescaled pure gauge curve, while the MILC data~\cite{BKT94,MILC96} 
remained well above it. There was then, as there is now~\cite{MILC97}, 
the problem that this data for the two light quarks did not go to a 
high enough value in the temperature range to make such comparisons clear.
Although this particular work~\cite{BoMi} never got published, 
it did, nevertheless, serve as a starting point in further works 
~\cite{Mil97,Mil99,Mil00}.\\
~\\
\indent
     An important result of the investigation of numerical lattice 
simulations was already stated by Leutwyler~\cite{Leut} with some 
of the previously cited numerical work. This was the fact that the 
dilatation current, given by $D^{\mu}=x_{\nu}T^{{\mu}{\nu}}$, and the 
four conformal currents $K^{\alpha {\mu}}(x)$ are not conserved quantities 
even at very high temperatures. The four conformal currents~\cite{Jack} 
are given by
\begin{equation}
K^{\mu\alpha}~=~(2{x^{\alpha}}{x_{\nu}}~-~g^{\alpha}_{\nu}x^2)T^{\mu\nu}.
\label{eq:specconfcurr}
\end{equation} 
\noindent
The form of the equations for these currents was first written down by Erich 
Bessel-Hagen~\cite{BesHag} as the result of a seminar series in 1920 at 
G\"ottingen led by Felix Klein . The equation for the dilatation current 
takes the form
\begin{equation}
\partial_{\mu} D^{\mu}~=~T^{\mu}_{\mu}.
\label{eq:dilcurr}
\end{equation} 
\noindent
A similar equation can be written for the divergence of the special conformal
currents
\begin{equation}
\partial_{\mu}K^{\mu\alpha}~=~2{x^{\alpha}}T^{\mu}_{\mu},
\label{eq:specconf}
\end{equation} 
\noindent
It is important to note that the nonconservative aspect of both these
currents $D^{\mu}(x)$ and $K^{\mu\alpha}(x)$ relate directly to the fact
that $T^{\mu}_{\mu}$ remains finite. We shall discuss some aspects of
these currents with special solutions as well as the corresponding
differential and integral forms in Appendix B. However, in this report 
it is our main interest to discuss these quantities  
from the point of view of lattice data.\\
~\\ 
\indent
     We shall show in the very next section how 
the actual computed thermodynamical functions
appear as a function of the temperature in physical units. Thereby
in the next part of this report we discuss how to use the lattice 
data for pure gauge theories\footnote[4]{In this report we shall only 
introduce the notation needed for the lattice evaluations. Here we will
not discuss the usual field theoretical notation for QCD. We shall follow
the standard texts on gauge fields such as~\cite{DoGoHo,Kugo,Muta,Poko}. For 
discussions of the ideas and their development see~\cite{GtH}.}. 
In particular the method for the evaluation~\cite{EKSM82,EFRSW89} 
of the lattice quantities as the ratios $p/T^4$, $\varepsilon/T^4$ 
and $\Delta(T)$ is the real starting point for the discussion of 
the thermodynamical quantities leading to the form of the equation of state. 
It symbolizes the change of scale breaking from the pure vacuum 
contributions to the dominance of the high temperature region 
in the thermodynamics. Furthermore, it is significant that just 
above the deconfinement temperature $T_d$ the relative breaking is 
the largest as seen from the peak in $\Delta(T)$ with the gradual decline
thereabove. The actual equation of state $\varepsilon(T) - 3p(T)$ shown in the 
next section demonstrates this fact much less dramatically just above $T_d$
since its rise is very sharp there which at much higher temperatures then 
simply slows down. In the following section we consider the inclusion of
dynamical fermions in the numerical computations. Although we shall generally
mention some of the studies with finite chemical potentials on the lattice
as well as later consider some special cases as examples, we will not take
up the details of the investigations here. Instead, in the next section we
shall provide numerical analyses of various cases for the finite temperature 
behavior of the quark and gluon condensates relating to the breaking of
chiral and conformal symmetries. The last major section attemps to put
together these numerical results with the contemporary understanding in
nuclear and high energy physics at high temperatures. 
A short section with concluding remarks is followed 
by two appendices with added details. 
~\\
\newpage
\noindent
{\Large{\bf{II. Thermodynamics of Gauge Theory on the Lattice}}} \\
~\\
Here we begin by defining the physical quantities in terms of the lattice
varibles~\cite{Wil74} used in the following parts for the lattice gauge 
computations. First we discuss the thermodynamics of the pure Yang-Mills 
fields as it is computed on the lattice~\cite{Creu83,MonM96} in the 
canonical ensemble in the formalism necessary for finite lattice sizes. 
A basic fact of the pure gauge theory is the existence of a transition 
temperature often known as the deconfinement temperature or $T_d$. 
For the $SU(2)$ invariant gauge theory this transition temperature is 
often called the critical temperature $T_c$ since this transition 
is of second order~\cite{EKSM82,EKR94}. However, for the $SU(3)$
invariant gauge theory it is a first order~\cite{EFRSW89,EFKMW90,Boyd} phase
transition which goes between the "confined" and the "deconfined" states.
In this case the critical temperature acts as the highest temperature
where there is a distinction between these phases\footnote[5]{The presence 
of a transition temperature to a high temperature phase was suggested many
years ago from renormalization group arguments. This phase was regarded as 
where the gauge coupling became weaker~\cite{ColPer}}. In this part we 
go more specifically into the numerical results of the actual lattice 
computations for the thermodynamical functions using lattice 
gauge theory at finite temperatures~\cite{EKR94,Boyd}.\\
~\\
\noindent
{\large{\bf{II.1 Lattice Thermodynamics}}}\\
~\\
\noindent 
As it is usually done in statistical physics, we start with the canonical 
partition function ${\cal{Z}}(T,V)$ for a given temperature T and spatial 
volume $V$. From this quantity we may define the pressure $p$ for large
homogeneous systems in thermal equilibrium through its relation to the 
free energy density $f$ as follows:
\begin{equation}
     p~=~-f~=~{T\over{V}}{\ln{\cal{Z}}(T,V)}.
\label{eq:freen}
\end{equation}
The volume $V$ is determined by the lattice size $N_{\sigma}a$, where $a$ is
the lattice spacing and $N_{\sigma}$ is the number of steps in the given
spatial direction. The inverse of the temperature $T$ is determined by
$N_{\tau}a$, whereby $N_{\tau}$ is the number of steps in the (imaginary)
temporal direction. Thus the simulation~\cite{EFKMW90,EKR94,Boyd} 
is done in a four dimensional Euclidean space with the given 
lattice sizes ${N_{\sigma}^3}\times{N_{\tau}}$, which gives the 
volume $V$ as $(N_{\sigma}a)^3$ and the inverse temperature $T^{-1}$ as 
$N_{\tau}a$ for the four dimensional Euclidean volume.\\ 
~\\
\indent
     In the early lattice simulations for the $SU(2)$ gauge theory one took 
for the actual lattice sizes up to $N_{\sigma}=24$ in the spatial directions 
and the temporal sizes $N_{\tau}=4,6$~\cite{EFRSW89,EFM92}. Later for the 
$SU(3)$ lattice simulations the spatial values were taken as $N_{\sigma}=16,32$ 
with the temporal sizes $N_{\tau}=4,6,8$~\cite{Boyd}, which we shall show here. 
In general for $SU(N_{c})$ gauge theory the lattice spacing $a$ is a function 
of the bare gauge coupling $\beta$ defined by $2N_{c}/g^2$, where g is the bare 
$SU(N_{c})$ coupling. Thereby this function fixes both the temperature and the 
volume at a given coupling. Now we write $P_{\sigma,\tau}$ as the expectation 
value of, respectively, space-space and space-time plaquettes in terms of the
link variables $U_{i}$
\begin{equation}
    P_{\sigma,\tau}~=~1~-~{1\over{N_c}}
     {Re{\langle{Tr(U_{1}U_{2}U^{\dagger}_{3}U^{\dagger}_{4})}\rangle}}
\label{eq:plaq}
\end{equation}
\noindent
for the usual Wilson action~\cite{Boyd}. These plaquettes $P_{\sigma,\tau}$ 
may be generalized to the improved actions on anisotropic 
lattices~\cite{EngKarSch} for SU(2) and SU(3). For the  
Wilson action we define the parts $S_{0}~=~6P_{0}$ on the 
symmetric lattice $N^{4}_{\sigma}$ and $S_{T}~=~3(P_{\sigma}+P_{\tau})$ 
on the asymmetric lattice ${N^{3}_{\sigma}}\times{N_{\tau}}$. We now 
proceed to compute the free energy density ratio defined above in the
equation (\ref{eq:freen}) by integrating these expectation values as
\begin{equation}
 {f(\beta)\over{T^4}}~=~-{N^{4}_{\tau}}{\int\limits_{\beta_{0}}^{\beta}}
                        {d{\beta'}[S_{0}~-~S_{T}]},
\label{eq:frenplaq}
\end{equation}
where the lower bound $\beta_{0}$ relates to the constant of normalization.
At this point we should add that the free energy density is a fundamental
thermodynamical quantity from which all other thermodynamical quantities
can be gotten. Also it is very important in relation to the phase structure
of the system in that the determination of the transitions for their order
and critical properties as well as the stability of the individual phases 
are best studied. The integral method~\cite{EFKMW90} for the computation of 
the pressure ratio yields as a result from the equations (\ref{eq:freen}) and 
(\ref{eq:frenplaq}) given by
\begin{equation}
    {p\over{T^4}}~=~{p\over{T^4}}\Bigg|_{\beta_0}~+~
{N^{4}_{\tau}}{\int\limits_{\beta_{0}}^{\beta}}{d{\beta'}[S_{0}~-~S_{T}]}.
\label{eq:pressplaq}
\end{equation}
\noindent
The integral method provides an approach for the evaluation of the pressure
ratio which is free from many of the problems arising in the evaluation 
of the ratios of other thermodynamical functions as well 
as assuring that the value of the pressure ratio always be 
positive as distinguished from many of the earlier evaluations.
For a general discussion see reference~\cite{MonM96}.\\ 
~\\
\indent
     Next we define the lattice beta function in terms of the lattice 
spacing $a$ and the coupling $g$ as follows:
\begin{equation}
  {\tilde{\beta}(g)}~=~-2{N_c}{a\frac{dg^{-2}}{da}}.
\label{eq:betlat}
\end{equation}
\noindent
The dimensionless interaction measure $\Delta(T)$ discussed in the Introduction
~\cite{EKSM82} is then given by
\begin{equation}
  \Delta(T) = {N^{4}_{\tau}}~{\tilde{\beta}(g)}~{\left[S_{0}~-~S_{T}\right]}.
\label{eq:intactmeas}
\end{equation}
\noindent
The crucial part of the more recent lattice gauge calculations is the use 
of the full lattice beta function, $\tilde{\beta}(g)$ in obtaining
the lattice spacing $a$, or the scale of the simulation, from the 
coupling $g^{2}$. Without this accurate information on the temperature 
scale in lattice units it would not be possible to make any claims about 
the behavior of the gluon condensate discussed in detail in Part IV. The 
interaction measure $\Delta(T)$ is the thermal ensemble expectation value given 
by $(\epsilon - 3p)/T^4$. Thus because of equation (\ref{eq:epstemp}) above 
the trace of the temperature dependent part of the energy momentum tensor,
here denoted by $\theta^{\mu}_{\mu}(T)$ is equal to the expectation value
of $\Delta(T)$ multiplied by a factor of $T^4$. This physical quantity may 
be directly computed~\cite{BoMi,Mil97} as a function of the temperature as
\begin{equation}
  \theta^{\mu}_{\mu}(T)~=~\Delta(T){\cdot}T^4.
\label{eq:trace}
\end{equation}
\begin{figure}[b!]
\begin{center}
\epsfig{bbllx=127, bblly=265, bburx=451, bbury=588,
file=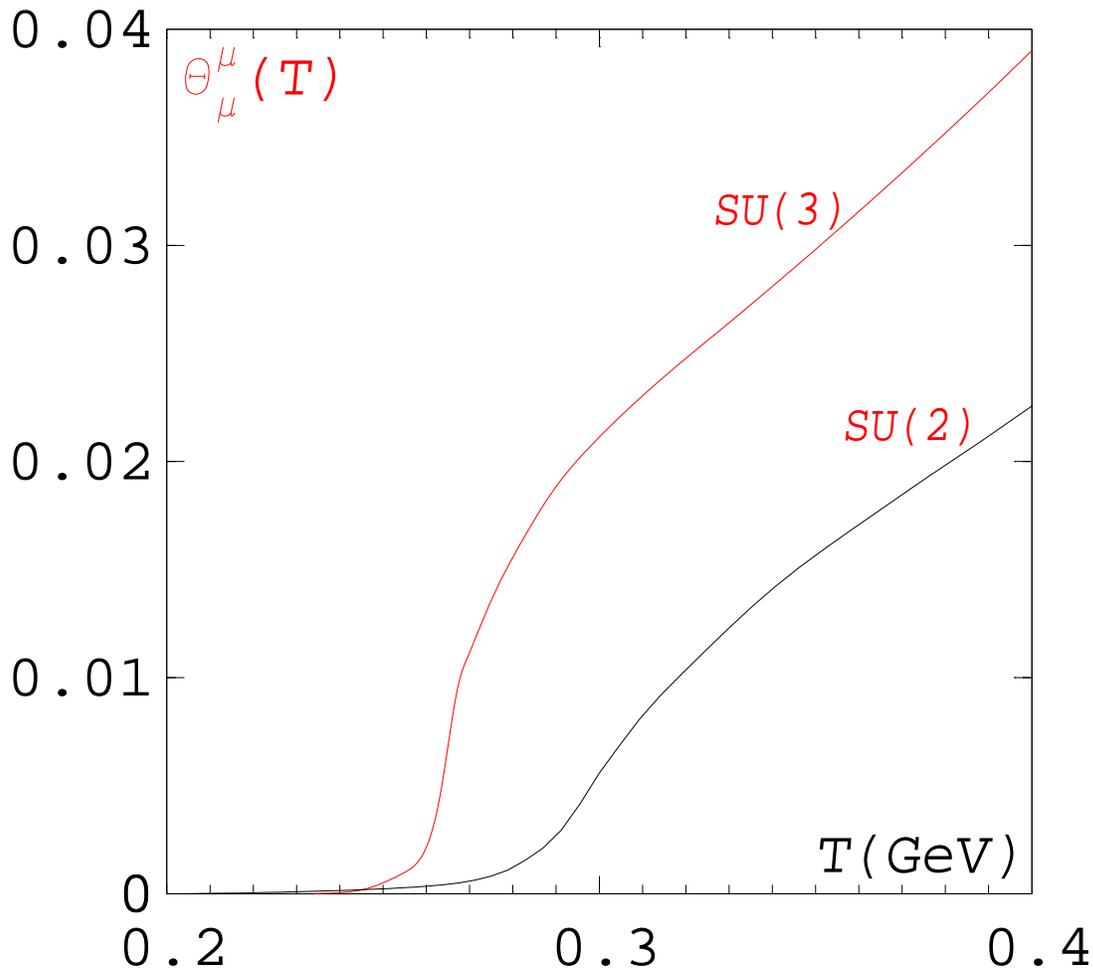, width=11.5cm}
\\[1.0cm]
\caption{\it We compare the computed values of trace of the energy momentum tensor
in the units $[GeV^4]$ for the lattice gauge theories  $SU(2)$ and $SU(3)$ as 
indicated. The values for the deconfinement temperature are $T_d = 0.290, 
0.264$~GeV for $SU(2)$ and $SU(3)$, respectively.}   
  \label{fig:eqstgauge}
\end{center}
\end{figure}
\noindent
There are no other contributions to this trace for the pure gauge fields on 
the lattice. The heat conductivity is zero. Since there are no other finite 
conserved quantum numbers and, as well, no velocity gradient in the lattice 
computations, hence no contributions from the viscosity terms appear. For
a scale invariant system, such as a gas of free massless particles, then the 
trace of the energy momentum tensor in equation (\ref{eq:trace}) is clearly zero.
However, any system that is scale variant, perhaps from a particle mass, 
has a finite trace, whereby the value of the trace then measures the magnitude 
of scale breaking. These results are shown in figure~\ref{fig:eqstgauge}.\\
~\\
\indent
     We now discuss a few other important properties of the pure gauge 
theories on the lattice. The order parameter is often taken as the lattice 
average of the Polyakov loop~\cite{MonM96} , which in many cases turns out 
to give more information as the order parameter than does the general Wilson 
loop. The defining operator for the Polyakov loop is given by
\begin{equation}
  L(\vec{x}) = {{1}\over{3}}{\prod}^{N_{\tau}}_{x_4=1}U_4(\vec{x},x_4),
\label{eq:polyop}
\end{equation}
where the index 4 stands for the euclidean time direction. Then the actual
Polyakov loop is defined by the expectation value as
\begin{equation}
  L = {{1}\over{{N_{\sigma}}^3}}\langle{\sum}_{\vec{x}}L(\vec{x})\rangle.
\label{eq:polyloop}
\end{equation}
Since a vanishing expectation value is generally a signal of an exact symmetry,
one oftentimes defines the absolute value of the quantity $\bar{L}$ as
\begin{equation}
\bar{L} = {{1}\over{{N_{\sigma}}^3}}\langle|{\sum}_{\vec{x}}L(\vec{x})|\rangle,
\label{eq:polyloopabs}
\end{equation}
which serves as an approximate order parameter. For the pure gauge 
theories one can determine the critical couplings $\beta_c$ from 
the analysis of the Polyakov loop susceptibility $\chi_L$,
\begin{equation}
{{\chi}_{L}}~=~N^{3}_{\sigma}{\left[ \langle L^2 \rangle~
-~{\langle L \rangle}^2 \right]},
\label{eq:poloopsusc}
\end{equation} 
~\\
\noindent
which we shall later discuss in more detail in conjunction with the 
dynamical quarks in the next section. From this calculation one can 
accurately determine the value of the deconfinement temperature $T_d$.\\
~\\ 
~\\
\noindent
{\large{\bf{II.2 Thermal Field Theoretical Evaluation}}}\\
~\\
\noindent 
     In a very recent work Zwanziger~\cite{Zwanz} has carried out an 
analytical determination of the properties of the equation of state 
for the pure gluon plasma at high temperatures well above the
deconfinement temperatures. He uses the Gribov dispersion relation~\cite{Grib}
as a means of suppressing the infrared modes when gauge equivalence is 
imposed at the nonperturbative level, which takes the form
\begin{equation}
  E(k)~=~\sqrt{k^2+M^4/k^2}.
\label{eq:gribdisp}
\end{equation}
\noindent
After a brief analysis of the trace anomaly at finite temperature~\cite{Leut}
as well as comparison with the $SU(3)$ lattice data~\cite{Boyd} Zwanziger 
derives a form of the trace of the energy momentum tensor
\begin{equation}
  \theta^{\mu}_{\mu}(T) = LT~+~A,
  \label{eq:zwantrace}
\end{equation}
\noindent
where $A$ is a temperature independent term of order $O\left(1\right)$ in 
the units $[GeV^4]$ and $L$ is an integral of the form~\cite{Zwanz}
\begin{equation}
   L~=~{{3({{N^2}_c}-1)}/{\pi^2}}{\int^{\infty}_0}dkk^2\ln[{{E(k)}/{k}}],
  \label{eq:zwanint}
\end{equation}
\noindent
This integral can be exactly evaluated for the Gribov dispersion relation
(\ref{eq:gribdisp}) yielding the following result:
\begin{equation}
   L~=~{{(N^2_c-1)M^3}/{\pi\sqrt{2}}}.
  \label{eq:zwanval}
\end{equation}
\noindent
Although this result with a linear growth in the temperature for 
$\theta^{\mu}_{\mu}(T)$  had been already observed~\cite{BoMi,Mil97} 
in the lattice data, it had not been previously derived analytically 
for this energy spectrum~\cite{Zwanz}. If one looks carefully at the 
figure~\ref{fig:eqstgauge} for $\theta^{\mu}_{\mu}(T)$, one sees well 
above the deconfinement temperatures $T_d$ that both the curves begin 
to slow up to an almost linear behavior as a function of temperature T
even for the temperatures of around $2T_d$. Nevertheless, there still 
are some quite gradual changes of curvature upwards until almost $3T_d$ 
or about $T=0.8 GeV$, whereupon, it appears for higher temperatures to 
be quite linear~\cite{BoMi}. Furthermore, one sees a new scale $M$ 
in the equation (\ref{eq:gribdisp}) which has the units $[GeV]$. 
According to Zwanziger~\cite{Zwanz} it arises from the infrared 
regulator, which can be estimated to be around  $0.7GeV$. 
From a more general standpoint what is very interesting about
Zwanziger's analytical calculation of $\theta^{\mu}_{\mu}(T)$ is that 
the linear result for the temperature arises from the regulation of
the infrared behavior of QCD which dominates over this high temperature
domain. It is by now very well known that there are serious infrared 
problems in the high temperature perturbation theory~\cite{Lind}, which 
are not easily dealt with using the standard perturbation theory. These 
problems indicate that the asymptotic freedom may only be strictly 
valid in the region of short distances and times~\cite{MonM96}.
We already know that for QCD the asymptotic freedom has appeared 
in the beta function as a property of the ultraviolet structure. 
This property of QCD provides a stable ultraviolet fixed point in the 
beta function in contrast to quantum electrodynamics (QED) which has 
a stable infrared fixed point~\cite{Muta,Kugo,Poko}. However, this
fact means that QED is infrared stable in contrast to QCD. \\
~\\
\noindent
{\large{\bf{II.3 Numerical Evaluations of Physical Quantities}}}\\
~\\
\noindent
The numerical evaluation of the equation of state  at finite temperature
for strongly interacting quarks and gluons has long been the main objective 
for lattice simulations~\cite{Creu83,MonM96,Roth92}. The computed pressure 
ratio $p(T)/T^4$ for the pure SU(3) lattice gauge theory~\cite{Boyd} 
is shown as a function of temperature in their Figure 4 with 
three different values of $N_{\tau}$ in the Euclidean time direction.
They remark concerning the sizes of the lattices for the different cases 
${N_{\sigma}}^3\times{N_{\tau}}$. In what follows we take the spatial
sizes ${N_{\sigma}}a$ and temporal sizes $N_{\tau}a$ whereby $a$ is the
lattice spacing between points on the lattice. In these particular
computations~\cite{Boyd} the spatial step numbers have the two values 
${N_{\sigma}}=16,32$ while the temporal number have the three values
$N_{\tau}=4,6,8$, each of which has been explicitly studied~\cite{Boyd}.
We should note that in their Figure 4 the value for the "critical" deconfinement 
temperature called $T_c$ for the different lattice sizes has been been 
evaluated from the renormalization group beta function on the lattice.
The attained value for $T_c$ from an extrapolation to the continuum limit
is given by $T_c/\sqrt{\sigma}=0.629(3)$. The string tension is taken
for their simulations as $\sqrt{\sigma}=420MeV$, which results in a
"critical" temperature of about $T_c=264MeV$.\\
~\\
\indent
     The actual simulations~\cite{EFKMW90} for the pressure ratio in 
$SU(N_c)$ around $T_c$ was proposed by using the integral method, which 
prevented the undesirable effects in some earlier evaluations of the
pressure, in which the pressure ratio appeared to be negative near the
critical coupling~\cite{Deng,BCDGW}. Furthermore, the computational effect 
of the lattice anisotropy can be very accurately computed by using
the integral and differential methods for the anisotropy coefficients 
leading to a much higher resolution. This program has been discussed 
in detail more recently~\cite{EngKarSch} by showing how these lattice 
computations for both $SU(2)$ and $SU(3)$ are carried out. Similarly
the energy density $\varepsilon(T)$ is calculated from the pressure $p$
and the interaction measure $\Delta(T)$ to arrive at the form  
$\Delta(T){\cdot}T^4~+~3p$ as a function of temperature.
This method was briefly mentioned earlier. These numerical 
results are shown in a similar plot in their Figure 6 with the 
same basic parameters as those for the pressure ratio~\cite{Boyd}.
In their Figure 6 we can see that the energy density ratio
$\varepsilon(T)/T^4$  as a function of the temperature $T$ for pure $SU(3)$ 
gauge theory rises much faster around the critical temperature $T_c$ than 
does the pressure ratio $p(T)/T^4$ in the previously mentioned Figure 4. 
It is evaluated from the interaction measure as a function of the coupling. 
In the sense of the thermodynamics  the internal energy is generally used
to thermally describe the state of the system. Thus we may well expect 
that its derived form as a density, $\varepsilon(T)$, should be more
sensitive to the change of phase at $T_c$. From a casual view of the
number scale we are able to see that the energy density ratio curves in each 
case for $N_{\tau}$ always lie considerably above that of the pressure-- 
even above {\it three} times the corresponding numerical values~\cite{Boyd}.\\
~\\
\indent
     Now we look explicitly at the equation of state of the pure gauge theory 
which is gotten directly from $\Delta(T){\cdot}T^4$. Thus this quantity which 
we have discussed extensively in the Introduction is simply related 
to the trace of the energy momentum tensor $\theta^{\mu}_{\mu}(T)$ 
given above in the equation (\ref{eq:epstemp}). Here we plot\footnote[6]{The 
author thanks J\"urgen Engels for pointing out errors in earlier graphs 
and replacing these plots in a corrected form from the original data~\cite{Boyd}.} 
the equation of state $\varepsilon - 3p$ as a function of $T/T_c$. 
\begin{figure}[thb]
\begin{center}
\epsfig{bbllx=127, bblly=264, bburx=451, bbury=588,
file=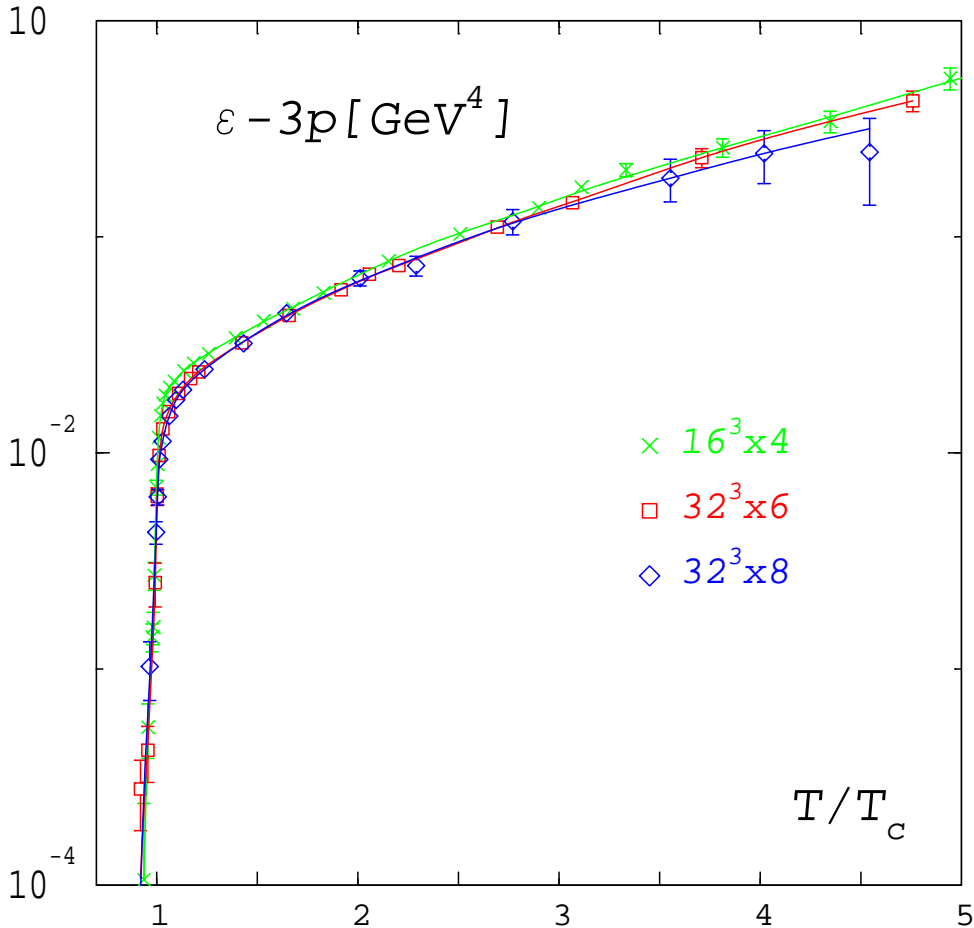, width=11.5cm}
~\\
~\\
\caption{\it The equation of state $\varepsilon - 3p$ for pure SU(3) 
gauge theory with the three lattice sizes in the physical 
units $[GeV^4]$. The temperature is given as $T/T_c$. }   
  \label{fig:eqnstate}
\end{center}  
\end{figure}   
\noindent
Furthermore, we want to point out carefully the continual growth of the 
equation of state for pure $SU(3)$ gauge theory as shown in the 
Figure~\ref{fig:eqnstate}. This fact arising from these numerical
simulations shows a behavior that is quite contrary to many of the 
common speculations on the equation of state for the quark-gluon 
plasma. We can see here no obvious signs from any of these 
computations that the dependence of this quantity $\varepsilon(T) - 3p(T)$
decreases to zero at any temperature above the "critical" deconfinement temperature 
$T_c$. In fact, one can reaffirm here with only very minor variations for 
the sizes of the different finite lattices that the high temperature linear 
$T$ dependence theoretically calculated~\cite{Zwanz} is still quite well 
upheld. Thus it is safe to conclude that the pure $SU(3)$ gauge theory in 
the computed range of temperatures remains a {\it{strongly interacting}} 
system of gluons-- {\bf{not}} an {\it ideal} ultrarelativistic gas! \\
~\\
~\\
\noindent
{\large{\bf{II.4 Comparison of the Physical Quantities}}}\\
~\\
\noindent 
In this paragraph the results of all these computations are summarized in 
the last figure in this part. In this Figure ~\ref{fig:comparison} we make 
an actual comparison using the physical units
which are written below the curves
in the square brackets. Here we can clearly see the differences between 
these computed thermodynamical quantities as functions~\cite{Boyd} 
of the temperature $T[MeV]$. In addition to the above analysed quantities 
we have included the entropy density $s[fm^{-3}]$ which is obtained 
from $(\varepsilon + p)/T$, all of which are known. However,
the entropy density $s[fm^{-3}]$  shown as the upper curve on the right 
cannot be directly compared with the others because of the units. 
The equation of state $\varepsilon - 3p[GeV/fm^3]$  as the lower curve
on the right can be compared with the energy density $\varepsilon$ and the
pressure $p$, since all are in the same units. One can see that
for these  physical quantities the relative growth of 
each at the deconfinement temperature $T_d$. Above $T_d$ each
one flattens out at different rates. This effect we can 
clearly see in the comparison for pure gauge theory in  
Figure~\ref{fig:comparison}. The growth of the equation of state
is obviously very much slower than all the other thermodynamical
quantities. This difference amounts to the comparison of a 
linear increase in the temperature to those with cubic or
quartic powers when taken on a logarithmic scale. \\
\begin{figure}[thb]
\centerline{\includegraphics[width=16.cm]{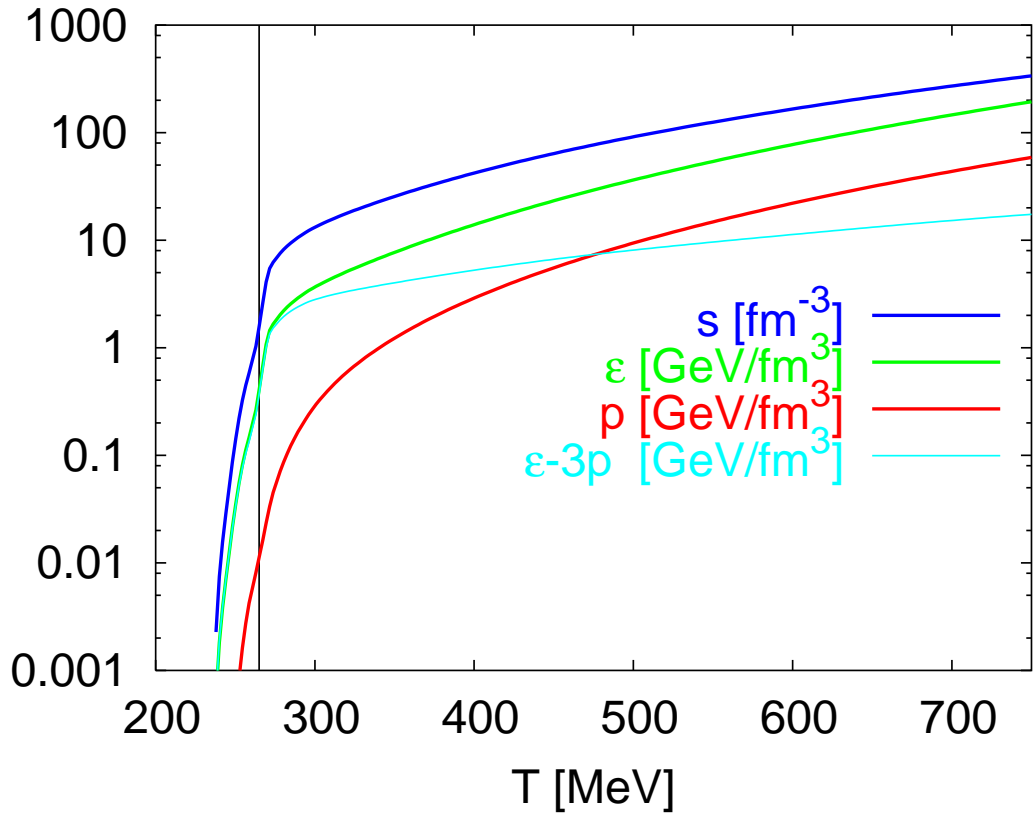}}
\caption{\it The different thermodynamical functions of the temperature 
for pure gauge theory. The vertical line is the deconfinement temperature.
Here we take for these comparisons the more useful density units 
rather than the pure energy units as in the previous figures. }   
  \label{fig:comparison}  
\end{figure}
~\\
\indent
     There has been some recent high temperture work in perturbation 
theory~\cite{KaLaRuSchr03} to higher orders in the coupling $g$, for which
the pressure ratio to the Stefan-Boltzmann limit has been compared to 
the lattice results for the pure gauge theory~\cite{Boyd}. Furthermore,
a comparison between the perturbative results~\cite{MikYor} up to the
order $O\left(g^6\right)$ and these lattice gauge simulations in 
Figure~\ref{fig:eqnstate} for $\varepsilon - 3p$ shows a very good
agreement above about $800MeV$. This perturbative evaluation
\footnote[7]{The author thanks Mikko Laine for showing to him these 
numerical results from finite temperature\\ perturbation theory.} 
for $\varepsilon - 3p$ continues to increase well beyond the range of 
the lattice data in roughly the same way as the simulations~\cite{Boyd}.\\
~\\ 
\noindent
{\large{\bf{II.5 Discussion of Conformal Symmetry}}}\\
~\\
\noindent 
Specific investigations of the effective measure of conformal symmetry
have been recently carried out on the lattice at high temperatures
for pure $SU(3)$ gauge theory. As we have already discussed above in 
the Introduction, the presence of a nonzero trace for the energy
momentum tensor relates with the breaking of scale and conformal
invariance. We shall discuss more thoroughly in the Appendix B
the mathematical nature of the related physical currents $D^{\mu}(x)$
and $K^{\mu\alpha}(x)$ which are relate directly to the fact
that the trace of the energy momentum tensor remains finite.\\
~\\
\indent
     Then the question clearly arises concerning how close the given
thermodynamical system is to achieving the scale and conformal invariances.
A measure of the deviation from conformality is given by the expression
$\mathcal{C}=(\varepsilon - 3p)/{\varepsilon}$, which has been proposed 
by Rajiv Gavai, Sourendu Gupta and Swagato Mukherjee~\cite{GavGupMuk}. 
Clearly if the value of $\mathcal{C}$ were identically zero, 
then the conformal invariance of the theory would be fully upheld. 
However, they clearly find from their simulations~\cite{GavGupMuk} 
at both the temperatures of $2T_d$ and $3T_d$ the regions 
with finite (nonzero) values for $\mathcal{C}$. Thereby they 
show in a plot of the pressure ratio $p/T^4$ against the energy 
density ratio ${\varepsilon}/T^4$ that all the numbers lie 
significantly below the line of conformality in the pressure where 
$\mathcal{C}=0$, which corresponds to the ideal ultrarelativistic 
gas. Then from their work we are able to see that all of the 
values of $p/T^4$ and ${\varepsilon}/T^4$ in their simulations 
are visibly removed from their ideal gas values for a lattice 
with those given quantum numbers of the spin and color.
Thus for pure gluon system we would expect to find the
Stefan-Boltzmann limiting values for these ratios to be 
$8\pi^2/45$ and $8\pi^2/15$, respectively. Furthermore,
the higher temperature points at $3T_d$ are distributed in a 
cluster which appears to be further from the line of conformality 
$\mathcal{C}=0$ than were those at $2T_d$.\\
~\\
\indent 
     Finally we conclude this part on the thermodynamics of the pure
lattice gauge theory. In summary, for the gluon gas with strong 
interactions simulated for both $SU(2)$ and $SU(3)$ symmetries all 
the thermodynamical functions in {\it physical units} including 
the equation of state grow monotonically in the temperature. 
This clear statement from the computed numerical results is 
obviously quite contrary to the usual expectations from the 
ideal gas oriented theories. In some earlier investigations and
collaboration with Graham Boyd~\cite{BoMi,Mil97} we have 
shown further properties of these thermodynamical functions 
to possess a steady growth in the presently considered range 
of temperatures continuing to, at least, $1.5[GeV]$.\\
~\\
\newpage
\noindent
{\Large{\bf{III. Dynamical Quarks at finite Temperature}}}\\
~\\
\noindent
In this part of the report we look at the full lattice QCD including
the thermodynamical contributions to the equation of state arising
from the thermal properties of the dynamical quarks. The presence of 
these quarks with multiple flavors changes radically the evaluation 
of the thermodynamical quantities. The changes are largely due to the 
presence of the broken chiral symmetry in the hadronic ground state 
of the colored quark fields. Furthermore, the restoration of the 
chiral symmetry at finite temperatures radically restructures the
high temperature quark-gluon phases. The critical temperature $T_c$ 
for the chiral restoration is considerably lower than the 
deconfinement temperature $T_d$ for the pure $SU(N_c)$ lattice gauge 
theories discussed in the last section. The order parameter can be
defined similarly to the Polyakov loop $L$ in equation (\ref{eq:polyloop}) 
for the pure gauge theory except that one takes only the real part $\Re{e~L}$
when the dynamical quarks are present~\cite{Holt, EngHolSchu}.\\
~\\
\indent
     Before we consider the actual lattice data, we mention the 
thermodynamics of the ideal relativistic gas of $n_f$ flavored quarks 
with a given rest mass $m$ at a temperature $T$. The quantities 
like the pressure and the energy density are well known, which  
gives the equation of state
\begin{equation}
   \varepsilon(T) - 3p(T) = g{m^3}T{K_1(m/T)},
  \label{eq:relgaseqst}
\end{equation}
where $K_1(m/T)$ is the modified Bessel function of the second kind. The
statistical degeneracy factor $g$ has the value $3n_f/{\pi}^2$. For a fixed 
quark mass $m$ and at temperatures large compared to the mass energy then 
the right side of (\ref{eq:relgaseqst}) becomes just $g{m^2}{T^2}$. 
Then the equation of state of a pure massive gas grows quadratically 
in both the temperature and the mass in the high temperature limit.\\
~\\ 
\noindent
{\large{\bf{III.1 Equation of State for two light Quark Flavors}}}\\
~\\
\noindent
First we consider the equation of state for two light 
dynamical quarks which appears as a plot similar to the 
figure~\ref{fig:eqnstate} for the pure $SU(N_c)$ gauge 
theories in terms of the trace of the energy momentum 
tensor for the equation of state 
${\theta}^{\mu}_{\mu}(T)~=~\epsilon(T)~-~3p$. In this
new figure for the MILC97 data~\cite{MILC97} we have shown
the equation of state using equation (\ref{eq:trace}) 
in terms of the physical units $[GeV/fm^3]$.
\begin{figure}[b!]
\begin{center}
\epsfig{bbllx=127, bblly=265, bburx=451, bbury=588,
file=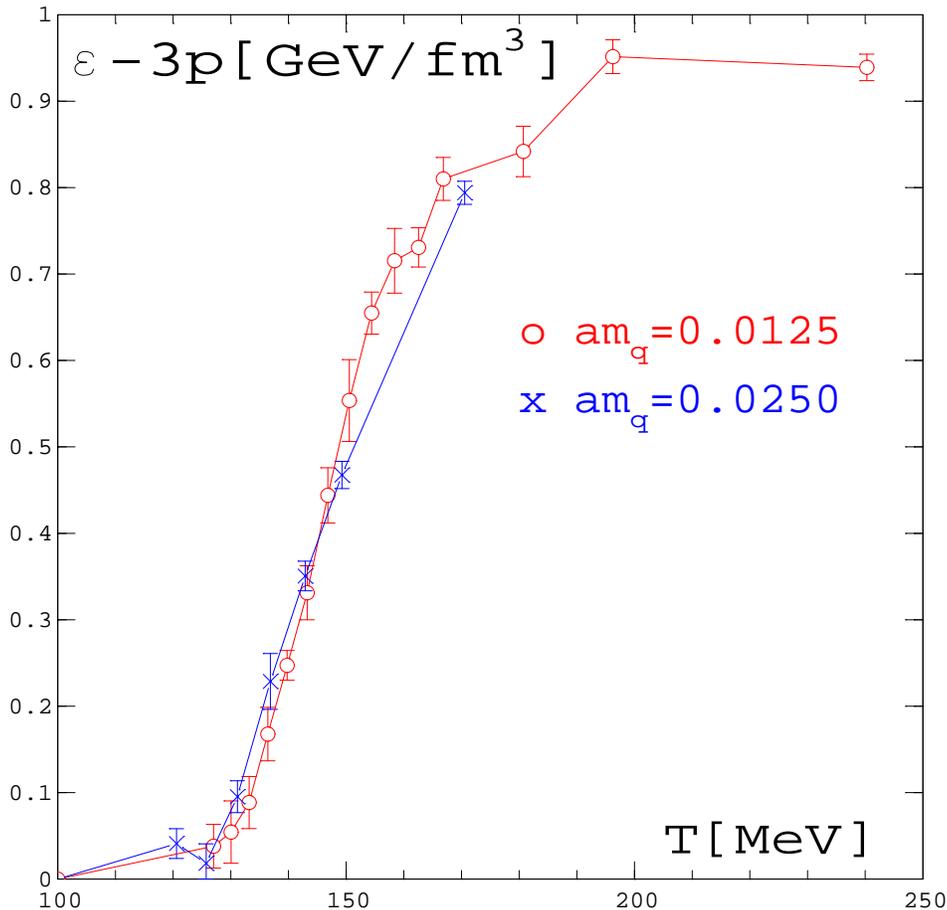, width=11.5cm}
\caption{\it The equation of state for two light quarks given by the 
two different lattice masses. }   
  \label{fig:eqstqu}
\end{center}
\end{figure}
Furthermore, in this representation we are able to compare 
our results directly with those in their Figure 7 for the MILC 
data~\cite{MILC97} which shows the interaction measure written 
as $(\varepsilon-3p)/T^4$ against the QCD lattice coupling 
$6/g^2$ over the range of values from 5.36 to 5.50 with the 
two mass values $am_q=0.0125$ and $am_q=0.0250$, which
are expressed in terms of the lattice spacing.\\
~\\
\indent
     It is important that we note here for the case of the light 
dynamical fermions that even though the pressure ratio is computed 
in a similar way to the pure lattice gauge theory starting with 
the equation (\ref{eq:freen}), it, nevertheless, has the
predominant effect of the average of the product of the antiquark 
$\bar\psi_q$ and quark $\psi_q$ fields written as 
${\langle \bar{\psi}_q{\psi}_q \rangle}$. Thus the equation
for the pressure ratio (\ref{eq:pressplaq}) gets modified to contain
these contributions. Since we do not present here the numerical
values for the pressure, one should go to the original MILC97 
data~\cite{MILC97} for the exact form of the pressure equation, 
which, however, uses slightly different notation in the 
integration method  as applied to the free energy ratio. For this
reason the interaction measure has an additional contribution
beyond just the plaquette terms of the pure lattice gauge theory
as given in equation (\ref{eq:intactmeas}). This new term
arises from the mass renormalization on the antiquark-quark
terms ${\langle \bar{\psi}_q{\psi}_q \rangle}$. These two 
contributions together give the interaction measure $\Delta_m(T)$ 
with the the quarks of mass $m_q$
\begin{equation}
  \Delta_m(T) = {n^{4}_{\tau}}~\{{\tilde{\beta}(g)}~{\left[S_{0}~-~S_{T}\right]}
~+~\tilde{\gamma}(am_q)\left[{\langle \bar{\psi}_q{\psi}_q \rangle}_0~-~
{\langle \bar{\psi}_q{\psi}_q \rangle}\right]\}.
\label{eq:intactmeasm}
\end{equation}
\noindent
The beta function ${\tilde{\beta}(g)}$ is given similarly to that
in equation (\ref{eq:betlat}).However, the renormalization
group gamma function for QCD~\cite{Muta} can be defined on the 
lattice $~\tilde{\gamma}(am_q)$ as
\begin{equation}
  {\tilde{\gamma}(am_q)}~=~-{\frac{d(am_q)}{da}}.
\label{eq:gamlat}
\end{equation}
\noindent
The presence of both these terms in $\Delta_m(T)$ carries very 
important consequences in the equation of state as can be seen 
in the figure~\ref{fig:eqstqu}. The behavior between around
$100MeV$ up to just below $250MeV$ for the MILC97 data~\cite{MILC97},
that is near to and just above $T_c$ shows a decisively different
change in the equation of state from that of the pure $SU(3)$
gauge theory just above $T_d$ as seen in figure~\ref{fig:eqnstate}.
Although the pure gauge theory shows some variation in its rise
above $T_d$ depending upon the lattice sizes~\cite{Boyd}, after
about $300MeV$ its increase is reduced to an approximately linear 
rate~\cite{Zwanz}. In the later paragraphs we shall contrast
both these results with some newer lattice computations for 
different numbers and masses of quarks. First, however, we shall 
discuss a little more carefully the structure of chiral symmetry 
breaking.\\ 
\newpage
\noindent
{\large{\bf{III.2 Chiral Symmetry and Dynamical Quarks}}}\\
~\\
\noindent
In the presence of dynamical quarks another symmetry becomes important--
the chiral symmetry. When the quarks have masses, this symmetry is 
automatically broken. The chiral symmetry is a property of the two different
representations of $SL(2,\bf{C})$ denoted by $\bf{2}$ and $\bf{2^{*}}$
arising for the Dirac spinors in the Weyl representation~\cite{Kugo}. 
It is the presence of the quarks' mass terms in the 
Dirac equation that formally breaks the chiral symmetry. This comes 
formally out of the nonconservation of the axial current $j^{\mu}_{5}$
as discussed~\cite{Jack,Kugo} relating to the triangle 
diagrams, such that the chiral anomaly for QCD takes the form
\begin{equation}
\partial_{\mu}j^{\mu}_{5}~=~j_{5}~+
{{k_{1}}\over{8\pi^2}}{{\bar{G}^{\mu\nu}_{a}}G^{a}_{\mu\nu}},
\label{eq:chiranomqcd}
\end{equation} 
\noindent
where $k_{1}$ is a constant and $j_{5}$ is a pseudoscalar contribution. 
This situation has important implications in
the case for finite temperatures where for $T$ sufficiently high the chiral
symmetry is restored in the small mass or chiral limit, $m_q \rightarrow 0$.
We shall discuss the implications of this both from the theoretical 
side and the numerical side where a finite small mass is present.\\
~\\
\indent
    We now look at the chiral condensate at finite temperatures using chiral
perturbation theory. The low temperature expansion for two massless
quarks can be written~\cite{Leut} in the following form:
\begin{equation}
\label{eq:chirpert}
{{\langle \bar{\psi}_q{\psi}_q \rangle_{T}}\over{\langle \bar{\psi}_q{\psi}_q
  \rangle}_{0}}~=~1~-~{1\over{8}}{T^{2}\over{F^{2}_{\pi}}}~-~
{1\over{384}}{T^{4}\over{F^{4}_{\pi}}}~-~{1\over{288}}
{T^{6} \over {F^{6}_{\pi}}}{\left\{ln{{\Lambda_{q}}\over{T}}\right\}}
~+~O(T^{8})~+~O(\exp{-M\over{T}}),
\end{equation}
\noindent
where $F_{\pi}$ is the above mentioned pion decay constant and the scale
$\Lambda_{q}$ is taken as approxamately $0.470GeV$. Leutwyler has shown
that this expansion up to three loops remains very good at least up 
to around $0.100GeV$. At very low temperatures the probability of 
finding any given excited mass state is related to the exponentially
small correction, which in this case has a very small value. As the 
temperature becomes higher, the number of different states begins to grow
\footnote[8]{We note that here the behavior of the quark condensate
$\langle \bar{\psi}_q{\psi}_q \rangle$ is no longer dominated by the low
energy meson states like the pions and kaons. As the energy increases 
the number states begins to grow exponentially as would be indicated by 
the Hagedorn spectrum~\cite{Haged}, which leads to an expected problem 
with this type of series at high temperatures. We will mention this 
situation later in the report.}. 
Nevertheless, at sufficiently low temperatures the excited states 
may be regarded as a dilute gas of free particles since the chiral 
symmetry supresses the interactions by a power of $T$ of this gas 
of excited states with the primary pionic component~\cite{Leut}.\\
~\\
\indent
     Upon approaching the chiral symmetry restoration temperature 
$T_{\chi}$ the picture changes drastically. At this point the 
ratio $T/F_{\pi}$ is considerably greater than unity. It is here 
where one expects the chiral condensate to be very small or to have 
totally to have vanished. This effect has been studied recently numerically
~\cite{Laer} for two light flavors at finite temperature on the lattice.
The results of this simulation is shown for ${
{\langle \bar{\psi}_q{\psi}_q \rangle_{T}}/{\langle \bar{\psi}_q{\psi}_q
  \rangle}_{0}}$, which we simply write as the quark condensate ratio,
$\langle \bar{\psi} {\psi}\rangle$, in the 
following two plots for the restoration of chiral symmetry.
We show this quark condensate ratio as a function of the coupling $\beta$
for the range where the chiral symmetry is mostly restored~\cite{Laer}.
The  Figure~\ref{fig:chir1}  shows this ratio for two light quarks 
with a mass in lattice units of $0.02$ on a lattice of size~$16^{3}\times 4$.
We remark that at the value for the coupling 5.24 the chiral symmetry for the 
two light quarks with the lattice mass of 0.02 is already about $20\%$ restored.
While at the upper value of 5.33 it is still only about $80\%$ restored. Thus 
we see that the chiral limit has not in this case been reached.
\begin{figure}[b!]
\begin{center}
\epsfig{bbllx=127, bblly=265, bburx=450, bbury=588,
file=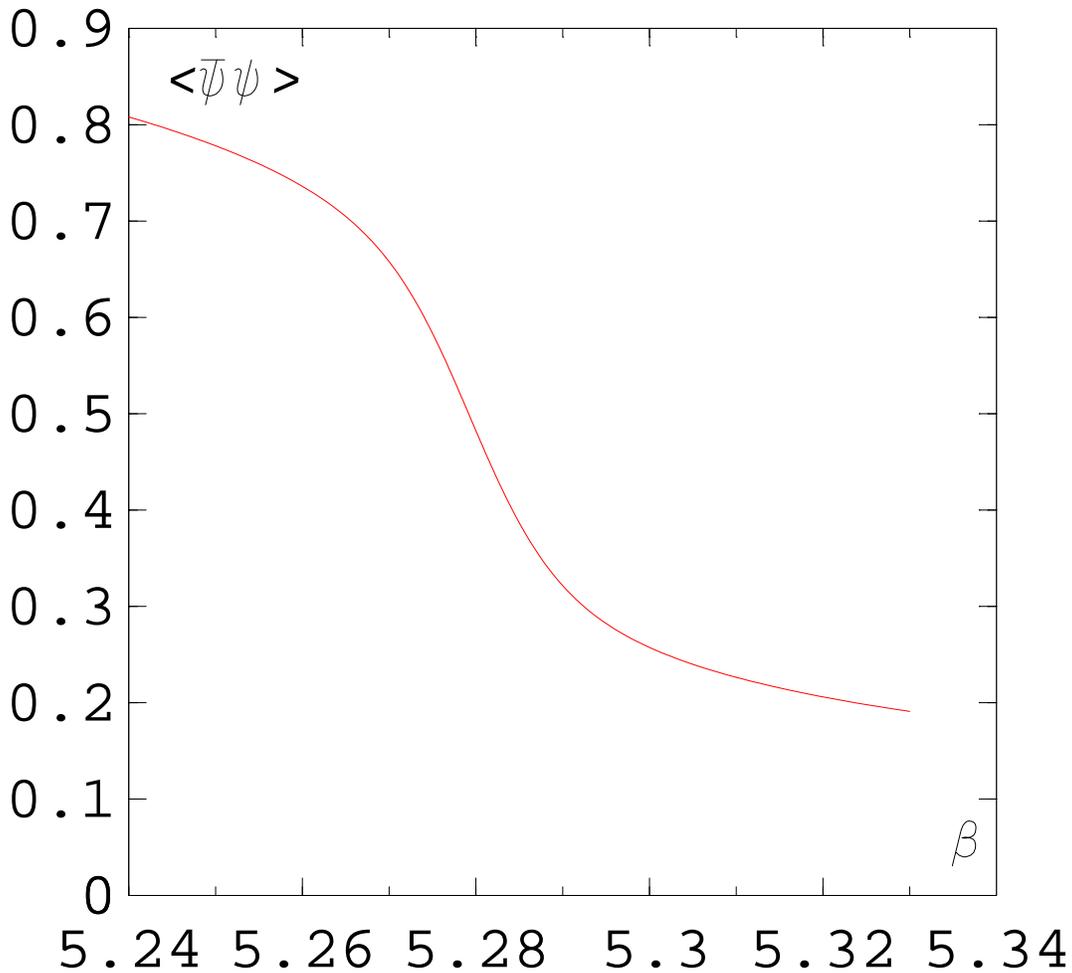, width=11.5cm}
\\[0.7cm]
\caption{\it We show  $<\bar\psi\psi >$ as a function of the coupling $\beta$
for the quark mass in lattice units $ma$= 0.02, which is normalized to the
vacuum value of the chiral condensate. }   
  \label{fig:chir1}
\end{center}
\end{figure}
\noindent
The Figure~\ref{fig:chir2} comparess how the different mass values shown from 
left to right of $0.02$, $0.0375$ and $0.075$ depend upon the given lattice 
sizes, which are $8^{3}\times 4$, $12^{3}\times 4$ and $16^{3}\times 4$.
\begin{figure}[b!]
\begin{center}
\epsfig{bbllx=127, bblly=265, bburx=450, bbury=588,
file=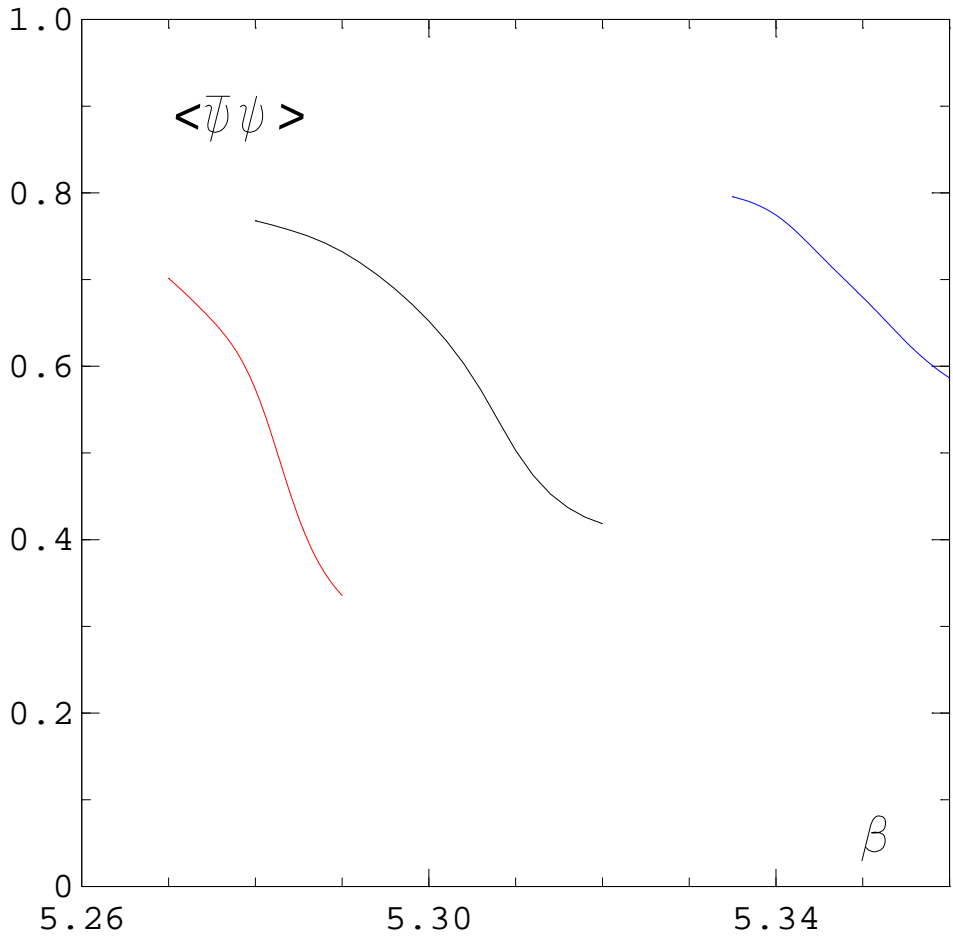, width=11.5cm}
\\[0.7cm]
\caption{\it We show the same physical quantity with the different values of the 
quark masses 0.02, 0.0375 and 0.075 from left to right~\cite{Laer}. }   
  \label{fig:chir2}
\end{center}
\end{figure}
\noindent
We should also notice how the larger mass values slow the restoration down, 
which corresponds to moving the transition~$T_{\chi}$ to higher temperatures
or even eliminating it altogether as indicated by the flatness of the curves.\\ 
~\\
\noindent
{\large{\bf{III.3 Thermodynamics with Dynamical Quarks}}}\\
~\\
\noindent
The main quantities which were analyzed here were the various susceptibilities:
~\\
\noindent
1. The Polyakov loop susceptibility we have defined earlier in the last
section in the equation (\ref{eq:poloopsusc}). We  will use now in 
connection with two other forms of the susceptibilities which are to follow:\\
~\\
\noindent
2. The magnetic or chiral susceptibility;
\begin{equation}
{{\chi}_{m}}~=~{T\over V}~{\sum\limits_{i=1}^{n_{f}}}
{{\partial^2}\over{\partial m^{2}_{i}}}{\ln{\cal{Z}(T,V)}},
\label{eq:magsusc}
\end{equation}
~\\ 
\noindent
3. The thermal susceptibility;
\begin{equation}
{{\chi}_{\theta}}~=~-{T\over V}~{\sum\limits_{i=1}^{n_{f}}}
{{\partial^2}\over{\partial m_{i} \partial(1/T)}}{\ln{\cal{Z}(T,V)}}.
\label{eq:thermsusc}
\end{equation}
~\\ 
\noindent
One compares the critical properties of ${{\chi}_{L}}$,
${{\chi}_{m}}$ and ${{\chi}_{\theta}}$ in order to establish 
the value of $T_{\chi}$ and its critical properties in the 
chiral limit where $m_{q} \rightarrow 0$. For the moment we 
use $T_{\chi}$ for the chiral restoration temperature in contrast 
to the critical temperature $T_c$ for the dynamical quark simulations.  
However, in numerical simulations $m_{i}$  must be taken to be finite-- 
this means that one must use various different small values of $m_{i}$ 
on the different sized lattices $N^{3}_{\sigma}\times N_{\tau}$. This 
procedure uses the lattice data to find the values around the peak of the 
susceptibility ${{\chi}_{m}}$ at $T_{\chi}$ for the smallest masses, with 
which one can determine the critical structure. A careful determination
of the topological susceptibility relating to the chiral current 
correlations can be related to the square of the topological charge 
$Q^{2}_{T}$ in the chiral limit~\cite{HanLeut}, such that
\begin{equation}
{n_{f}\over{m}}\langle Q^{2}_{T} \rangle~=~
      V{\langle \bar{\psi} {\psi}\rangle}_{m \rightarrow 0},
\label{eq:topcond}
\end{equation}
\noindent
where $n_{f}$ is the number of flavors. Thus from these susceptibilities 
one can arrive at the quark condensate ratio $\langle \bar{\psi} {\psi}\rangle$. 
However, in this computation it is a major problem to properly 
set the temperature scale for small lattices with finite masses. 
The plots in the figure~\ref{fig:chir1} and~\ref{fig:chir2}  are made with 
the coupling $\beta$ which may be compared 
with pure $SU(3)$ on one side and the two flavor
dynamical quark simulations on the other~\cite{MILC97}. In the case of
pure $SU(3)$ the critical coupling $\beta_{c}$ for a ${16^3}\times 4$
lattice has the value~\cite{Boyd} of about 5.70, which is considerably 
larger than the values of $\beta$ shown in the figure~\ref{fig:chir1} 
and ~\ref{fig:chir2}. However, for the two light 
flavored dynamical quarks~\cite{MILC97} the value 
of ${\beta}_{c}$ is around 5.40, which is still somewhat above these 
values shown in the two figures.\\
~\\
\indent
     Here we have investigated the properties of the chiral symmetry
restoration for the quark condensate $\langle \bar{\psi}_q {\psi}_q\rangle$ 
alone at various values of the coupling $\beta$. We have also shown 
in the figure~\ref{fig:chir2} how the coupling shifts to higher numbers
for larger values of the mass. These results can be compared to the MILC97 
data~\cite{MILC97} with much lighter quark masses which, nevertheless,  
stays mostly in the same range of the couplings (see in the reference
~\cite{MILC97} their figure 4.) We can see there that this data is 
extended into a higher range of coupling to get considerably smaller
values of the quark condensate. These values we shall use in the next
section for the mass contribution of the quark condensate at higher values
of the temperature. We can then compare this effect in the chiral limit. 
However, here it is very difficult to immediately go over to a physical
temperature scale in the same way as in the previous section for the pure
gauge or gluon system. In what follows we shall look into the gluon condensate
in the presence of dynamical quarks. Here we know that the presence of the
quark masses are an immediate cause of scale symmetry breaking which of course
change the scale of the system. This in turn changes the beta function as well
as adds a term due to the mass renormalization. Thus the renormalization group
equations are changed accordingly. This effect we shall discuss more thoroughly
in the following.\\
~\\
\noindent
{\large{\bf{III.4 Equation of State with Different Quark Flavors}}}\\
~\\
\noindent
Next we look at the thermodynamical functions including the
pressure $p(T)$, energy density $\varepsilon(T)$ and the equation 
of state $\varepsilon(T)-3p(T)$ all in the physical units $[GeV/fm^3]$ 
for the case of somewhat heavier dynamical quarks with different numbers
and types of flavors. These different cases have been worked out
in the doctoral thesis of Andreas Peikert~\cite{Peik} at the
Universit\"at Bielefeld for the lattices of sizes $8^{3}\times 4$, 
$16^{3}\times 4$ and for comparison the symmetrical lattice  $16^{4}$.
In general throughout these numerical simulations
the masses of the different quarks were considerably heavier
than the MILC97 data~\cite{MILC97}, which were taken between
$7.5MeV \le m_q \le 15MeV$ . The lighter quarks had masses
in these newer simulations with the values between 
$40MeV \le m_q \le 60MeV$, while the heavier one was more
in the range of $100MeV \le m_q \le 150MeV$, which are considerably
higher than the light up and down quarks, but are not so far off
for the strange quarks. The presented data was simulated using the
p4 action for which comparisons were made to the more usual 
actions~\cite{Peik}.The resulting figures taken from these simulations 
we shall contrast the various different arrangements of flavors 
$n_f=2$, $n_f=2+1$ and $n_f=3$ given as functions of the temperature. 
In each of these flavor arrangements the quarks are at least 
an order of magnitude heavier than those in the previous 
figure~\ref{fig:eqstqu} for the MILC97 data~\cite{MILC97}.\\
~\\
\begin{figure}[thb]
\centerline{\includegraphics[width=16.cm]{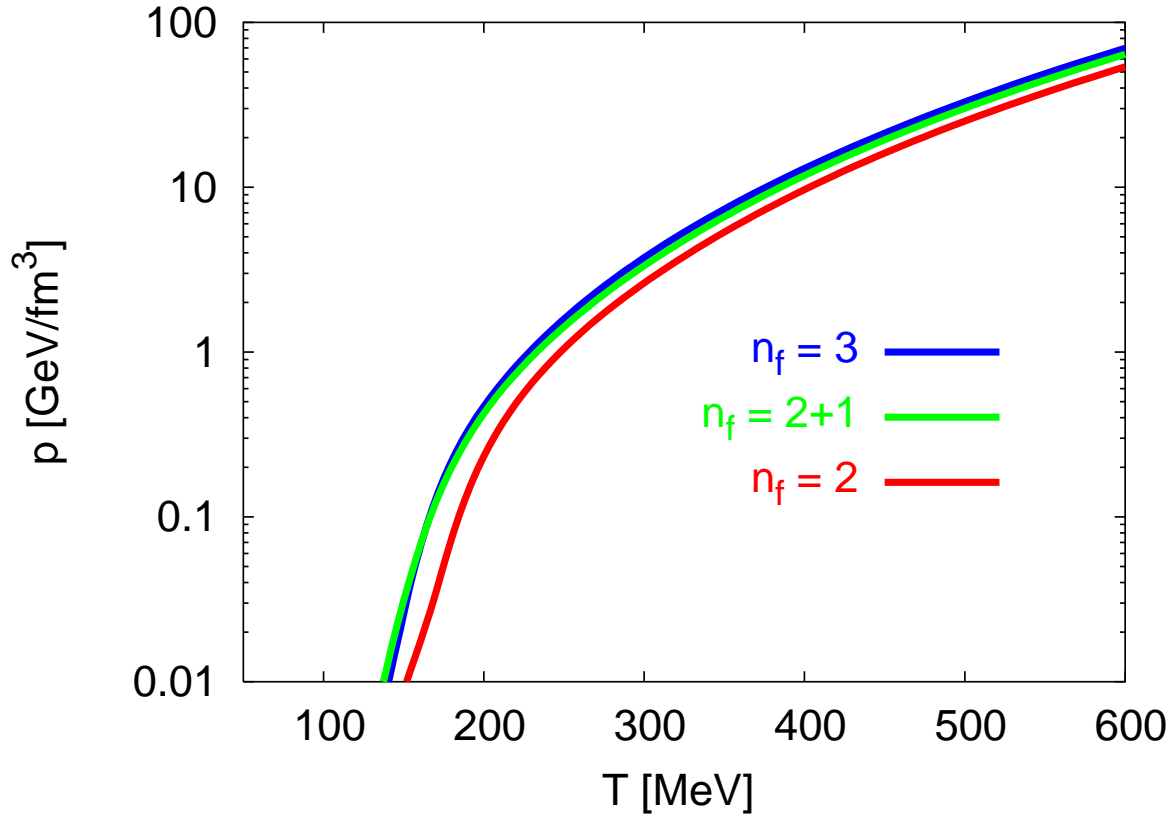}}
\caption{\it The pressure for the three values of $n_f=2,2+1,3$ with
massive dynamical quarks. }   
  \label{fig:pressquark}  
\end{figure}
\indent
     We now look into the properties of each thermodynamical quantity
more specifically. The Figure~\ref{fig:pressquark} shows the 
pressure $p(T)$ for the three different flavors~\cite{KLP,KLP1,Peik} in 
a way similar to that of the pure $SU(3)$ gauge theory. 
Although the general shape of these curves are quite 
similar in appearance, the values of the temperatures 
for the transitions are noticably lower in the theory
with the dynamical fermions. Again the pressure was computed in the
same manner as that of the pure lattice $SU(N_c)$ using the integration
method~\cite{EFKMW90,EKR94,Boyd} as was discussed in the last part. 
In the case of the pressure $p(T)[GeV/fm^3]$ the shape of all three curves 
is almost the same except for the starting point for the case of $n_f=2$.
This is because of the difference in the value of  $T_c$.  For the case 
of $n_f=2$ the value of $T_c=175MeV$ is somewhat higher than $n_f=2+1$ and 
$n_f=3$ of around $155MeV$. However, when one plots the pressure ratio $p/T^4$, 
as was originally done~\cite{KLP,KLP1,Peik}, one notices a large difference
in the different curves arising from the different number of degrees 
of freedom in each case, which changes the rate of approaching the
Stefan-Boltzmann limit as well as the actual value of number itself.
In our figure~\ref{fig:pressquark} the pressure in physical units
represented on a logarithmetrical scale appear to start very near to
the actual value of $T_c$. However, in the form of the ratio $p/T^4$
the small values for the different flavors start well below $T_c$ 
with a very gradual increase until reaching the critical temperature.
Thereabove the increase in each curve is at different rates due
to the different number of degrees of freedom and the corresponding
masses. In the actual original plot~\cite{Peik} (see figure 4.8) 
of $p/T^4$ against $T/T_c$,  one sees that the curve for  $n_f=2$ 
lies well under that of $n_f=2+1$, which itself is significantly 
under that of the cases of  $n_f=3$. Furthermore, the placing of 
these curves is quite different on the approach to the Stefan-Boltzmann 
limit. At the temperature of about $4T_c$ the case of  $n_f=2+1$
has only achieved around $75\%$, whereas both the cases  $n_f=2$ 
and  $n_f=3$ have arrived at about $80\%$ of their respective limiting 
values for high temperatures. Whether this limiting value is actually
attained in the high temperature limit, still remains as an open question.
Finally we remark that in the case of the pressure the errorbars are
generally very small so that their presence is not essential to the figure.\\ 
~\\
\begin{figure}[thb]
\centerline{\includegraphics[width=16.cm]{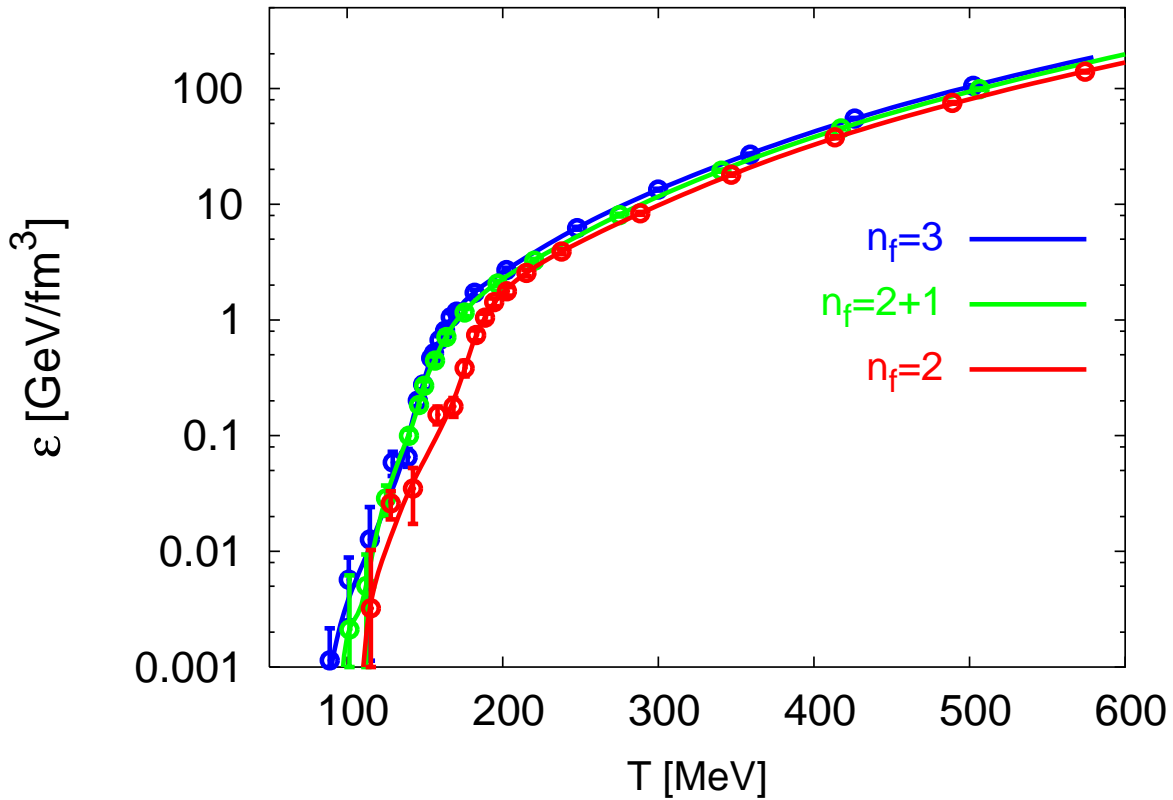}}
\caption{\it The energy density for the three values of $n_f=2,2+1,3$  
with massive dynamical quarks. }   
  \label{fig:endenquarks}  
\end{figure}
\indent
     We next turn our attention to the energy density $\varepsilon(T)$ 
which presents quite a different problemmatic in the computation. The
corresponding accuracy of the evaluations is considerably lower, which
may be apparent by the presence of the errorbars in figure~\ref{fig:endenquarks}. 
There we see that $\varepsilon(T)[GeV/fm^3]$  shows a quite different 
behavior as a function of the temperature $T[MeV]$ from that of the 
pressure in the Figure~\ref{fig:pressquark}. Even well below $T_c$ there 
are small values of $\varepsilon(T)$ which then grow much more rapidly 
at $T_c$. In contrast the pressure really starts to take on sizable 
values only at temperatures very near to $T_c$. Furthermore, one can 
see quite different rate of rise in $\varepsilon(T)$ depending upon the 
masses and the number of flavors. Nevertheless, its values start off
at lower temperatures and rise more slowly than the corresponding 
values of $p$. Then just above $T_c$ the energy density $\varepsilon(T)$ 
rises very rapidly as a function of $T[MeV]$. We notice that the 
curve for $n_f=2$ starts later because of the larger $T_c$ and 
rises more slowly just above $T_c$. For temperatures above 
about $200MeV$ $\varepsilon(T)$ for all the three flavors rise at 
approximately the same rate--practically on top of eachother. 
However, the $n_f=2$ the values always remain somewhat below 
the others as it was also the case for the pressure curves.\\
\begin{figure}[thb]
\centerline{\includegraphics[width=16.cm]{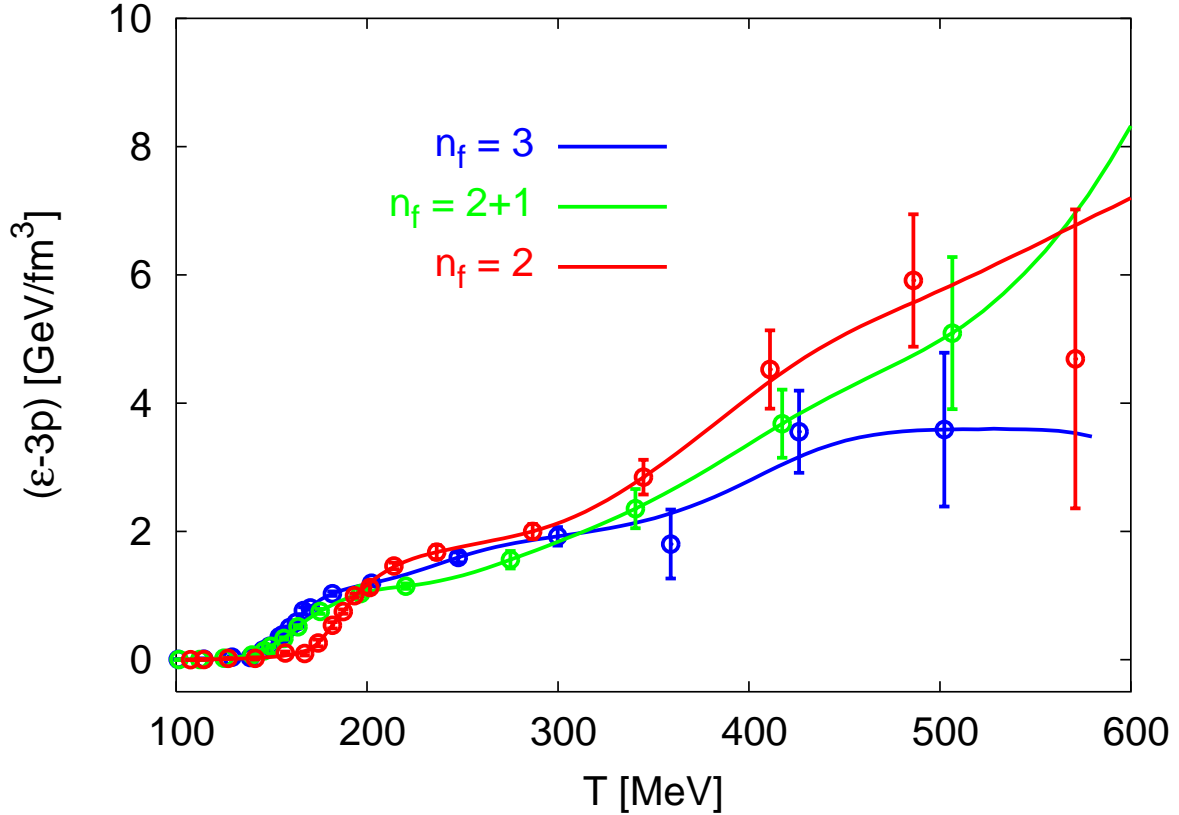}}
\caption{\it The equation of state is evaluated for the three values of $n_f=2,2+1,3$ 
including the massive dynamical quarks. }   
  \label{fig:equostaquark}  
\end{figure}
~\\
\noindent
     From the knowledge of the pressure $p(T)$ as well as the energy density 
$\varepsilon(T)$ we are able to calculate the equation of state in terms of 
$\varepsilon(T)-3p(T)$. However, even as a difference between these basic 
physical quantities the equation of state still shows another type of 
behavior in relation to the flavors $n_f$ and the quark masses $m_q$.  
The values of the masses in the stated ranges have always held the $n_f=2$
values of these thermodynamical functions of the temperature to be smaller
than $n_f=2+1$ and $n_f=3$. However, we see in figure~\ref{fig:equostaquark}
that this statement holds only up to about $200MeV$, above which temperature 
the $n_f=2$ values remain within the errorbars clearly larger than the others.
Furthermore, above about $300MeV$ the case with $n_f=2+1$ takes on larger
values than $n_f=3$, but still remains smaller than  $n_f=2$. Furthermore,
in this range between $200MeV$ and $300MeV$ the change is very slow when 
compared with the region between $150MeV$ and $200MeV$ near the critical
points. Thus we can clearly see here the types of contrasts between the 
parameters and the physical quantities. We remark also that for the 
equation of state the problem with the error bars at higher temperatures 
becomes very significant so that in all cases above $500MeV$ the curves 
give no real physical predictions.\\
~\\
\indent
     In this section we have numerically evaluated using the lattice gauge
simulations~\cite{KLP,KLP1,Peik}. We have shown the thermal properties 
in the three separate figures~\ref{fig:pressquark},\ref{fig:endenquarks} and
~\ref{fig:equostaquark} the basic thermodynamical quantities-- 
the pressure $p(T)$, the energy density $\varepsilon(T)$ and the 
equation of state in the form $\varepsilon(T) - 3p(T)$ as functions
of the temperature $T$. Furthermore, we have looked at these quantities 
in terms of the input parameters like the number of quark flavors 
and the quark masses, for which earlier in this part we have brought in the
properties of chiral symmetry through the chiral condensate ratio. In the
next section we shall briefly discuss some very recent data appearing within
the last year, which we can compare with the above shown results for different 
values of the lattice mass parameters.\\
~\\
~\\
\noindent
{\large{\bf{III.5  Discussion of Recent Massive Quark Data}}}\\
~\\
\noindent
Since the start of the actual writing of this work, there have appeared some 
newer data~\cite{MILC2005,LAT2005} for the three quark equation of state. 
In this section we shall present a discussion of some of these newer 
numerical simulations for 2+1 flavors with a brief analysis of the lattice 
results. In particular, we look at the simulations~\cite{LAT2005} plotted 
as a function of the ratio $T/T_c$ in their Figure 1, in which they include
the data points for the following lattice quantities: the interaction measure,
here written as $I/T^4$, the pressure ratio $p/T^4$ and the energy density 
ratio $\varepsilon/T^4$. In their simulations they use a Symanzik improved 
gauge action and the Asqtad $O(a^2)$ improved staggered quark action for lattices
with the temporal extents $N_{\tau}=4$ and $N_{\tau}=6$. They set their value
of the heavy quark mass near to the physical strange quark mass $m_s$. Then
they choose the two degenerate light quark masses to have the values of $0.1m_s$ 
and $0.2m_s$, for which they compute these quantities in the temperature range 
from $0.7T_c$ to $2.15T_c$. For the computation of these thermodynamical ratios
the integral method~\cite{EFKMW90} has been used as discussed in previous sections.
Furthermore, the estimated value~\cite{MILC2005} of the critical temperature $T_c$ 
is somewhat higher than for the Bielefeld 2+1 flavor data for this system with much 
lighter quarks, which is found to be about $169\pm12\pm4MeV$.\\ 
~\\
\indent 
     These more recent results may be properly compared to the Bielefeld 
data~\cite{KLP,KLP1,Peik} only for the simulations with  $N_{\tau}=4$. 
In general we can compare the values for the interaction measure in the 
2+1 flavor case~\cite{Peik} used in the above figure~\ref{fig:equostaquark} 
to compute the equation of state to those values of $I/T^4$ plotted for 
$N_{\tau}=4$ in the Figure 1 of this newer work~\cite{LAT2005}. This comparison
shows that the actual difference between the does not result in a large change 
in the interaction measure. Therefore, we could expect to have rather small 
changes in the properties of the equation of state over the temperature range
in common to both cases. A simple pointwise comparison shows that for temperatures
below and just above $T_c$ the values for the interaction measures are close to
the same for both cases within the respective errorbars. Above $1.5T_c$ the newer 
values of the interaction measure are actually somewhat larger. In the case of 
$N_{\tau}=6$ the numerical values for the interaction measure are considerably 
larger than the  corresponding values for the Bielefeld data~\cite{KLP,KLP1,Peik}
just above $T_c$ and remain so throughout the higher temperatures.\\
~\\
\indent
     In summary for this new data we notice that the $N_{\tau}=4$ is generally
closer to the numerical values used here. The larger values for the light quark mass
provide the bigger numbers for the interaction measure especially around the
critical temperature. We should also mention some values for the quantity
$\varepsilon/T^4$ approach the Stefan-Boltzmann limit for the $N_{\tau}=6$
simulations near to $1.25T_c$, after which the changes in value are quite small.
This fact accounts for the larger growth of the interaction measure near to $T_c$.
This new data  describes the QCD thermodynamics for three flavors of improved
staggered quarks quite consistently with the data use here~\cite{KLP,KLP1,Peik}. 
In the next part we will consider the results from our previous analyses of
the equations of state in the figures~\ref{fig:eqstqu} 
and~\ref{fig:equostaquark} in more detail in relation to the gluon and 
quark condensates at finite temperature.\\
\newpage
\noindent
{\Large{\bf{IV. Gluon and Quark Condensates at Finite Temperature}}}\\
~\\
In this part we start our discussion of the thermal properties of the gluon
condensates by using the approach that we described in the Introduction which
includes both the  pure $SU(2)$ and $SU(3)$ gauge theories~\cite{Boyd,BoMi}
as well as the massive quarks~\cite{Mil97}. We shall first discuss how the 
pure gluon condensate looks with no dynamical quarks present, for which 
the results for the equation of state in Part II can be directly used. 
The effects of the chiral phase transition with massive dynamical 
quarks are then discussed in various cases, for which we discuss 
the different relationships to the quark condensates. \\
\begin{figure}[b!]
\begin{center}
\epsfig{bbllx=127, bblly=265, bburx=450, bbury=588,
file=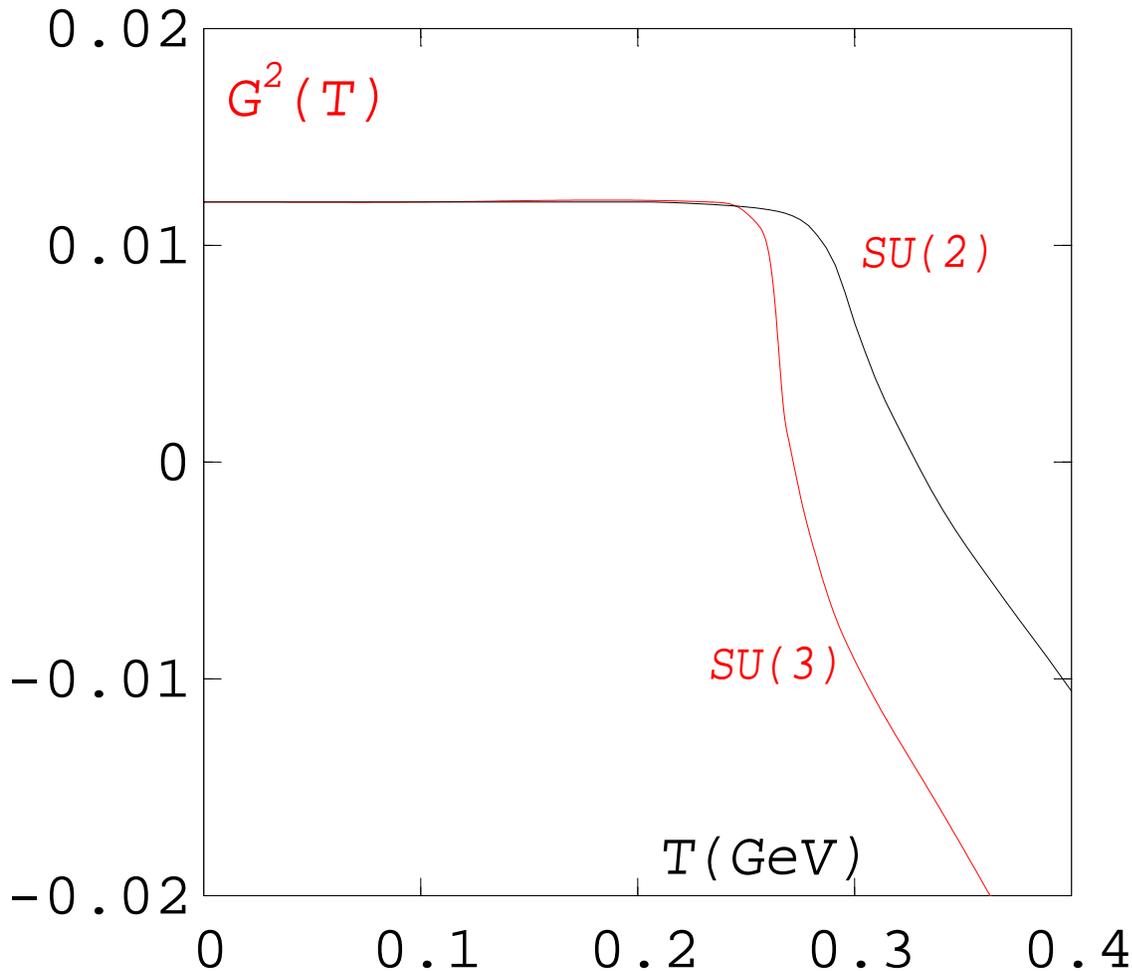, width=11.5cm}
\\[0.7cm]
\caption{\it We show the gluon condensates for the $SU(2)$ and $SU(3)$
invariant gauge theories, where both the ordinates are in $[GeV^4]$. }   
  \label{fig:glucondgauge}
\end{center}
\end{figure}
~\\
\noindent 
{\large{\bf{IV.1 Pure Gluon Condensate}}}\\
~\\
\noindent
The results for the gluon condensation in the cases of the pure 
gauge theories  are shown in the figure~\ref{fig:glucondgauge} 
for the lattice data~\cite{EKR94,Boyd}. In both these cases 
we have taken the zero temperature gluon condensate~\cite{SVZ1} 
to be $\langle G^2 \rangle_0~=~0.012[GeV^4]$ as the starting 
value for the gluon condensate at finite temperature, which we 
shall write for simplicity $G^2(T)$. We can see that this value 
remains very constant for temperatures up to nearly the deconfinement 
temperature~\cite{Leut,BoMi}, which are given as $T_d = 0.290~and
~0.264$~GeV for the color $SU(2)$ and $SU(3)$, respectively. The basic 
equation comes from the finite temperature trace anomaly which remains 
much the same as the one considered above used by Leutwyler~\cite{Leut}
except that it is applied to the pure gauge theory not massless quarks. 
In the course of this section we shall reconsider the values in order 
to include the massive quarks.\\
~\\
\noindent
{\large{\bf{IV.2 The Effect of Quark Condensates}}}\\
~\\
\noindent 
In the presence of massive quarks the trace of the energy-momentum tensor
takes the altered form~\cite{tracanom} from the trace anomaly as given by
the equation (\ref{eq:thetaquark}) for which we recall that  
$m_q$ is the light (renormalized) quark mass and ${\psi}_q$,
$\bar{\psi}_q$ represent the quark and antiquark fields respectively.
In the last section we have discussed the temperature dependence of
the chiral condensate in relatiion to chiral perturbation theory~\cite{Leut}.
We recall that the second term in equation (\ref{eq:chirpert}) is
quadratic in the temperature. This term corresponds to the process of
pion absorption and emission given by the matrix element
${\langle \pi| \bar{\psi}{\psi}|\pi \rangle}$. This term as well as the
succeeding even powers in the expansion of ${\langle \bar{\psi}{\psi}\rangle}_T$ 
allow the chiral condensate to melt quite gradually. Furthermore,
we should note that in equation (\ref{eq:chirpert}) that all the
terms subtract off the zero temperature condensate. In contrast to
the multitude of processes in the chiral condensate melting,
Leutwyler clearly points out~\cite{Leut} that such terms
are absent in the pure gluonic processes since the field strength
operators in the color combination given in equation (\ref{eq:glucondop}) 
in the form $ G^{{\mu}{\nu}}_{a}G_{{\mu}{\nu}}^{a}$ is a chiral singlet
so that the corresponding single pion matrix element vanishes~\cite{Leut}.
When one calculates the corresponding trace of the energy momentum tensor,
one finds for two massless flavors
\begin{equation}
 \theta^{\mu}_{\mu}(T)~=~{\pi^{2} \over {270}}
{T^{8} \over {F^{4}_{\pi}}}{\left\{ln{{\Lambda_{p}}\over{T}}~-~{1\over4}\right\}}
~+~O(T^{10}),
\label{eq:tracexp}
\end{equation}
\noindent
where the logarithmic scale factor ${\Lambda_{p}}$ has the approximate value
of $275MeV$~\cite{Leut,GerLeut}. Thus in the framework of chiral
perturbation theory the single and double loop contributions both vanish
in the gluon condensate at finite temperatures. The first actual term
present in the equation (\ref{eq:tracexp}) has the power $T^{8}$ arises 
from the three loop graph. The result of equation
(\ref{eq:tracexp}) can be used in Leutwyler's equation (\ref{eq:condef}) to
evaluate the finite temperature color averaged gluon condensate, 
which was discussed~\cite{Leut}.
In the following we shall use these averages together in a single
expression, both of which are evaluated from the numerical lattice
gauge simulations with massive dynamical quarks. Furthermore, we
shall include with these averages the renormalization group functions
$\beta(g)$ and $\gamma(g,m)$, which appear in the trace of the energy 
momentum tensor even in the vacuum~\cite{ColDunJog,Niel} from the
renormalization process.  
~\\
\indent
     Now we need to discuss further the changes in the computational procedure 
which arise from the presence of dynamical quarks with a finite mass.  
There have been recently a number of computations of the thermodynamical 
quantities in full QCD with two flavors of staggered quarks~
\cite{BKT94,MILC96,MILC97}, and with four flavors~\cite{edwinqmfklat96,Eng5}.
We have already mentioned the problemmatic of the simulations for the
dynamical quarks in the equation of state in the last section.
Furthermore, we mentioned there the problem of the quark condensate in
relation to the restoration of chiral symmetry. Now we bring these two
aspects of dynamical fermions together. Nevertheless, we first remark that 
these calculations are still not as accurate as those in pure gauge theory for
several reasons. The first is the prohibitive cost of obtaining statistics
similar to those obtained for pure lattice gauge theory. So the error on the 
interaction measure is considerably larger. 
The second reason, which is perhaps more serious, lies 
in the effect of the quark masses currently simulated. They are still, in 
most of the cases in the present simulations relatively heavy, which 
unduely increases the contribution of the average quark condensate part in
the interaction measure. In fact, it is known that the vacuum
expectation values for very heavy quarks is proportional to the gluon 
condensate $G^2_0$, written in simplified notation, which in the first 
approximation is given by~\cite{SVZ1}
\begin{equation} 
{\langle \bar{\psi}_q{\psi}_q \rangle}_{0}~=~
{-1 \over{12m_{q}}}{ G^2_0}.
\label{eq:quarktogluecond}
\end{equation}
Furthermore, there is an additional 
difficulty in setting properly the temperature scale even to the extent of 
rather large changes in the critical temperature have been reported in the 
literature depending upon the method of extraction. For two flavors of quarks 
the values of $T_c$ lie between $0.140GeV$~\cite{MILC97} and about $0.170GeV$
~\cite{Edwin} which is considered presently a good estimate of the physical 
value for the critical temperature. \\
~\\
\indent
     We are now able to write down an equation for the temperature 
dependence of the thermally averaged trace of the energy momentum tensor 
including the effects of the light quarks with a mass $m$ from $\Delta_{m}(T)$ 
so that
\begin{equation}
 \theta^{ \mu}_{m \mu}(T)~=~\Delta_{m}(T){\cdot}T^4.
\label{eq:fulltheta}
\end{equation}
The thermally averaged gluon condensate is computed including the
light quarks in the trace anomaly using the equation~\eqref{eq:thetaquark}
and the interaction measure in $ \theta^{\mu}_{m \mu}(T)$ to get
\begin{equation}
 G^2(T)~+~{\sum}_{q}m_q{\langle{\bar{\psi}_q{\psi}_q}\rangle}_{T}~= 
 G^2_0~+~{\sum}_{q}m_q{\langle{\bar{\psi}_q{\psi}_q}\rangle}_0
~-~{\theta}^{\mu}_{m \mu}(T), 
\label{eq:quarcondef}
\end{equation}
\noindent
It is possible to see from this equation that at very low temperatures 
the additional contribution to the temperature dependence of the gluon
condensate from the quark condensate is rather insignificant and disappears 
completely at zero temperature. However, in the range where the chiral  
symmetry is being restored there is an additional effect from the term
$\langle \bar{\psi}_q{\psi}_q\rangle_T$, which lowers $G^2(T)$.
Well above $T_c$ after the chiral symmetry has been mostly restored 
the only remaining effect of the quark condensate is that of 
$m_q\langle\bar{\psi}_q{\psi}_q \rangle_0$. It is then known~\cite{SVZ1} how
this term shifts the gluon condensate of the vacuum. Thus we may well expect
~\cite{BoMi}~that for the light quarks the temperature dependence can only be  
important below and near to $T_c$. In the case of the chiral limit
~$m_q \rightarrow 0$ the equation~\eqref{eq:fullcond} takes the form of 
Leutwyler's equation~\eqref{eq:condef} as, of course, it should 
because Leutwyler used two massless quarks~\cite{Leut}. For the smaller 
values of the simulated quark masses in lattice units of 0.01 to 0.02 
$\langle \bar{\psi}_q{\psi}_q\rangle_T$ has mostly disappeared in the 
range where $G^2(T)$ differs from $\langle G^2 \rangle_0$.\\ 
~\\
\noindent
{\large{\bf{IV.3 Gluon and Quark Condensates with two light Flavors}}}\\
~\\
\noindent
In the way we mentioned above for equation (\ref{eq:condef}) with 
the change of the pure gluon condensate to the sum of the quark contributions 
$\langle G^2 \rangle~+~{\sum}_{q}m_q\langle{\bar{\psi}_q{\psi}_q}\rangle$.
First we discuss the finite temperature gluon condensate in the "chiral 
limit" $m_q \rightarrow 0$ as shown in the figure~\ref{fig:glucondqu} with 
the two $m_q = 0$ curves. From the MILC light quark data~\cite{MILC97} 
we are able to subtract the quark condensates at finite temperatures 
multiplied by the two quark masses $m_q = 7.5MeV$ and 
$m_q = 15MeV$ for the two sets of lattice data. The effects 
of the temperature as well as the restoration of the chiral symmetry 
are included with the decreasing value of $m_q<\bar{\psi}\psi>$ with increasing 
temperature are gotten from the original lattice data~\cite{MILC97}. We use for 
the vacuum value of $<\bar{\psi}\psi>_0=-(259\pm27MeV)^3$ from a recent 
estimate~\cite{McN} taken from newer MILC dynamical quark data~\cite{MILC01}. 
Now we can reformulate the above equation (\ref{eq:quarcondef}) by replacing the
sum over the flavors simply by the number {\it two}. The the finite temperature
gluon condensate is given by
\begin{equation}
 G^2(T)~=~ G^2_0
~+~2m_q\langle\bar{\psi}_q{\psi}_q \rangle_0 \left( 1~-~
\langle\bar{\psi}_q{\psi}_q\rangle \right)
~-~ \theta^{\mu}_{m \mu}(T).
\label{eq:fullcond}
\end{equation}
The ratio of the finite temperature quark condensate to its vacuum value,
$\langle\bar{\psi}_q{\psi}_q\rangle_{T}/{\langle\bar{\psi}_q{\psi}_q\rangle}_0$,
we write simply as $\langle\bar{\psi}_q{\psi}_q\rangle$, which is just the
\begin{figure}[b!]
\begin{center}
\epsfig{bbllx=127, bblly=265, bburx=450, bbury=588,
file=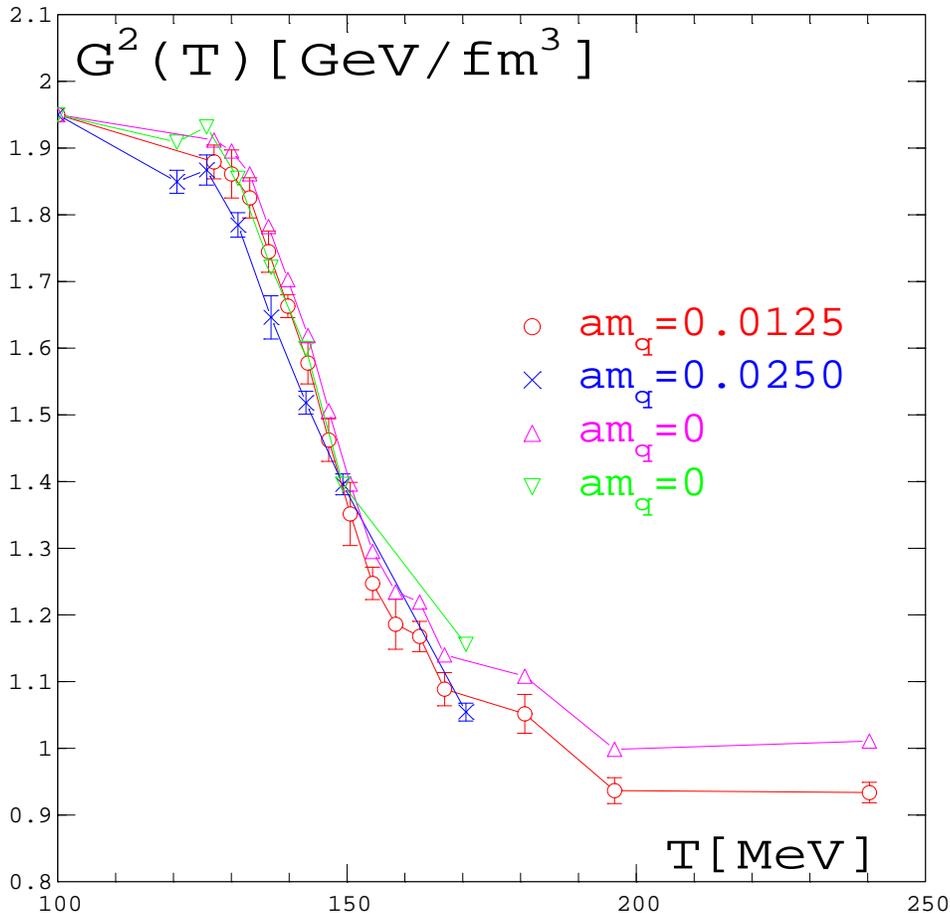, width=11.5cm}

\caption{\it The finite temperature gluon condensate is shown for two 
light quarks given by the two different lattice masses. 
We compare these with the corresponding data in the chiral limit
shown with the triangles. }   
  \label{fig:glucondqu}
\end{center}
\end{figure}
chiral order parameter. The values can be seen in the figure 4 of the MILC
data~\cite{MILC97}. We show these results in the figure~\ref{fig:glucondqu} 
for the two different stated masses including the errorbars arising from
both the interaction measure and the quark condensate data. We notice that
with increasing temperatures that the finite quark mass lowers the gluon
condensate. Furthermore, it can be seen in figure~\ref{fig:glucondqu}
that between the temperatures of about $140MeV$ and $160MeV$, which is
just below $T_c$, both with and without the quark condensates lie
very mush together-- almost within mutual errorbars. Below and above
these values the heavier quark values appear to digress although there 
are simply too few points to allow for a firm conclusion. This fact 
clearly shows an additional effect from the restoration of the chiral 
symmetry on the thermal gluon condensate in the presence of massive quarks. 
At the highest temperature around $240MeV$ the chiral symmetry for 
the smaller mass value is about $95\%$ restored. From these results we
can understand how the effect of the quark masses together with the 
restoration of the chiral symmetry changes the size of the gluon condensate. 
Whereupon we may conclude this discussion of the gluon condensate with 
two light quark from the MILC97 data~\cite{MILC97} by saying that the 
decondensation of gluons begins considerably below the critical temperature
and then slows down for a temperature range above it. Finally we remark
that we have taken a compromise value for the vacuum expectation value 
$G^2_0=1.95[GeV/fm^3]$. This choice for the simulations with dynamical 
quarks is somewhat larger than the earlier value~\cite{SVZ1} used for 
the pure gauge theory in the figure~\ref{fig:glucondgauge}. 
However, it is well within the range of many of the later 
estimates (for the literature see~\cite{Shur,DoGoHo}).\\
~\\ 
\noindent
{\large{\bf{IV.4 Gluon and Quark Condensates with more massive Flavors}}}\\
~\\
\noindent
After this evaluation of the gluon condensate for the very light quarks 
we present the results for two and three moderately light quarks 
using the Bielefeld data~\cite{KLP,KLP1,Peik}. Thereafter we look at 
the case with two lighter quarks together with one heavier quark.
We shall in general compare the effects of the pure gluon condensate 
with the maximally estimated effects arising from the quark mass terms, 
from which we can derive the changes in the thermal properties 
of the gluon condensate which are due to the dynamical quarks.\\
\begin{figure}[thb]
\centerline{\includegraphics[width=16.cm]{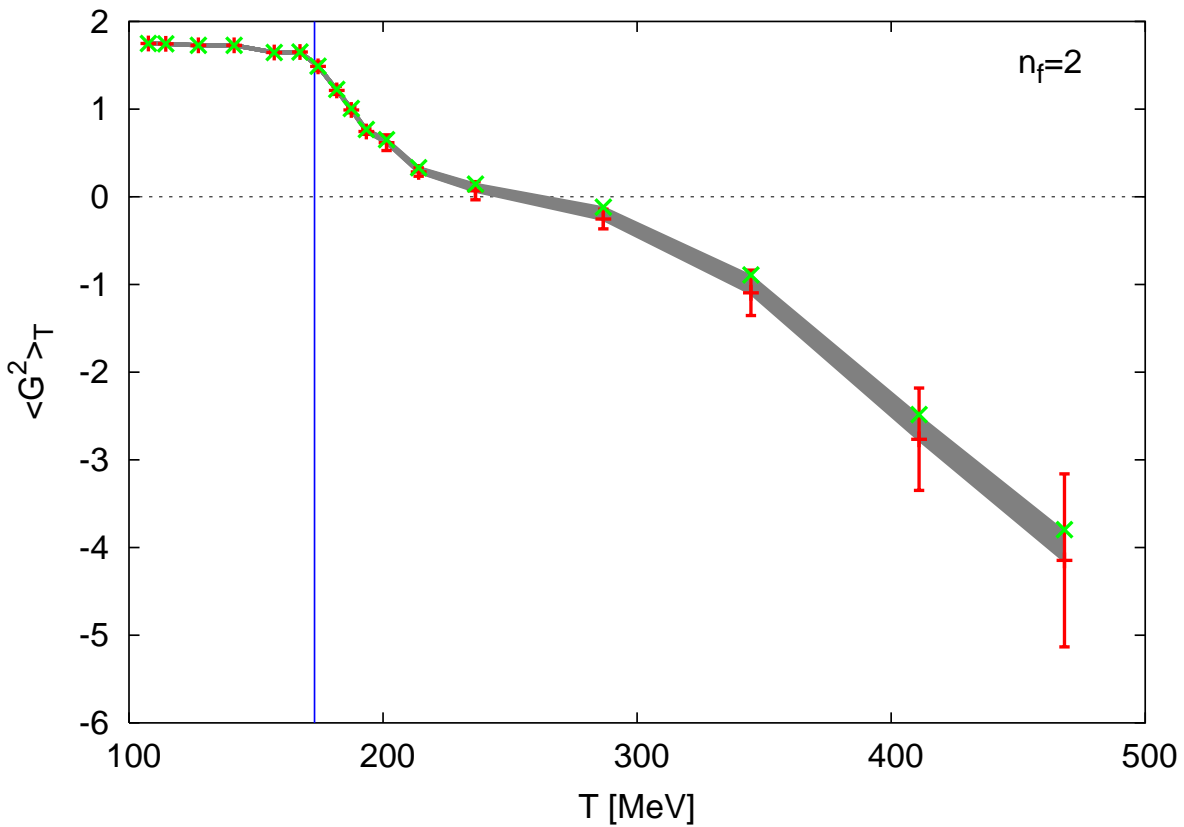}}
\caption{\it The pure gluon condensate $<G^2>_T$ in units of $GeV/fm^3$
for two moderately light quarks is indicated by the vertical/horizontal lines. 
The crosses show the estimated maximal effects of the (thermal) quark masses. }   
  \label{fig:glucond2f}  
\end{figure}
\indent
     We see in Fig.~\ref{fig:glucond2f} the effects of decondensation
of colored quark-gluon fields containing two fairly light 
dynamical quarks. Here we can clearly notice that above $200~MeV$ the 
difference between the pure gluon condensate with the errorbars and
the mass terms appearing with the chiral restoration given by
the crosses. The darkened region between the horizontal line and
the cross $\times$ is the immediate effect of the thermal mass term caused by 
the chiral restoration. We can see here that the thermal effects of the 
mass terms coming from the quark condensate do partially 
slow down the decondensation of the gluons. This effect we can 
compare with one of our earlier analyses~\cite{Mil99,Mil00} 
of the data for $n_f = 2$ using the even earlier MILC 
data~\cite{BKT94,MILC96,MILC97}. For the MILC data the quarks were 
much lighter-- almost an order of magnitude less--  from less than 
$5MeV$ to around $15 MeV$. We can now compare the numerical data in 
the last section for the equation of state shown for different flavor 
numbers in the figure~\ref{fig:equostaquark}. The gluon condensate 
we shall denote here as $\langle G^2 \rangle_T$. In the following figures 
we shall leave out the data for the temperatures above $500MeV$ because 
of the very large errorbars. Unfortunately, in the following analyses 
we have lacked the comparable data for the chiral averages for the 
quark condensate $\langle \bar{\psi}_q{\psi}_q\rangle_T$ which 
may be associated with the temperature as was the case for the MILC 
data~\cite{MILC97}. Thus the shaded region for the curve above $T_c$ gives 
the estimated maximal range of the effect of the chiral condensate, 
which in some cases may reach beyond the errorbars.\\
\begin{figure}[thb]
\centerline{\includegraphics[width=16.cm]{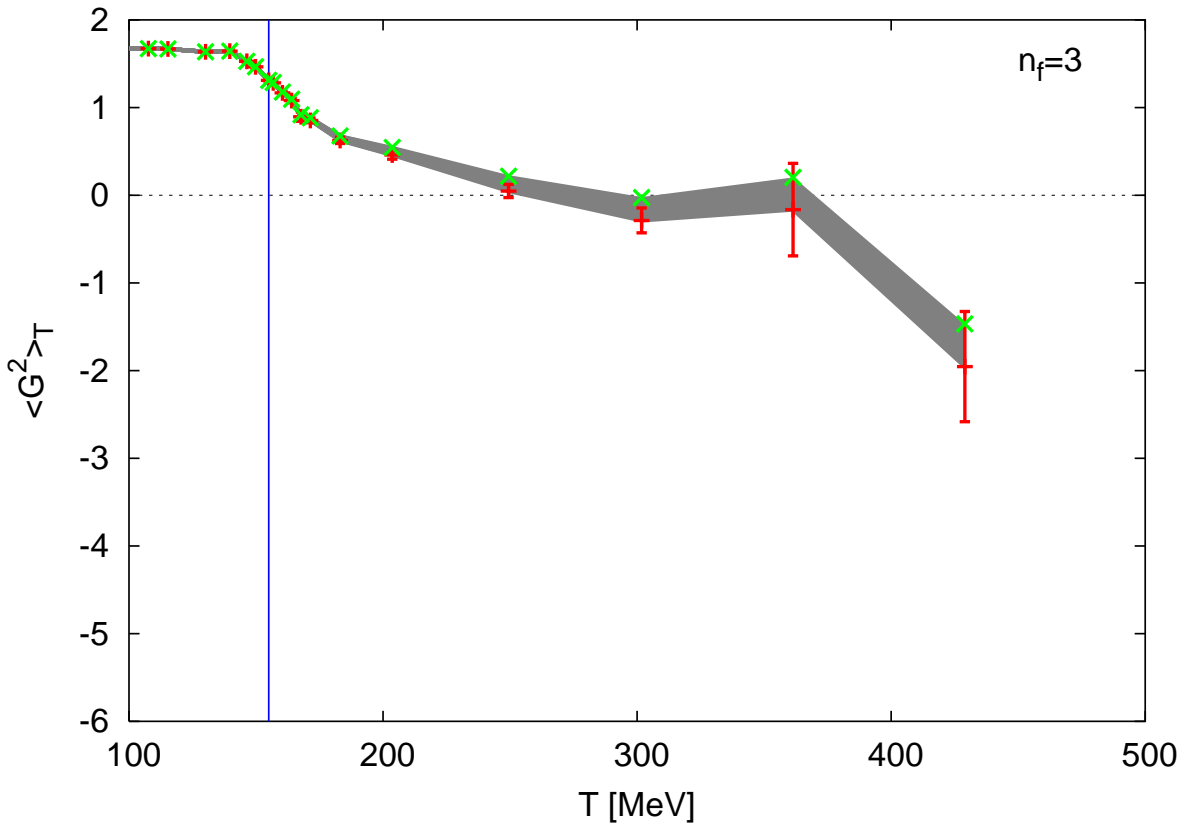}}
\caption{\it The gluon condensate$ <G^2>_T$ for three middle light quarks
in physical units $[GeV/fm^3]$.}   
  \label{fig:glucond3f}  
\end{figure}
~\\
\indent
     Next  we show in the figure~\ref{fig:glucond3f} the effects of three 
moderately light quark flavors on the gluon condensate $\langle G^2 \rangle_T$. 
A quick examination of this plot shows that the change of $\langle G^2 \rangle_T$ 
is not so  rapid over the same temperature interval as it was the case 
for the two quark flavors with similar values for the masses. Again we 
also show here the maximal estimated  effect coming out of the masses 
which when it is taken into account raise the points in the chiral limit. 
This same effect we have already just seen in the MILC data in
figure~\ref{fig:glucondqu}. The small cross $\times$ marks the estimated 
highest value for  $\langle G^2 \rangle_T$ due to the chiral restoration.\\
~\\
\indent
     We show in the next figure~\ref{fig:glucond2+1f} the computed results 
due to the effects of two lighter quark flavors together with one much heavier quark.
A quick examination of this drawing shows that the large masses of the dynamical
\begin{figure}[thb]
\centerline{\includegraphics[width=16.cm]{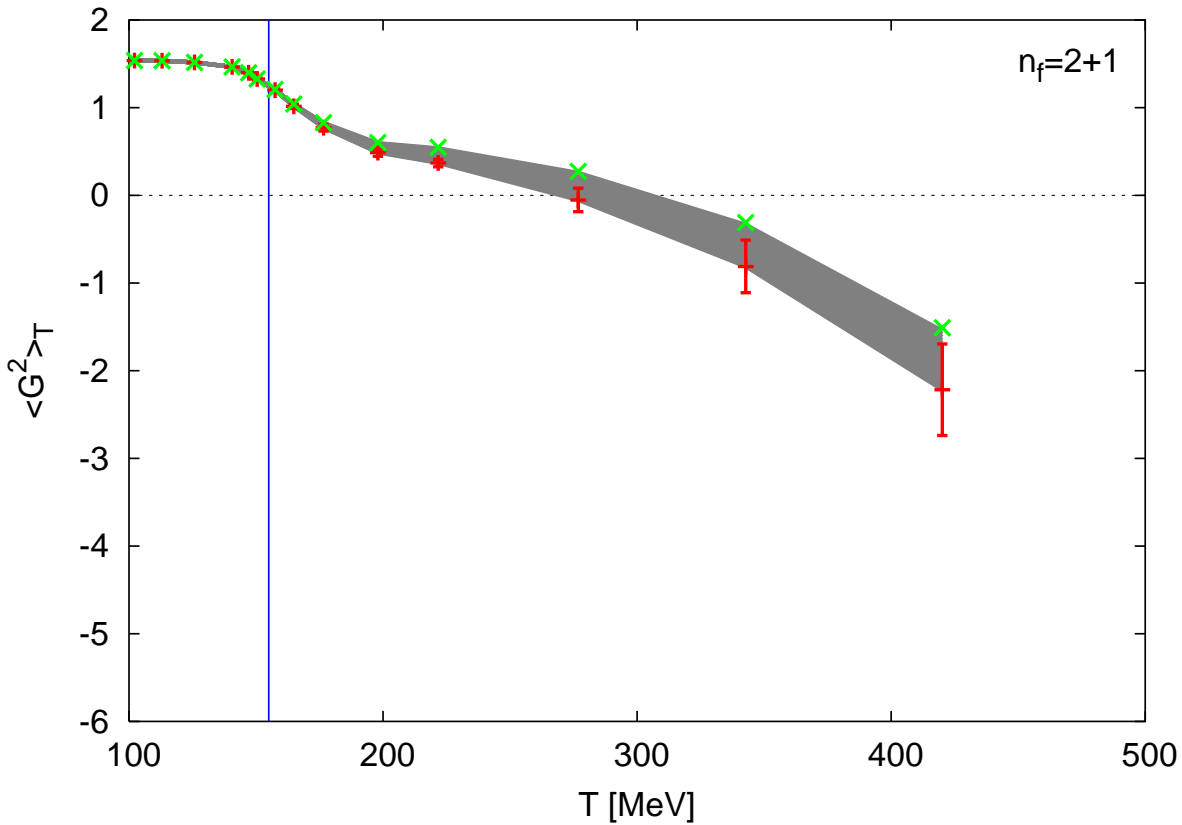}}
\caption{\it The gluon condensate$ <G^2>_T$ for two moderately light quarks
and one much heavier one. }   
  \label{fig:glucond2+1f}  
\end{figure}
quarks flattens the descent of the decondensation curve even further
in comparison to the other two cases. Clearly in this case the mass 
effects are much larger, which is primarily due to the much more
massive "strange" quark. We see that at higher temperatures the crosses
$\times$ are generally above the errorbars for the pure gluon condensate.  
It is obvious in this case with a heavier quark that we are very far 
from the situation of the chiral limit. Thus the restoration of the
chiral symmetry has much less effect on the gluon condensate in the
temperature region above the critical temperature $T_c$ and below
the deconfinement temperature $T_d$ of the pure $SU(3)$ gauge theory.
This remark is quite consistent with what we had noted in the last
part in the figure~\ref{fig:chir2} where we compared the effects of
the quark masses on $<\bar{\psi}\psi>$. We saw there that the larger
mass value caused the curves for $<\bar{\psi}\psi>$ to move towards 
higher values in the coupling $\beta$. Furthermore, the descent of
the curves at higher masses is slower, which means that the resoration
is less rapid at $T_c$ for the more massive dynamical quarks.\\ 
~\\
\noindent
{\large{\bf{IV.5 Comparisons of Properties of Gluon Condensates}}}\\ 
~\\  
\noindent
In this segment we discuss some previous work~\cite{Mil00} 
on gluon condensates. In particular, we shall take 
some special extreme examples out of the cases for 
which we have numerical and analytical evaluations. In the 
previous part of this report we have presented some numerical results
for the gluon condensates-- first for the pure lattice gauge theory, 
then for the light dynamical quarks with two flavors and finally for 
the somewhat heavier dynamical quark flavor combination. Therein 
we had also made some comparitive remarks between the various cases. 
However, in the preceeding part we have not tried to carefully compare 
the results for each system. Some years ago such a comparison had been 
discussed between the gluon condensates arising from the pure 
$SU(3)$ lattice gauge simulations~\cite{Boyd,BoMi} with those
coming from the light dynamical quark data from the MILC collaboration
~\cite{MILC97}. As an additional point of comparison~\cite{Mil00} a gluon 
condensate arising from a pure ideal gas of gluons was also introduced 
with the same ground state value as the QCD vacuum gluon condensate $G^2_0$,
for which the $0.012GeV^4$ in often taken~\cite{SVZ1}. For the ideal 
gluon condensate the condensation temperature $T_0$ was chosen to be 
$T_c$, the deconfinement or critical temperature of the pure lattice gauge theory.
We show the results of these comparisons in the figure~\ref{fig:gluconquar}.    
\begin{figure}[tb]
   \begin{center}
\epsfig{bbllx=127, bblly=265, bburx=450, bbury=588,
file=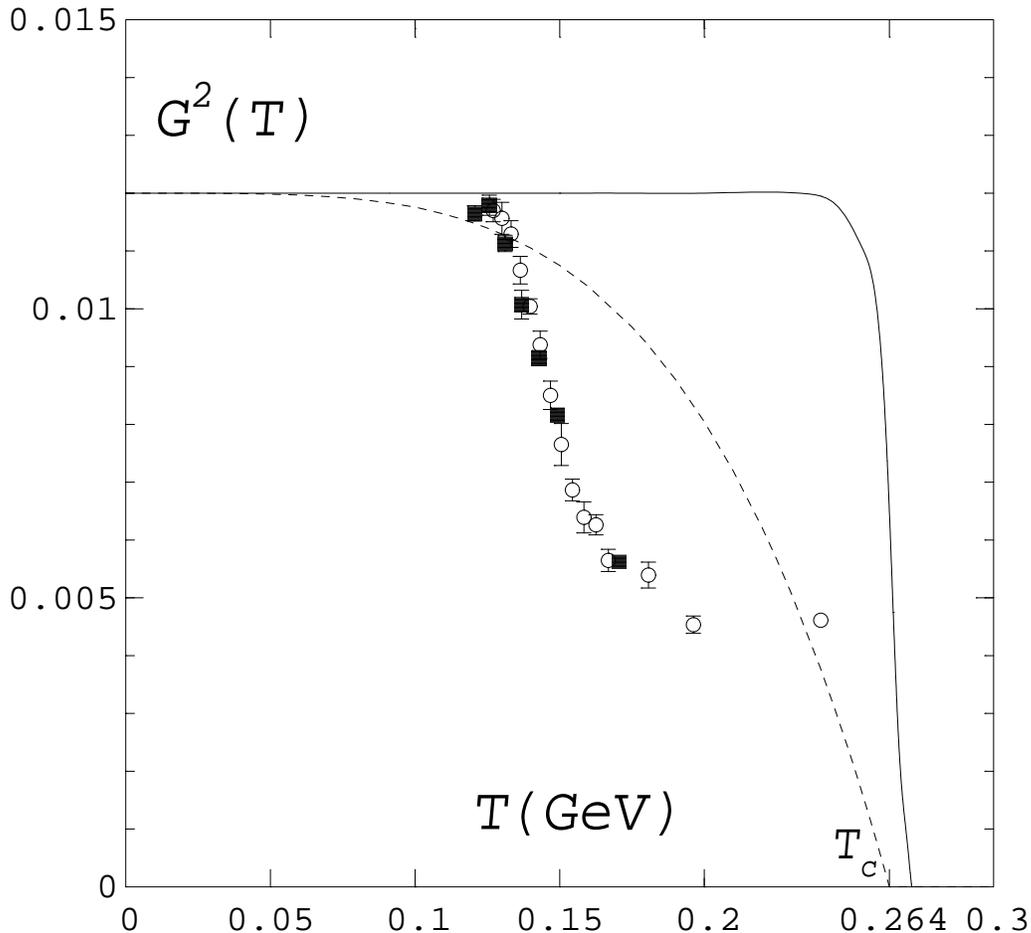, width=11.5cm}
\\[0.5cm]
     \caption{The lines show the gluon condensates for SU(3) (solid) and the 
ideal gluon gas (broken) in comparison with that of the light dynamical quarks 
denoted by the open circles and the heavier ones  with filled squares. The 
error bars are included when significant. The physical units are for $G^2(T)$ 
$[GeV^4]$ and T $[GeV]$.}
     \label{fig:gluconquar}
   \end{center}
\end{figure} 

~\\
\indent
     We now briefly discuss the condensation of an ideal gas of gluons, 
which we simply write as $G^2(T)$ for a finite temperature T. 
The vacuum expectation value of the pure gluonic system we denote 
by $G^2_0$, which has the above known value at zero temperature~\cite{SVZ1}. 
From dimensional considerations of the structure of the gluon condensate 
we can write down the ideal gluon condensate for $T\leq T_0$  
\begin{equation}
  \label{eq:idglucon}
   G^2(T)~=~G^2_0~{\left[1~-~{\left({T \over T_0}\right)}^4 \right]},
\end{equation}
where $T_0$ is the temperature at which the condensation process ceases. 
For $T\geq T_0$ then $G^2(T)~=~0$. This means that for the gluon 
condensate of an ideal gas system in equilibrium with its groundstate 
contribution must vanish above the condensation temperature. 
This form is just a generalization of the relativistic 
Bose-Einstein condensation to a four dimensional Euclidean 
space with $T_0$ acting as the critical temperature. Therefore, 
this equation gives a simple relation between the gluon 
condensate at a finite temperature $T$ and that at zero 
temperature. Thus we relate $G^2_0$ to the QCD vacuum 
condensate and $T_0$ to the deconfinement temperature, which 
are the corresponding quantities in the previous parts. Although 
this form for $G^2(T)$ as an ideal gluon gas seems at the 
present time very highly oversimplified, many of the earlier 
analyses in QCD at finite temperatures with gas models had 
assumed free gluons. This includes some earlier collaborative 
work of the author~\cite{BiMi} where this type of thermal behavior 
was assumed as a first approximation to the decondensation of 
gluons. The extent to which this assumption is valid we are 
now able to evaluate from the pure $SU(3)$ data~\cite{Boyd}. 
Furthermore, we will make no attempts here at the inclusion  
of the hadrons in the thermodynamics of the gluon condensate.\\
~\\  
\indent
     At the beginning of the previous part we have shown the 
results~\cite{BoMi} for the gluon condensates of the $SU(2)$ 
and the $SU(3)$ invariant lattice gauge theories over the whole 
temperature range up to $0.4GeV$. In both cases the uppermost 
temperature is well above the deconfinement temperature, 
for which the values of $G^2(T)$ are all less than zero as is shown
in the figure~\ref{fig:glucondgauge}. In the following paragraph
we used the argument of Leutwyler~\cite{Leut} in order to explain 
why the pure gluon condensate remains constant while retaining 
its vacuum value  $G^2_0$ almost until the temperature
\footnote[9]{ H. Leutwyler, "Quark and Gluon Condensates", 
plenary lecture at QUARK MATTER '96, Heidelberg, Germany on 
24. May 1996. Some further details on this approach were given by 
Heinrich Leutwyler on this subject in a private discussion with 
Graham Boyd and the author~\cite{BoMi} at Heidelberg. Here we take
$T_c$ as the {\it{critical}} temperature for the pure gluon gas.} 
reaches $T_c$. This effect clearly comes from the strong QCD
attractive forces between the gluons. Just below $T_c$ the drop 
in $G^2(T)$ signifies the start of the decondensation process. 
Not only do the gluons coming from the interactions in the 
vacuum condensate decondense, but the further gluons which 
are created at these high temperatures also take part in 
the decondensation process. We now can understand the 
difference in the behavior for the condensate of the ideal 
gluon gas which is shown in the figure~\ref{fig:gluconquar}. 
This gas of free gluons shows a continual decrease even at 
the low temperatures well under $T_c$. At and above $T_c$ 
the ideal gluon gas condensate vanishes so that $G^2(T)~=~0$ 
for all $T\geq T_0$. However, the pure $SU(N_c)$ gluon 
condensate continues into negative values of $G^2(T)$ 
as it was shown before in the Figure~\ref{fig:glucondgauge}. 
Thus the difference between the pure lattice gauge 
thermodynamics and the ideal gluon gas is apparent 
over the whole range of finite temperatures.\\
~\\
\indent
     The comparison of the gluon condensate from the two light
dynamical quarks brings new issues into these considerations. 
As we have discussed in the preceeding two parts of our work, 
the presence of the chiral restoration transition changes 
the thermodynamics completely. In the equation of state 
$\varepsilon-3p$ for the case of the two light quarks 
we can see in figure~\ref{fig:eqstqu} how around 
$140MeV$ both the curves rise very rapidly. However, at
about $200MeV$ the lighter quark mass slows up and even
appears to drop a little. Not only is this change at much
lower temperatures than the pure gauge results in the
figure~\ref{fig:eqstgauge}, which show a steady rise 
beginning at a much higher temperature. The presence of the
errorbars in the figure~\ref{fig:eqstqu} show that the
apparent decline of the last point at about $240MeV$ is
well within the errorbars of the preceeding point at
about $196MeV$. The MILC data is then again plotted for
the gluon condensate $G^2(T)$ in the figure~\ref{fig:glucondqu} 
for which the units are taken as $[GeV/fm^3]$ in contrast to the 
above figure~\ref{fig:gluconquar}. In figure~\ref{fig:glucondqu} 
we have connected the points together in order to provide
a better view of the changes. The two curves with the errorbars
have included the effects of the masses with the quark condensates, 
which adds some further errors. Nevertheless, we see that the
gluon condensate then continues to fall slightly. The unconnected
points in the above figure~\ref{fig:gluconquar} correspond to
the unconnected points given in the chiral limit. This fact has 
been recently pointed out~\cite{BroRho02,BGLR} in relation to 
the "soft" glue, which disappears much earlier than the hard glue 
or "epoxy". It can also be related to the problem of mass and 
the presence of mesonic bound states~\cite{PaLeBr} above $T_c$, 
which here denotes the actual critical temperature for the two
flavor light quark thermodynamical system. 
We shall return to this topic in relation to chiral 
resoration models in a later section of the next part.\\
~\\
\indent
     Finally we are now in a position to discuss the interrelationship
between the process of the gluon decondensation and the restoration of
chiral symmetry. In the previous three figures~\ref{fig:glucond2f},
~\ref{fig:glucond3f} and \ref{fig:glucond2+1f} we can see the varying
extents that the dynamical quark properties affect the amount the
two different broken symmetries are altered with the change of the
temperature. As it has been clearly stated by Leutwyler~\cite{Leut}, 
the quantity $<\bar{\psi}\psi>$ acts like an order parameter for 
the breaking of the chiral symmetry. Actually it is the term 
$1~-~<\bar{\psi}\psi>$ in the equation (\ref{eq:fullcond}) which
acts as the {\it restoration quantity} for the effects of the quark 
condensate on the gluon decondensation. In the figure~\ref{fig:glucondqu} 
we can see the direct effects of this term which clearly lowers the 
gluon condensation curve with increasing temperatures due to the
restoration of the chiral symmertry. Naturally the bigger effect of
the decondensation of the gluons arises in the interaction measure 
$\Delta_m(T)$ in equation (\ref{eq:intactmeasm}) from the mass 
renormalization of the quarks. We see that the antiquark-quark 
average over the lattice or site average clearly lowers the plaquette
average from the links in the interaction measure. This we had
already seen in figure~\ref{fig:eqstqu} for the equation of state 
for the light dynamical quarks. The equation of state for the 
dynamical quarks still provides the most important contribution
to the decondensation of the gluons. However, the extent to 
which this last statement is true varies with the number of
flavors and the masses of the quarks. Thus we can clearly see the 
main points of contrast between the decondensation of the gluons 
in the pure gauge theory and that of the theory with the
dynamical quarks.\\  
\newpage
\noindent
{\Large{\bf{V.Discussion of Physical Results}}}\\
~\\
\noindent
In this part of our work we shall describe how we are able to calculate
some physical quantities from the structure of the equation of state
which relate to such phenomena as the possible phase transitions
between the hadronic states and the much sought after quark-gluon plasma. 
First we discuss some models relating to the relativistic high energy
collisions at high temperatures and large energy denstiies~\cite{Shur}. 
Thereafter we shall look at the low energy and low temperature limits where 
the effective Lagrangians are very useful. We discuss a model for scalar 
meson dominance with the coupling to the trace of the energy-momentum tensor
~\cite{FreuNam}. In the last section of this part we look at some other 
aspects of the equation of state which arise at temperatures well below
the critical temperature. In our last comments at the end we mention how 
these considerations may relate to some newer lattice simulations not included
in the presented data.\\
~\\
~\\
\noindent
{\large{\bf{V.1 Phenomenological Models of Quark and Gluon Properties}}}\\ 
~\\
\noindent
One of the most successful models for the highly relativistic 
nucleus-nucleus collisions was proposed almost a quarter of a 
century ago by Bjorken~\cite{Bjork}. He suggested a senario where 
a quark-gluon plasma should survive on a time scale of $5fm/c$, 
where c is the speed of light which we have usually taken to be unity. 
During this short time the collision region maintains a longitudinal 
flow which should be described according to the laws of relativistic 
hydrodynamics~\cite{LanLif6}.\\ 
~\\
\indent
     The dynamical structure of the Bjorken model can be reduced 
to that of a 1+1 dimensional Lorentz boost invariant system in 
the time and space coordinates t and z. Bjorken, however, then 
chose to represent the flow in terms of the "light cone" variables, 
for which he took the propertime $\tau=\sqrt{t^2-z^2}$ and the rapidity 
$y={1\over2}\ln{{t+z}\over{t-z}}$. These variables are useful in 
maintaining the boost invariance. In addition to the usual energy 
momentum conservation Bjorken further postulated~\cite{LeBel96,Shur} 
a conservation of the entropy density $s=\beta(\varepsilon+p)$, where 
$\beta$ is the inverse temperature. Out of these assumptions one gets 
for the fourvector entropy density $s^{\mu}=su^{\mu}$ the conservation law
\begin{equation}
 {\partial}_{\mu}s^{\mu}=0, 
\label{eq:entdencon}
\end{equation}
\noindent
where $u^{\mu}$ is the fluid fourvelocity with $u_{\mu}u^{\mu}=1$.
After setting the initial propertime $\tau_{0}$ with the entropy density
$s(\tau_{0})$, one finds simply the solution~\cite{Bjork}
\begin{equation}
 {s(\tau)\over{s(\tau_0)}}={{\tau_0}\over{\tau}}. 
\label{eq:entdensol}
\end{equation}
\noindent
In terms of the variables $\tau$ and $y$ the entropy density stays on the
lightcone. This result may be interpreted as the entropy content per
unit rapidity is a constant of the motion~\cite{Bjork}. Furthermore,
this result may be related to the Bjorken equation of state, which 
may be written as
\begin{equation}
\varepsilon=p(3+\Delta_1). 
\label{eq:bjeqnost}
\end{equation}
Comparably the entropy density for the Bjorken model may be written as
\begin{equation}
s={\beta}p(4+\Delta_1) . 
\label{eq:bjorkeqnost}
\end{equation}
\noindent
The new quantity $\Delta_1$ is defined in the Bjorken model as follows:
\begin{equation}
 \Delta_1={T\over{n}}\left(dn\over{dT}\right). 
\label{eq:bjorkentden}
\end{equation}
\indent
     In the rest of this very original work Bjorken~\cite{Bjork} 
showed how these equations can be used for the effective number 
of plasma degrees of freedom $n(T)$ which he had originally
proposed to have a sharp rise for the temperatures above 
the transition temperature around $200MeV$. An important 
application of an extension of the Bjorken equation of 
state~(\ref{eq:bjorkeqnost}) has been found in the calculation
of the radiative energy loss of high energy partons traversing 
an expanding QCD plasma~\cite{Baieretal}.\\
~\\
\indent
     For us in this work what is really significant here is that the 
Bjorken equation of state leads to corrections beyond the ideal gas 
as well as the simple bag model pictures. Bjorken~\cite{Bjork} also 
pointed out that the general features of the hydrodynamical expansion
follow from the positivity condition on the trace of the energy 
momentum tensor. In fact, this statement would imply that the  
positive definiteness condition on the trace of the energy 
momentum tensor is such that the trace itself would only be 
zero in the absence of all interactions.\\
~\\
\indent
     More recently the hydrodynamical models~\cite{Shur} have been brought in 
relation to the Bjorken model. In this context the well known ideas of
relativistic hydrodynamics~\cite{LanLif6} have been extended for use
in the explanation of the data from heavy ion collisions undergoing 
a hydrodynamical expansion~\cite{Dirky}. At the present time there is a  
considerable amount of phenomonological work in this direction. \\
~\\          
\noindent
{\large{\bf{V.2  Theoretical Models for the Quark and Gluon Properties}}}\\ 
~\\
\noindent
In this section we shall look into some of the theoretically important 
statistical models for strong interaction physics. Some of these relate 
to the recent RHIC data~\cite{Shur}. First we mention the Statistical 
Bootstrap Model (SBM) which was earlier regarded for providing 
the limiting temperature for hadronic physics~\cite{Haged}. 
In the following paragraphs we shall discuss here some further models 
involving the quarks and gluons in large statistical systems.\\
~\\
\indent
     Now we discuss the properties of resonances at high temperatures 
and densities which are typical of high energy  particle collisions. 
The particle creation process at such high energies is the starting point 
for the SBM first introduced in the midsixties by Rolf Hagedorn~\cite{Haged}. 
Earlier pure statistical models based on ideal gas theory with the proper 
statistics related the temperature directly to the center of mass energy.
These models had in common an exponential energy distribution relating to 
the probabilities of each of the hadronic states. However, Hagedorn had 
noted that in high energy proton-proton collisions that the masses $m$
are distributed in what appeared to be asymptotically an exponentially
rising mass spectrum $\rho(m)$. This spectrum included the entire known
resonance structure of hadronic particle physics. Furthermore, he then
postulated that the mass spectrum approached asymptotically the energy 
level density $\sigma(E)$ in the high energy limit. After having
calculated $\sigma(E)$ from the partitiion function with  $\rho(m)$ 
for a gas of particles and resonances, he invoked the statistical
bootstrap condition relating $\rho(m)$ asymptotically to $\sigma(E)$. 
However, this condition only held true up to a particular temperature 
$T_0=158\pm3[MeV]$, when certain statistical assumptions were made on 
the pararmeters of $\rho(m)$. Nevertheless, this temperature $T_0$ had 
found its special place in the particle phenomenology of that time 
through its place as the limiting temperature. A more contemporary 
understanding of this resonance formation is found from an examination of
this Hagedorn temperature and the partition thermodynamics~\cite{BFS}. The 
hadronic phase with all the included resonances can only exist below $T_0$. 
This statistical theory was clearly modeled on the particle states which were 
known at that time. The sense of such a model with a limiting temperature 
has changed drastically with the inclusion of quarks as the basic 
constituents of the hadrons. More recently some further implications in 
relation to the quark-gluon transition have been discussed~\cite{BJBP}.
The SBM serves as a starting point of a statistical treatment of the
hadrons with their constituent quarks and interacting gluons.\\
~\\
\indent
     In the early nineteen seventies not long after the initial 
development~\cite{DoGoHo,Kugo,Muta,Poko} of QCD as the gauge theory 
for the strong interactions between the quarks, there was a great 
deal interest in finding analytical solutions for particular cases. 
The discovery of the instanton solution~\cite{Beletal} in 1975 of the 
$SU(2)$ Yang-Mills field and its further development~\cite{Shur,GtH} 
were important starting point for an approximate analytical approach 
to nonperturbative QCD. After a brief initial excitement 
giving hope of a discovered particle in the years directly 
following these discoveries, there began an era of long and 
systemmatic investigation of these rather intriguing mathematical 
objects~\cite{Raja,Schwarz} in the name of the instanton gases and 
liquids with various forms~\cite{Shur}. More recently the importance 
of the bound states and the instanton molecules~\cite{GerryLeeRhoEd}
has become clearer. Furthermore, it was noted that the fermionic (quark) 
zero modes can be related to the axial anomaly~\cite{NoRhoZah}. This 
leads to the properties of the chiral symmetry breaking and its 
restoration~\cite{BGLR} as we had mentioned previously. A more lengthy 
discussion of these ideas can be found~\cite{Shur,DoGoHo} which 
give details on the explicit structure. The relation to some earlier lattice 
computations is also quite important (see~\cite{Roth92} in section 17.6).
Furthermore, the thermodyanmical properties of the instantons 
relating to the topological charge $Q^{2}_{T}$ 
can be evaluated on the lattice using the thermal 
susceptibility in equation (\ref{eq:thermsusc}) as indicated above in 
the equation (\ref{eq:topcond}). There are also related
lattice simulations on the flux tubes (see~\cite{Roth92} section 17.7) 
for further discussions on the newer evaluations. Further work involving 
the instanton has been presented by  Ilgenfritz and Shuryak~\cite{IlgenShur} 
in a simple model at finite temperature. The role of the instantons 
has been generally discussed by Shuryak~\cite{Shur} in relation to the 
transition from hadrons to the quark gluon plasma. As we mentioned in the
last part, there was an attempt~\cite{BiMi} to formulate a mean field model 
of an instanton (caloron) gas at finite temperature in contact with the
condensate of an ideal gluon gas.\\
~\\
\indent
     The early finite temperature simulations~\cite{Creu83,Roth92} for 
lattice gauge theory provided information of the thermodynamical functions 
around the critical temperature $T_c$. The discussion of some previous 
results for the gluon and quark condensates were derived from some of
the earlier simulations. In this relationship we must mention that 
the values of the vacuum condensates were calculated out of sum 
rules~\cite{SVZ1,Nari}. The early use of the finite temperature 
lattice results~\cite{AdHatZah,kochbrown} has had a significant 
part in thedevelopment in such computations. These authors had 
evaluated the gluon condensate in the separate parts arising from the 
chromoelectric and chromomagnetic fields with the thermally averaged 
condensates as seen in reference~\cite{kochbrown} in their Figure 4.
Unfortunately in the earlier lattice data the temperature scale was very 
hard to determine. In 1994 this changed significantly through a better 
evaluation of the lattice beta function~\cite{EKR94} for the SU(2) gauge 
theory\footnote[10]{The author the made an analysis of the trace anomaly 
relating to the lattice data from the interaction measure to the gluon 
condensate, under the title "The Trace of the Energy Momentum Tensor
and the Lattice Interaction Measure at Finite Temperature", Bielefeld
Preprint, BI-TP 94/41. This first calculation was extended in later 
works~\cite{BoMi,Mil97,Mil99,Mil00} using the same approach in which  
data for SU(3) pure gauge theory as well as for dynamical quarks were 
included.}. Even in these earlier times the use of these ideas was 
apparent in the study of the hadron to quark-gluon transition~\cite{BJBP}.\\ 
~\\
\indent
     Finally we mention some other work involving QCD at finite temperature
and density. First we remark on the models involving quasiparticles~\cite{Pesh, RebRom}
which take into consideration the lattice properties of the high temperature states.
These models generally reproduce the lattice data above about $2T_c$. Some recent 
lattice work~\cite{EngHolSchu} using a QCD model with two flavors of adjoint quarks 
provides a very unique picture of both the deconfinement and chiral phase transitions. 
It shows both a chiral second order transition and a first order deconfinement phase 
transition. However, it can be seen in their Figure 12 showing the chiral condensate 
in terms of the coupling $\beta$ and the lattice quark mass ${m_q}a$ that in this 
model the deconfinement transition appears at a lower temperature than 
that of the chiral restoration. In fact they estimate~\cite{EngHolSchu} 
that $T_c/T_d\approx7.8(2)$, which means that the critical temperature
remains very far above the deconfinement temperature.\\
~\\
\indent
     In this section we have related the results of numerical simulations 
which are usually represented as ratios of the quantities to powers of the 
temperature to the actual physical quantities like the pressure, energy density,
entropy density. We can arrive at a form of the equation of state which can be
used to find other physical quantities.
We look again at the first pure gauge theory simulations at finite
temperature, which provided for us a way to arrive at the thermodynamical 
quantities in the gluon condensate.\\
~\\
\noindent
{\large{\bf{V.3 Scaling Properties in Matter}}}\\ 
~\\
\noindent
There are a number prominent examples where the parameters in the effective 
actions for physical processes need be determined by some further rules.
Important among these is the Brown-Rho Scaling properties for finite densities 
$\rho$, which we have mentioned in the introduction~\cite{BroRho91,BroRho02}.
It is commonly given by a scaling function ${\Phi}(\rho)$ for the different 
hadrons (for a general description see reference~\cite{BroRho02}) in the 
following form:
\begin{equation}
{\Phi}(\rho)~\approx~{{f^{*}_{\pi}(\rho)}\over{{f_{\pi}}}}~\approx~
{{m^{*}_{\sigma}(\rho)}\over{{m_{\sigma}}}}~\approx~{{m^{*}_{V}(\rho)}\over{{f_{V}}}}
~\approx~{{M^{*}(\rho)}\over{{M}}}. 
\label{eq:brorhoscal}
\end{equation}
\noindent
The expressions $f^{*}_{\pi}(\rho)$ and $f_{\pi}$ are the pion decay constants in 
the medium with the density $\rho$ and that of the vacuum respectively. In this same 
notation (with the $*$ above) the masses of the scalar $\sigma$ and vector $V$ 
particles are compared with the corresponding values in their different media. 
In the last ratio there appears the nucleon mass $M$.  Nevertheless, it is 
important in all cases to stress that the Brown-Rho Scaling is a 
{\it{mean field relation}} which emerges at the tree level from the 
effective Lagrangian for the chiral fields~\cite{BroRho91,BroRho02}.
This scaling rule may be further approximately generalized 
to the quark condensates are related to the hadron masses.
\begin{equation}
{{m^{*}}\over{m}}~\approx~{{\langle{\bar{q}q}\rangle^{*}}\over{\langle{\bar{q}q}\rangle}}. 
\label{eq:quarcondscaling}
\end{equation}
\noindent
This case is referred to as the "Nambu scaling" which usually 
includes the temperature dependence~\cite{kochbrown,BroRho02}.
This property is a generic feature of the linear sigma model. 
It is then supposed to be upheld as the temperature rises towards 
the chiral restoration temperature. It was noted in a recent work 
how the matter induced modifications of certain resonances fitting 
the masses of various particles~\cite{EdGerry} causes a shift as 
well as a change of shape of various particles and resonances. Further 
work on the nature of the chiral restoration~\cite{BGLR}, used an earlier
work~\cite{Mil00} of the author to discuss how the mixed phase~\cite{PaLeBr} 
appeared after the chiral restoration has started, but before the deconfinement
temperature in the pure gauge theory. We have already seen in the last part
that the Figure~\ref{fig:gluconquar} can be interpreted in terms of the 
soft glue, which disappears rather quickly with the chiral restoration, 
while the hard glue (epoxy) remains well above $T_c$. There these properties
were associated with the slowing up of the decondensation processes of the
light dynamical quarks. The heavier quarks gave a much flatter decondensation
curve for which the two different properties are harder to recognize. 
Furthermore, it is expected that the correlations above $T_c$ are, indeed, 
very significant and have an important relationship to the recent 
RHIC experements~\cite{BroRho02,BGLR}.\\
~\\
\noindent
{\large{\bf{V.4 Scalar meson dominance }}}\\
~\\
\noindent
In this section we look into the model of Freund and Nambu~\cite{FreuNam} 
for the dominance of the scalar meson\footnote[11]{The actual scalar meson 
states are discussed in the "Review of Particle Physics"~\cite{PDG} in a 
"Note on Scalar Mesons" pp. 506 -510 and 522 -526. The known low lying scalar 
meson states go under the names $f_0(600)$ (or $\sigma$) and $f_0(980)$ for 
the isoscalar as well as $a_0(980)$ for the isovector. These states are all 
known to decay into various pairs of pseudoscalar mesons and, of course, 
secondarily into pairs of photons.}. This property is deeply rooted in the 
empirical observations of the meson and baryon mass splittings. Already in 1968 
they had proposed that the classical meson field $\varphi(x)$ be coupled to the
trace of the energy momentum tensor $\theta^{\mu}_{\mu}$. A more detailed
discussion of the relation of these (pseudo)scalar fields to the energy momentum
tensor one finds in section three of Bogoliubov and Shirkov~\cite{BogolShir}. In
this model the nonhomogeneous Klein-Gordon equation could be written in the form
\begin{equation}
      (\Box~-~m^2)\varphi(x)~=~g{\theta^{\mu}_{\mu}},
\label{eq:scalmesdom}
\end{equation}
\noindent
whereby $m$ is the scalar meson mass and $g$ its coupling to the trace. Here 
it is important for us to note that this coupling $g$  differs from the usual
QCD coupling in its place in the model, which was, of course, not known at
that time. In the following paragraphs we briefly sketch the basic arguments
for this model which we shall later relate phenomenologially to the process 
of gluon decondensation. \\ 
~\\
\indent
     From the usual properties of field theory one can derive the trace 
of the energy momentum tensor~\cite{BogolShir}${\theta^{\mu}_{\mu}}$ 
from the Lagrangian $\mathcal{L}(\varphi,{\partial}_{\mu}{\varphi})$. 
Out of the above form of the Klein-Gordon equation~(\ref{eq:scalmesdom}) 
for the scalar field $\varphi(x)$ Freund and Nambu derived~\cite{FreuNam} 
their effective Lagrangian by taking into account the appropriate 
boundry conditions. It took the following form:
\begin{equation}
  \mathcal{L}(g,\varphi)~=~{\frac{1}{2}}{\partial}_{\mu}{\varphi}R(x)^{-1}
{\partial}_{\mu}{\varphi}~-~{\frac{1}{2}}m^2{\varphi}^2,
\label{eq:lagrang}
\end{equation}
\noindent
where $R(x)$ is defined to be $(1+2g\varphi(x))$. After a careful consideration 
of the form of the Lagrangian they~\cite{FreuNam} 
introduced a new field $\psi(x)$, which is defined by
\begin{equation}
      \psi(x)~=~g^{-1}R(x)^{1/2}.
\label{eq:psidef}
\end{equation}
\noindent
This field transformation yielded a new Lagrangian $\mathcal{L}'(g,\psi)$
in terms of the new field in the form
\begin{equation}
  \mathcal{L}'(g,\psi)~=~{\frac{1}{2}}({\partial}_{\mu}{\psi})^2~+
~{\frac{1}{4}}m^2{\psi}^2~-~{\frac{1}{8}}m^2g^2{\psi}^4-~{\frac{m^2}{8g^2}}.
\label{eq:newlagrang}
\end{equation}
\noindent
Furthermore, it must be noted here that the sign is wrong for the usual 
"mass term" in the Lagrangian density. This situation provides the 
conditions~\cite{DesGriPend} of a degenerate vacuum which gives rise 
to Goldstone bosons upon quantization.
It can be easily seen after one further change to $\psi_{\pm}$ as $\psi{\pm}1/g$.
Then we find this change of the fields results in
\begin{equation}
  \mathcal{L}'(g,{\psi}_{\pm})~=~{\frac{1}{2}}({\partial}_{\mu}{\psi}_{\pm})^2~-
~{\frac{1}{2}}m^2{\psi}_{\pm}^2~\pm~{\frac{1}{2}}m^2g{\psi}_{\pm}^3-
~{\frac{1}{8}}m^2g^2{\psi}_{\pm}^4,
\label{eq:newlagrangpm}
\end{equation}
\noindent
which is equal~\cite{FreuNam} to the original Lagrangian $ \mathcal{L}(g,\varphi)$.
As we would expect, the dynamical properties of both of these fields $\psi_{\pm}$ 
correspond to the same mass $m$. Either one of the Lagrangians $\mathcal{L}(g,\psi)$
or $\mathcal{L}'(g,{\psi}_{\pm})$ yields after quantization the same physical
properties in the corresponding S Matrix as the original one $\mathcal{L}(g,\varphi)$
derived~\cite{FreuNam} in equation (\ref{eq:lagrang}). Thereby it is clear that
there is a twofold degeneracy in the vacuum of the quantized theory. This model
even in its unquantized form was recognized by Bruno Zumino to break the scale
invariance~\cite{DesGriPend} as well as allow for the Goldstone bosons. 
Furthermore, this situation leads to the fact that the coupling of the trace 
$g{\theta^{\mu}_{\mu}}$ in the nonhomogeneous term of the Klein-Gordon equation 
(\ref{eq:scalmesdom}) relates directly to the square of the mass times the strength
of the transformed scalar field. Thus the Lagrangian of Freund and Nambu provides
a simple model for the scalar field dominance of the trace of the energy momentum
tensor $\theta^{\mu}_{\mu}$.\\
~\\
\indent
     The quantized theory can be achieved from the classical one in the usual way 
from the Poisson brackets~\cite{BogolShir}. The  momentum $\pi(x)$ canonically 
conjugate to $\varphi(x))$ from the original Lagrangian $\mathcal{L}(g,\varphi)$ 
is gotten as the Poisson bracket ${\frac{1}{2}}\{R(x)^{-1},{\partial}_0{\varphi}\}$.
Because of this form of the Poisson brackets the ordinary equal-time commutation 
relations do not specify the canonical commutator but only the mixed commutator 
in the form
\begin{equation}
{\frac{1}{2}}\{R(x')^{-1},[\varphi(x),{\partial'}_0{\varphi(x')}]\}_{x_0=x_0'}~
=~i\delta^3(x-x').      
\label{eq:mixcom}
\end{equation}
\noindent
From this form and the specification that $[\varphi(x),\varphi(x')]_{x_0=x_0'}$
and $[{\partial}_0{\varphi(x)},{\partial}_0'{\varphi(x')}]_{x_0=x_0'}$ both
identically vanish allows the calculation of the equal time field momenta
$[\pi(x),\pi(x')]_{x_0=x_0'}$. With the definition that at the same point
\begin{equation}
[R(x)^{-1},R(x)^{-1}]\delta^3(0)~=~0.      
\label{eq:rinvcom}
\end{equation}
\noindent
it is possible to evaluate the Schwinger commutator~\cite{Bert,Jack} for 
$\theta_{00}(x)$. More important to us in the present work is the fact 
that Freund and Nambu have evaluated the corresponding commutator 
for the trace of the energy momentum tensor $\theta^{\mu}_{\mu}(x)$ as
\begin{equation}
[\theta^{\mu}_{\mu}(x),\theta^{\mu}_{\mu}(x')]_{x_0=x_0'}~=
~-4i[\theta_{0i}(x)~+~\theta_{0i}(x')]{\partial}_i\delta^3(x-x').      
\label{eq:tracecom}
\end{equation}
\noindent
This feature allows that the trace of the energy momentum tensor be treated as a 
quantum field. These authors~\cite{FreuNam} have suggested that this commutation
relationship be considered as a general feature of the theory of hadrons. Thus
the Lagrangian $\mathcal{L}(g,\varphi)$ in their formulation provides a rather
simple model for scalar field dominance from the trace of the energy momentum
tensor. These results can be understood from the transformation properties 
of the action integral under the scale transformations~\cite{DesGriPend}.\\
~\\
\indent
     Next we work out the special case of the scalar meson field $\varphi(x)$ with 
the mass $m$ for the ground state structure in the simple bag model. For simplicity
we assume a simple cubic region with the sides of length $1[fm]$ for the discrete
momentum representation~\cite{BogolShir}. The solution for the zero mode of the
nonhomogeneous Klein-Gordon equation~(\ref{eq:scalmesdom}) for the scalar field 
$\varphi_0(x)$ using the isotropic bag model value~\cite{BJBP} of the energy 
momentum tensor ${\theta}^{\mu}_{\mu 0}~=~4{\mathcal B}$. In 
this case we find a simple time dependent solution $\varphi(x_0)$ with the
zero spatial modes for the scalar meson ground state solution
\begin{equation}
      \varphi_0(x_0)~=~{4g{\mathcal B}\over{m^2}}\left(cos(mx_0)-1\right),
\label{eq:hadgrdsol}
\end{equation}
\noindent
where we have taken $\varphi_0(0)=0$ and ${\dot\varphi_0}(0)=0$. This special case
is so easily solvable since there is no spatial dependence coming into the 
field equation from the trace of the energy momentum tensor. Thereupon the value 
of the amplitude only involves the nonhomogeneous term divided by the mass squared.
Then this special case yields just the simple harmonic oscillator type of solution.
Now we see for the chiral limit for the solution in equation~(\ref{eq:hadgrdsol})
that the solution goes to the finite value
\begin{equation}
\lim m \rightarrow 0,~~~\varphi_0(x_0)~\rightarrow~{2g{\mathcal B}}{x_0}^2.
\label{eq:chirhadgrdsol}
\end{equation}
\noindent
Thereby the small mass limit for the zero momentum state has a finite value which 
is positive definite from ${x_0}^2$. Since this zero momentum state solution 
has no spatial dependence in it, we may replace $x_0$ by the propertime $\tau$ 
at $\vec{x}=0$. Then we find at the center for $\lim m \rightarrow 0$ that
\begin{equation}
        \varphi_0(\tau)~=~{2g{\mathcal B}}{\tau}^2.
\label{eq:chirprop}
\end{equation}
\noindent
Thus we would expect the scalar field for the hadron to grow quadratically in 
the proper time, for which only on the light cone does it totally disappear. 
Furthermore, after a little further investigation, we find that all the 
higher mode solutions in the discrete momentum representation would involve 
free spatial wave solutions set to zero at the boundries of the box. 
However, these solutions involve an infinite sum in each spatial dimension.\\
~\\
\indent
     As a concluding point to this discussion we comment on the interesting
possibilities of this model from the view of the equation of state. Next
we shall consider the trace of the total energy momentum tensor 
$T^{\mu}_{\mu}(T)$ as a function of the temperature. In the simple 
solution found above for $T=0$ we saw that the trace appeared only 
in the solution form the bag constant $4{\mathcal B}$. Then we would expect that 
even at finite temperatures this same type of dependence on the equation 
of state would directly enter into the solution. For the sake of 
simplicity we shall now only consider the effects of the decondensation 
for the pure $SU(3)$ gluons as shown in the Figure~\ref{fig:glucondgauge}. 
We recall that the decondensation of gluons starts around the
deconfinement temperature of about $270MeV$, where the equation of state
undergoes a rapid growth as shown in the first figure for the temperature 
dependence. If we assume for simplicity that the scalar meson state in the 
simplest case is only changed by the added temperature dependence of the trace, 
then we would expect a rapid growth in the scalar field $\varphi_0(x_0)$ as a
function of the temperature. Thus our boundstate solution for the nonhomogenous 
Klein-Gordon Equation (\ref{eq:scalmesdom}) given in (\ref{eq:hadgrdsol}), 
shows rapid changes in the classical scalar meson field just above the
decondensation temperature\footnote[12]{The author thanks Gerry Brown for his
suggestions on the subject of the mesonic bound states and the problem of the 
meson mass~\cite{PaLeBr}.}. Furthermore, if we take the lowest scalar meson,
the $\sigma$, with a mean mass $m_{\sigma}=600MeV$ and a full width at least
$400MeV$, this added energy from the decondensing gluons could perhaps be enough
to bring about a single decay of a virtual scalar meson state into two pions. 
In no sense are we advocating that this very rough picture should provide a 
realistic explanation for the effects of gluon condensation in QCD. However, 
it is generally consistent with the decay of the scalar meson in its very 
simplest case.\\
~\\
~\\ 
{\large{\bf{V.5 Entropy for the Hadronic Ground State}}}\\
~\\
\noindent
In this section we look at a different situation involving the equation
of state for the quarks making up the ground state structure of the hadrons. 
We start this investigation with the introduction of the entropy density $s(T)$
for a thermal quark system at very low temperatures $T$. Here we assume 
the singlet structure for the confined quarks in the hadronic bag which 
should hold for temperatures well below the critical temperature $T_c$. 
Herewith we are able to include some aspects of the groundstate structure
in the equation of state.\\ 
~\\
\indent
     We start with the energy density $\varepsilon(T)$ and pressure $p(T)$ of a 
confined colored quark gas for $T \ll T_c$. We use the First Law of Thermodynamics
to include the classical heat density contribution 
\begin{equation}
       s(T)T~=~\varepsilon(T)~+~p(T)~-~\mu_q{n_q(T)},
\label{eq:thermden}
\end{equation}
\noindent
where $\mu_q$ is the quark chemical potential and $n_q(T)$ is
quark density distribution function for a single quark flavor.
We may rearrange this equation using the fact that the thermal
average of the trace of the energy momentum tensor providess the 
equation of state as given in the equation (\ref{eq:epstemp}),
which now takes the form
\begin{equation}
    \theta^{\mu}_{\mu}(T) ~=~s(T)T~-~4p(T)~+~\mu_q{n_q(T)}.
\label{eq:tempenmomten}
\end{equation}
\noindent
For the bag model in the limiting case that the temperature 
approaches to zero we use $p=-\mathcal{B}$, where, as before, 
$\mathcal{B}$ is the bag constant~\cite{DoGoHo} independent
of the temperature. Clearly in this limit $~s(T)T~$ vanishes. 
Furthermore, in the above equation of state we replace $n_q(T)$ with 
the Fermi distribution function $f(\mu_q,T)$ times  $n_0$, 
which is the quark number density in the bag at zero temperature. 
\begin{equation}
    {\theta^{\mu}_{\mu}}_0 ~=~4\mathcal{B}~+~\mu_q{n_0}f(\mu_q),
\label{eq:eqstatea}
\end{equation}
\noindent
In the low temperature limit $f(\mu_q,T)$ is simply a step function so that
$f(\mu_q)$ becomes the normalized particle distribution at zero temperature.
As we have previously explained in the introduction, the trace of 
the energy momentum tensor can be related to the gluon and quark
vacuum expectation values arising from the operator product expansion
~\cite{DoGoHo, SVZ1, Leut}  following from equation (\ref{eq:thetaquark}) 
for ${\theta^{\mu}_{\mu}}_0$. Thus we have, as before, 
included the important vacuum contributions of both 
operator dimensions three and four to the equation of state, 
which are both clearly independent of the temperature.
~\\
~\\
\indent
     In relation to the previous section on scalar meson dominance we 
choose as a simple special case that of the meson with the same light 
quarks surrounded by nuclear matter as an example for this investigation. 
Then we have from the Fermi statistics using the quark-antiquark symmetry 
$~\mu_{\bar q}~=~-\mu_q~$ and for the antiquark quantum density distribution
function $~n_{\bar{q}}(T)~=~n_0(1~-~f(\mu_q,T))~$. Thus in the limit of zero 
temperature we have simply the expected value 
for the bag model~\cite{BJBP,Mil97}
\begin{equation}
   {\theta^{\mu}_{\mu}}_0 ~=~4\mathcal{B}.
\label{eq:eqstatebag}
\end{equation}
For the moment we take as a first approximation the case that  
we consider the value ~\cite{DoGoHo} of only
the gluon condensate as about $~1.95[GeV/fm^3]~$. Here we use the fact that 
the values of the quark condensate for the light quarks are almost negligible. 
Thus we note that for this special case the bag constant $~\mathcal{B}~$
is rather big-- just under $0.5GeV$. However, if we take T finite and small, 
the gluon condensate does not change very much~\cite{Leut, BoMi} until
we reach rather high temperatures.\\ 
~\\
\indent
     Nevertheless, the constituent entropy at very low temperatures due 
to the color degrees of freedom does have a very important contribution 
to the equation of state. In the singlet state it has been shown from the
structure of the hadronic density matrix~\cite{Mil04} that the quantum
ground state entropy for the colored constituent quarks and antiquarks
contribute an internal entropy with the value $\ln{3}$. Although the
hadrons themselves are pure states with {\it{zero}} quantum entropy, the 
constituents do have a finite entropy. In fact, in the singlet groundstate
the quarks(antiquarks) have the maximum entropy of pure (color) mixing.
Then the equation of state for the bag containing the constituents  can 
be written with the additional contribution from the mixing of the colors 
in the quarks and antiquarks treated as separate consitiuents. The total
constituent entropy density is then $~2s_M~$ for both the quarks and 
the antiquarks treated as separate entities.\\ 
~\\
\indent
     Now we consider the total equation of state $T^{\mu}_{\mu}(T)$
containing both the groundstate and the thermal states using the 
decomposition in Equation (\ref{eq:emtensor}) together with the above
Equations (\ref{eq:tempenmomten}) and (\ref{eq:eqstatea}) for the  
meson made up of the constituent quarks and antiquarks. The equation 
of state for the color singlet quarks in the mesonic bag is given by 
\begin{equation}
   T^{\mu}_{\mu}(T) ~=~(2s_M~+~s(T))T~-~4(p(T)~-~\mathcal{B})
                                          ~+~\mu_q n_0(2f(\mu_q,T)~-~1).
\label{eq:meseqnst}
\end{equation}
At very low temperatures we would expect that $~s(T)~$ and $~p(T)~$
both to remain insignificant so that the main change in the 
equation of state is the value of $2s_M~=~1.824[1/fm^3]$. Physically
we could relate this effect either to a lowered bag constant 
or a raised chemical potential. Thereby we may be able
to see how much the finite temperature changes the actual 
value of the entropy due to the color mixing with its corresponding
effect upon the equation of state.\\
~\\
\indent 
     Along this same line we have also investigated a baryonic model at finite
$T$ including the ground state entropy for colored quarks. Here we have studied 
the relation of the temperature and chemical potential to $\mathcal{B}$. In 
this previous work we calculated explicitly the effects of colored quark entropy 
on the bag pressure~\cite{MilTaw1} in terms of the temperature. The general 
equation of state is rather similar to Equation (\ref{eq:meseqnst}) with the 
exception that the groundstate contains the quark entropy with the factor of $3s_B$.   
Following this work we looked into the effects of the quark chemical 
potential~\cite{MilTaw2} on the entropy for a color superconductor.\\
%
\newpage
\noindent
{\Large{\bf{VI. Conclusions, Deductions and Evaluations}}}\\
~\\
\noindent
In this last part of our discussions of lattice calculations for the equation of 
state we bring together some of the main points in this report. An important aspect
investigated in the Introduction was the breaking of both the scale and conformal 
symmetries when the strong nuclear interactions are present. This situation
was very well stated by Stefan Pokorski in his book on {\it{Gauge Field Theories}}:
"Scale invariance cannot be an exact symmetry of the real world. If it were, all the 
particles would have to be massless or their mass spectra continuous."\footnote[13]
{This statement starts a "General Discussion" on the topic of {\it{Broken Scale 
Invariance}}~\cite{Poko} on page 237. A similar deduction was made on page 174 by 
Roman Jackiw~\cite{TreJaZuWit} where he declared that "scale and conformal symmetries 
are broken, as they must be in order to avoid a physically absurd mass spectrum."} 
This statement in itself is not so very surprising when we are only considering 
massive particles like quarks or hadrons. However, because of the trace or
conformal anomaly we saw in the second part from the simulation of pure gauge 
fields that this type of symmetry breaking continues for strongly interacting 
gluons even at very high temperatures. In the third part when the massive quarks 
are present in the simulations, the breaking of scale and conformal symmetries 
have a somewhat different relation to the phase transition because of the process
of the restoration of the broken chiral symmetry which changes the properties of 
the quarks in the transition from the confined phase to the deconfined one. It is 
known, in fact, that the Wilson or Polyakov loop as the order parameter of the pure 
gauge deconfinement transition ceased to provide this property in the presence of
dynamical quarks~\cite{Creu83,MonM96,Roth92}. In this context we looked explicitly
at the chiral anomaly in relation to the breaking of chiral symmetry in the presence 
of dynamical quarks. The thermodynamics is best related through the susceptibilies
to the chiral condensate, which becomes itself an order parameter in the limit of 
vanishing quark masses. These results led the way to the explicit numerical 
evaluations of the thermodynamical properties of the gluon and quark condensates 
in the following sections of the next part. Finally we presented models which 
more directly relate to physical results from experiments. Now we summarize and
expand upon some of the work which we have presented. \\
~\\ 
{\large{\bf{VI.1 Summary of Results and Related Ideas}}}\\
~\\
\noindent
Throughout this work we have mentioned the importance of the ranges of the 
temperatures computed by using lattice gauge theory with the $SU(N)$ color 
symmetry both without and with different numbers and masses of quarks present. 
By using the lattice data from numerous simulations we have plotted the various 
physical quantities in terms of the temperature. Without entering into the 
computational details we have investigated the numerical results in relation 
to the theoretical content of the simulation. We were in the above mentioned
cases able to determine the thermodynamical functions like the pressure, energy
density, entropy density and the resulting equation of state for the quarks and 
gluons as a function of the temperature. These results allowed us to see the
growth of these thermodynamical quantities above the transition temperature, 
from which we could examine the quark and gluon condensates as well as 
compute the thermal properties in the determined related phases.\\
~\\
\indent
     There were a number of other important lattice studies of various
different quark and gluon properties whose data and resuts we have not
directly used. The starting point was in 1974 the "Confinement of quarks" by
Kenneth Wilson~\cite{Wil74,WilEr} where he not only set out the formalism of the
lattice gauge theory by quantizing in Euclidean space-time, but also proposed
the method of evaluation using the Feynman path integrals with the 
strong-coupling approximation. This very original work also 
introduced the Wilson action including the quarks\footnote[14]
{For a description of the actual Wilson fermions see the book by Jan 
Smit~\cite{JanSmit}.}. An alternative approach~\cite{KogSuss} to the lattice 
fermions was proposed by John Kogut and Leonard Susskind known as the 
"staggered fermions." Most of the data presented above uses this formulation
for the dynamical fermions on the lattice since it is somewhat better
for the type of computations which have been carried out. The numerous
works of Michael Creutz were very essential to these numerical 
simulations~\cite{Creu83,Rebbi}. In the earlier times around 1980 
there were many very active groups which contributed greatly to the 
development of the lattice gauge computations with some very 
important results~\cite{KutPolSza,McLerSvet,EKMS81,Kogetal} for pure 
$SU(N)$ lattice gauge simulations. These works provided the starting
point for many of the later computations of the MILC and Bielefeld groups--
for further literature see~\cite{Satz,CleGS,Rebbi}.\\
~\\
\indent  
     In recent years there have been considerable progress in the simulations 
with finite chemical potentials on the lattice. The above discussions have
not included these many extensive computations~\cite{FodKat,BiSwan} with 
dynamical quarks who included the chemical potentials. Only in the above
very special example of the single flavored meson at low temperatures
have we directly included the effects the quark-antiquark chemical potential. 
This particular case has its sole importance for the analysis 
of the model of a meson for the equation of state at very low temperatures.
Others with dynamical quarks who included the chemical potentials were some works 
on the phase diagram by Fodor and Katz~\cite{FodKat} as well as the more 
recent work done in collaboration of Bielefeld and Swansea~\cite{BiSwan}. 
These results we have not included with our analysis. There are also 
further important contributions from the CP-PACS Collaboration with two 
heavier quark flavors using the Wilson action for which we were unable to 
set the temperature scale to find thermodynamical functions out of 
the computed ratios~\cite{JapPac}. Although all these works are
very interesting, the form of the data as well as the values are often not
in the temperature range where a comparison with the quantities are easily
carried out.\\
~\\
{\large{\bf{VI.2 Implications from the Analysis}}}\\
~\\
\noindent
     There has oftentimes been a problem with the interpretation of some of 
the results from the numerous lattice computations due to the many general 
statements arising out of the ideal gas models. In numerous papers and even in some 
very highly respected text books\footnote[15]{For example, at the end of a paragraph 
describing some numerical simulations~\cite{LeBel96} on page 10 it was written: 
"Even at $T=2T_c$, the energy density only reaches about $70\%$ of the ideal gas 
value. However, at least for $T \ge 2T_c$, one finds $\varepsilon \approx 3P$."} 
there have appeared statements where the expected approximate equality of 
$\varepsilon$ and  $3P$ becomes very nearly realized 
within a few integral mutiples of the critical temperature.
However, we have seen above that within the range of the present numerical 
data from lattice gauge simulations all of statements of this type are 
clearly not fulfilled. In the earlier parts of this report we have 
seen the results for the equations of state in the
figures~\ref{fig:eqstgauge}, \ref{fig:eqnstate}, \ref{fig:eqstqu}, and 
\ref{fig:equostaquark}. All of these evaluations clearly show directly 
that these equations of state in terms of the temperature remain finite 
well above the critical or deconfinement temperature. As we have already 
explicitly pointed out in the text, with the possible exception of our 
figure~\ref{fig:eqstqu}, all the other plots show clearly a monotonical 
growth of the equation of state as a function of the temperature. 
Perhaps one possible source of misunderstanding of the lattice data 
lies in the representation as ratios with the $T^N$ powers divided out.
From the computational side this representation is very useful for direct 
comparisons with the high temperature limits for lattices of different sizes. 
However, for the interaction measure, which is defined as a ratio there are 
thermal contributions to the energy density which are different from those 
arising in the pressure. The difference $\varepsilon-3p$ has contributions
of lower powers and logarithmic terms in the temperature which appear to go 
to zero in the high temperature limit when divided by the higher power. Thus
the interpretation coming from the appearance of the numerical results from
the lattice presented as ratios can be quite misleading for the curves with  
a finite or vanishing limiting behavior. Here, as we have stated above, it
is important that such terms be  written in terms of the well understood 
thermodynamical quantites, which must diverge in the infinite temperature 
limit in accordance with the leading powers. Furthermore, as one perhaps
could expect from the formulation of the laws of thermodynamics, we are 
also able to described the equation of state using exact differential 
forms which relate directly the known anomalies to the actual physical 
quantities~\cite{Mil97,Mil99}. In this general manner the action as a 
formal expression of quantum field theory is also clearly related to 
the general structure of the equation of state (see Appendix B).\\
~\\
\indent
     It was recently discussed how the lattice results show that even at 
temperatures much larger than the deconfinement temperature the thermodynamically 
related observables like the pressure ratio, the energy density ratio, the 
interaction measure and the baryon number density ratio at different chemical 
potentials still deviate by more than $20\%$ from their respective relativistic 
ideal gas values~\cite{KSTR}. These authors consider two special model cases: 
the two phase model and the mixed phase model, which they have quite 
appropriately named. The direct comparisons with the lattice data for the 2+1 
flavor cases~\cite{KLP,KLP1} and~\cite{FodKat} provided insight into both the 
critical behavior and the asymptotic approach at high temperatures. In the case 
of the interaction measure the two phase model had a much sharper peak at the 
critical temperature than provided by the lattice data. However, the mixed phase 
model the peak provided much smaller peaks in the interaction measure. For the
other ratios involving the pressure, energy density and the baryon number density
the results of the comparisons with the lattice data were generally quite similar.\\  
~\\
{\large{\bf{VI.3 Applications to Physical Processes}}}\\
~\\
\indent
     We have already mentioned in the previous part of this report
the properties of the equation of state at low temperatures for 
the study of the hadronic structure. In this temperature region 
the thermodynamical functions included in the equation of state
are much harder to compute using the numerical simulations of 
finite temperature QCD on the lattice. Nevertheless, there has 
recently been considerable progress on the numerical computations 
of correlations between quarks and antiquarks at these lower 
temperatures in certain model cases~\cite{KKPZ}. Certainly a 
part of this computational difficulty lies in complicated
quantum structure in the low temperature thermodynamics
providing for the basic properties of the hadronic ground state. 
In this context we have previously looked into the role of the 
quantum entropy~\cite{Mil04} in the discussion of the stability 
of the various quantum states-- particularly relating 
to the singlet hadronic states.\\
~\\
\indent
     The thermodynamics of quarks and gluons in the confined region
is of great interest in the examination of the formation of particle
states below the deconfinement temperature. The recent computations
of the quark-antiquark free energies~\cite{KKPZ} using the renormalized 
Polyakov loop has provided considerable new insight into the short 
distance interactions between the quarks and the antiquarks over a 
large range of temperatures. The results from these numerical
simulations evaluate the color averaged and the color singlet free
energies for static quark-antiquark sources placed in a thermal
gluonic heat bath. An important procedure~\cite{KKPZ} in these 
calculations is the renormalization of these free energies using 
the short distance properties of the zero temperature heavy quark 
potential. This procedure leads to the definition of the renormalized
Polyakov loop as an order parameter for the deconfinement phase
transition of the $SU(3)$ gauge invariant field theory. The color 
averaged free energy $F_{\bar{q}q}(r,T)$ of a static quark-antiquark 
pair acting as sources in a thermal medium is related up to a 
function of the temperature to the thermally averaged product 
of the color averaged Polyakov loops at two different spatial 
points separated by a distance $r$. The color singlet free energy
$F_1(r,T)$ is related to both the thermal and color averaged product 
of the gauge field variables. The significance of these results lies 
in the fact that at very short distances these two differently defined 
free energies become the same up to a factor of $T\ln 9$. Thus in
the limit of short distances it was found~\cite{KKPZ} that
\begin{equation}
\lim_{r\to 0} [F_{\bar{q}q}(r,T)~-~F_1(r,T)]~=~T\ln 9,~~~\forall{T}.
\label{eq:freenercorr}
\end{equation}
\noindent
However, at larger values of the distance $r$ there are additional
contributions to the colored averaged free energy coming from the
octet states, which vanish at small distances. More recently the actual 
effective running coupling at finite temperatures has been calculated 
for both the quenched~\cite{KKZP} and two flavor~\cite{KZ1} QCD from the 
derivative of the free energy with respect to the separation $r$ between 
the quarks and antiquarks. In a later work these latter authors~\cite{KZ2}
look at the singlet internal energy and entropy at large separations
and higher temperatures about $1.3T_c$.  However, in the case of 
small separations these authors note that both $U_1(r,T)$ and 
$TS_1(r,T)$ suffer from lattice artifacts which result in the 
same problems for the singlet free energy.\\
~\\
\indent
     Next we  mention some ideas closely related to these results but arise from 
quite different considerations for low temperature quantum critical systems. 
First we recall from thermodynamics that the free energy $F$ and the internal 
energy $U$ are related through the entropy $S$ by a simple transformation: 
$F = U-TS$, whereby we can write that $S =-{\frac{\partial{F}}{\partial{T}}}$. 
Thus we see that for finite temperatures we have a simple relation to the entropy. 
However, as mentioned in the last part, the entropy of the hadronic groundstate 
has the role of the mixing of the colors. It has been recently pointed 
out~\cite{FriKon} that for one-dimensional quantum critical systems the 
entropy near criticality can be clearly divided into two separate parts-- 
the bulk part arising from the partition function and the boundry terms relating 
to the groundstate degeneracy $d_g$. This approach had been successfully 
applied in the past~\cite{AffLud} to such important problems as the Kondo 
effect and the Heisenberg ferromagnet. Here we are interested in some different 
applications in relation to quark string models for the discussed computations 
on the lattice of Polyakov loop correlations~\cite{KKPZ}. We recall in our
above discussion of the quantum ground state of the hadronic singlet state
that the associated entropy of the quark and antiquark is $\ln 3$. If we regard
the quark and antiquark as the end of a one-dimensional string, then we may
associate the boundary entropy $\ln{d_g}$ as arising from the groundstate
degeneracy $d_g~=~3$ for the static quark and antiquark.
Thus by the addition of entropies we may write the total boundry entropy 
for the two static sources as $2\ln 3$. All other sources of entropy we may 
associate with the bulk entropy in the string~\cite{FriKon}, which we assume 
here is mainly due to the gluons. Finally if we assume that the internal
energy at zero temperature is just the static quark-antiquark potential, 
we are thereby able to identify the term $~T\ln 9$ as arising from the
groundstate entropy.\\    
~\\
\indent
     We now recall some of the general results of our report as written 
above. The light quarks first go through the chiral restoration transition 
at a temperature $T_c$ then the effects of deconfinement actually 
will appear at a higher temperature $T_d$. The nature including even 
the order of these transitions appears to depend greatly upon the
number of flavors as well as the masses. The heavier quarks show 
the separation between the transitions much less clearly. However,
in all cases both the scaling and the conformal symmetries are still
broken at much higher temperatures well above both of these transitions. 
We see from the physical equation of state that the very high temperatures 
provide a further symmetry breaking which is not present in the QCD 
vacuum contributions. The properties of the Brown-Rho scaling are 
important to the actual particle structure when surrounded by hot 
dense matter. Also recently the mass and the width of the sigma resonance
have been calculated~\cite{CCL} by locating the pole in the pion-pion
scattering amplitude with the quantum numbers of the vacuum\footnote[16]
{This work has been further clarified by another even more recent
investigation~\cite{Penn} of the sigma coupling to the photons in 
the process ${\gamma}{\gamma}\rightarrow{\pi}{\pi}$.}.
The quarks and gluons are strongly correlated even well 
above the transition temperatures. There is also an effect of the 
running coupling~\cite{KKZP,KZ1} involving the confining part of the 
quark-antiquark interaction well above $T_c$.  In a very recent article 
Gerry Brown and collaborators posed a critical question on the results
of the recent heavy ion experiments at Brookhaven National Laboratory.
Therein they indicate the nature of the high temperature state by the 
statement~\cite{BroGelRho}: "We suggest that {\it{the new form of matter}} 
found just above $T_c$ by the Relativistic Heavy Ion Collider is made 
up of tightly bound quark-antiquark pairs,..." Their observation 
gives further motivation for the present investigations of the 
correlations and interactions at temperatures above $T_c$.\\ 
~\\
\indent
     Finally in the two appendices we add some detail to the discussions
of the models and the physical currents. There we formally discuss 
the actual physical meaning of these currents as expansion parameters 
and forces arising from the breaking of the scale and conformal symmetry. 
We take the well known example described in Appendix A -- the MIT Bag model.
The bag constant itself arises in the trace to make it nonzero. In the 
Appendix B we discuss how the dilatation and conformal currents are 
related to both the equation of state and the gluon condensates. 
We show these relations are best understood from the integral or 
the dual forms of these currents.\\  
\newpage
\noindent
{\Large{\bf{Acknowledgements}}}
The author wants to thank all the colleagues in Bielefeld, especially
Rolf Baier, J\"urgen Engels, Frithjof Karsch, and  Helmut Satz for 
their help on many parts of this work. He would also like to recognize 
important discussions with Dietrich B\"odecker, Rajiv Gavai, 
Sourendu Gupta, Olaf Karczmarek, Peter Kolb, Edwin Laermann, 
Mikko Laine, Andreas Peikert, Bengt Petersson, Peter Petreczky, 
Paul Romatschke, York Schr\"oder, Edward Shuryak, Ismail Zahed 
and Felix Zantow. Further he is very grateful to Carleton DeTar 
for providing the MILC97 data. He thankfully recognizes the 
efforts of Graham Boyd for his earlier work in an essential 
collaboration on this subject. The author wants to acknowledge
Abdel-Nasser Tawfik for his help with the preparation of some
useful figures and some earlier collaborative discussions. To 
Ivan Andri{\'c}, Neven Bili{\'c}, Ivan Dadi{\'c}, 
Jerzy Lukierski and Krzysztof Redlich he wishes to express 
his thanks for their help and friendship as hosts in Zagreb 
and Wroclaw in the Fulbright Scholar Program a well as many 
scientific discussions. He is very indebted to Gerry Brown 
both as a teacher and presently as the editor for his 
interest, ideas and many suggestions on this subject. Finally 
he is grateful to the Pennsylvania State University for the 
Research Award from the University Commonwealth College.\\
~\\
~\\
\noindent
{\large{\bf{APPENDIX A: Phenomenological Models for Confinement}}}\\
~\\
\noindent
In this added section we sketch the properties of some of the most prominent
phenomenological models used in the theory of strong interactions to explain
the observation of quark comfinement. It is meant as an extension to the
introduction for the added properties.\\
~\\
\noindent
1. The {\it{MIT bag model}} was proposed as a model for the extended hadrons which 
contain free quarks inside a small volume. The action~\cite{ChJJThW} with the internal
energy density ${\mathcal B}$ is written in the following form:
\begin{equation}   
\mathcal{W}~=~{\int}^{t_2}_{t_1}dt{{\int}_R}{d^3}r[{\frac{1}{2}}\dot{\phi}^2~-
~{\frac{1}{2}}(\vec{\nabla}\phi)^2~-~{\mathcal B}],
\label{eq:bagaction}
\end{equation} 
\noindent
where ${\phi}(t,r)$ is the prototype of a hadronic field, which are 
the partonic or hadronic constituents, and $R$ and $(t_1,t_2)$ are 
the space and time regions for the "bag". This action provides the field 
equations inside the bag
\begin{equation}   
-{\Box}\phi(x)~=~0
\label{eq:baginside}
\end{equation} 
\noindent
and on the surface
\begin{equation}   
{\hat n}\cdot{\frac{\partial{\vec{R}}}{\partial t} }\dot{\phi}~+
~{\hat n}\cdot{\vec{\nabla}\phi}=~0.
\label{eq:bagsurface1}
\end{equation} 
\noindent
Then the bag condition at the surface becomes
\begin{equation}   
{\frac{1}{2}}\dot{\phi}^2~-~{\frac{1}{2}}(\vec{\nabla}\phi)^2~=~{\mathcal B}.
\label{eq:bagsurface2}
\end{equation} 
\noindent
These three equations are basic to the {\it{MIT bag model}}. We note
that the structure of this model is not directly in correspondence with
QCD because of the time of development of the fields\footnote[17]
{We shall use only the static properties of ${\mathcal B}$ for the 
special examples discussed in Appendix B}.\\
~\\
\noindent
2. There have been many further developments on the equation of state for 
many cases involving bag models in a more general form. One can propose a form 
for the energy $E(T,V)$ containing the bag term as ${E_0}(V)={\mathcal B}V$.
As a correction one often adds the thermal radiation terms of the Stefan-
Boltzmann type ${{\sigma}_{SB}T^4}V$. However, this form may be extended to
a more general energy equation of the following type:
\begin{equation}   
~E(T,V)~=~{E_0}(V)~+~E'(T,V).
\label{eq:energyeq}
\end{equation} 
\noindent
The pressure is defined by the usual relation $p=-({\partial}E/{\partial}V)_T$.
When the free energy density $f'(T)$ of the finite temperature part is scaled
with a temperature $\vartheta$ so that $f'(T)=T{\phi}({\vartheta}/T)$, where 
$\phi$ is only a function of ${\vartheta}/T$, then the equation of state is just
\begin{equation}   
~p~=~-{\mathcal B}~+~\gamma{\varepsilon}'.
\label{eq:eqosta}
\end{equation} 
\noindent
The constant is defined $\gamma=-d(\ln{\vartheta})/d(\ln{V})$ for the energy
density at finite temperature ${\varepsilon}'$. For the ultrarelativistic gas
$\gamma=1/3$, which yields the known relationship to the bag constant
~\cite{ChJJThW,BJBP}. The form of this equation for the relativistic gas 
is similar to the Debye equation of state for solids at low temperatures.
In the Debye theory of solids $\gamma$ is often referred to as the
"Gr\"uneisen constant." This approach of separating the groundstate and 
thermal states allows us to write a very general form of the equation of 
state\footnote[18]{The application of this equation of state to 
nuclear matter is described in the book of John Dirk Walecka~\cite{JDW}.}.\\
~\\
\noindent
3. The chiral bag model~\cite{NoRhoZah} has additional properties in relation to QCD.
This new structure is often referred to as "The Cheshire Cat Mechanism." For simplicity
one usually looks at a model in 1+1 dimensions. The action $S$ is then decomposed
into three parts
\begin{equation}   
~S~=~S_V~+~S_{\tilde{V}}~+~S_{\partial{V}}
\label{eq:chirbagaction}
\end{equation} 
\noindent
where for the fermions
\begin{equation}   
~S_V~=~\int_V d^2x{\bar\psi}i{\gamma}^{\mu}{\partial}_{\mu}{\psi}~+...,
\label{eq:chirbagferm}
\end{equation} 
\noindent
while for the bosons
\begin{equation}   
~S_{\tilde{V}}~=~\int_{\tilde{V}} d^2x{\frac{1}{2}}({\partial}_{\mu}{\phi})^2~+....
\label{eq:chirbagbos}
\end{equation} 
\noindent
The additional surface term becomes with the value of $f~=~{\frac{1}{\sqrt{4\pi}}}$
and $d{\Sigma}^{\mu}$ is the surface area with the normal vector ${n}_{\mu}$ so that
\begin{equation}   
~S_{\partial{V}}~=~\int_{\partial{V}} d{\Sigma}^{\mu}\{{\frac{1}{2}}{n}_{\mu}{\bar\psi}
\exp{(i{\gamma}_5{\phi}/f)}{\psi}\}~+....
\label{eq:chirbagsurf}
\end{equation} 
\noindent
~\\
~\\
\noindent
{\large{\bf{APPENDIX B: Mathematical Forms and Physical Currents}}}\\
~\\
In this appendix we provide some additional information on some particular 
differential forms relating to the various discussed broken symmetries due
to the properties of the strong interacions. There are two different types
forms relating to the structures of the anomalies, which are related to the
axial and conformal anomalies respectively. In the former case there are
already a large number of works~\cite{Bert,Kugo}  relating to the chiral 
symmetry breaking and the axial anomaly, for which there is a very well 
known differential four-form $\mathcal{A}d{\Omega}$. Furthermore, 
there is the related Chern-Simons Form $d{\mathcal{B}}$, which can be
represented as a three-form on a three dimensional closed surface.
One can also discuss the physical interpretation of this form which is 
related to the topological charge. These ideas together with the properties 
of the chiral symmetry breaking in relation to the instanton solutions 
is discussed in the book of Edward Shuryak~\cite{Shur}.\\
~\\
\indent
     The scale and conformal symmetry breaking can be similarly related to 
the conformal or trace anomaly. In the above work we have found the presence 
of a finite trace of the energy-momentum tensor relating to the equation
of state and the QCD sum rules. It is known in quantum field theory that
classically the trace of the energy-momentum tensor is zero. In the following
we derive (introduce) the corresponding three-forms under the names 
{\it dyxle} for the scale breaking and the {\it fourspan} for the 
special conformal symmetry breaking. The actual physical meanings 
for these currents and charges have been previously  discussed~\cite{Jack}.
The actual interpretion of these as energy flux and shearing forces has been 
given not so long ago by the author~\cite{Mil99}, upon which we shall 
now briefly expand.\\
~\\
\indent
     The dilatation current $D^{\mu}$ has already been defined above 
in terms of the position four-vector $x^{\mu}$ and the energy momentum 
tensor~$T^{\mu \nu}$ as simply the product $x_{\alpha}T^{\mu \alpha}$ 
as moments of the energy density. In the case of the general 
energy-momentum conservation~\cite{TreJaZuWit,Jack} one can find
quite simply a relation to the equation of state~\cite{Mil97,Mil99}. 
We now look into a volume in four dimensional space-time ${\Omega}$ containing 
all the quarks and gluons at a fixed temperature $T$ in equilibrium.
The flow equation holds when the energy momentum and all the 
(color) currents are strictly conserved over the closed bounding surface 
$\partial {\Omega}$ of the properly oriented four-volume ${\Omega}$, 
which yields
\begin{equation}
\oint_{\partial {\Omega}}{\cal D}_{\mu}dS^{\mu}~=
~\int_{{\Omega}} T^{\mu}_{\mu}d{\Omega},
\label{eq:dyxleform}
\end{equation} 
We have already introduced~\cite{Mil99} the {\it dyxle} three-form 
as ~${\cal D}_{\mu}dS^{\mu}$ on the closed three dimensional surface 
$\partial {\Omega}$, which is simply just the divergence in the four 
dimensional Minkowski space. The dyxle is the dual to the 
dilatation current written as the one-form~$D_{\mu}dx^{\mu}$  
in the usual four dimensional space-time\footnote[19]{The relation
of a differential form to its dual form is important here~\cite{Flan},
see especially sections 4.6, 5.9 and 10.6.}. In this context the 
dilatation current $D_{\mu}dx^{\mu}$ acts as if it were a
{\it{streaching force}} over the space-time infinitesimal $dx^{\mu}$.
Whereas its dual form in space-time represents the divergence 
of this three dimensional closed surface acting as the 
boundry for the four dimensional volume ${\Omega}$. On the 
right hand side of~(\ref{eq:dyxleform}) the integrated four-form
~$\int_{{\Omega}} T^{\mu}_{\mu} d{\Omega}$ is an action or energy moment
integral involving the equation of state. Since we assume here the
positivity of the trace so that $T^{\mu}_{\mu} > 0$, we can conclude that 
this action integral is a positive quantity. It acts as the color averaged 
source of the energy flux.  This action integral gets quantized with 
the fields through the renormalization process.\\
~\\
\indent
     The analogous three-forms can be defined for the four special conformal 
currents which we shall collectively call the {\it fourspan}. The analogous
dual forms are derived from the equation in a similar manner to that 
previously done for the dyxle~\cite{Mil97,Mil99}. The results for the four conformal 
currents yield the four three-forms ${\cal K}^{\alpha}_{\mu}dS^{\mu}$ is 
derived similarly from the following equation: 
\begin{equation}   
\oint_{\partial {\Omega}}{\cal K}^{\alpha}_{\mu}dS^{\mu}~=
~\int_{{\Omega}}{2{x^{\alpha}}}T^{\mu}_{\mu}d{\Omega},
\label{eq:fourspanform}
\end{equation} 
\noindent
We point out here that for the {\it fourspan}  the source terms are 
the first moments in space-time of the equation of state. Furthermore, 
the physical nature of the quantities ${\cal K}^{\alpha}_{\mu}$ as the 
special conformal currents can be seen to be related to the shearing 
forces which destroy the conformal symmetry in all the four directions.
The effect of the {\it fourspan} is analogous to the {\it dyxle} 
along the radial line in the four dimensions. In summary these five
three-forms provide the mathematical structure for the known breaking
of the scale and conformal symmetries for the strong (nuclear) interaction.\\
~\\
\indent
     As an example of the formal statements above we work out the analytical 
properties of these forms may be exactly calculated in the special case of 
the well known~\cite{ChJJThW} {\it{MIT bag model}}. The scaling properties associated 
with the trace of its energy-momentum tensor we have noted above\footnote[20]
{We have discussed in the Introduction the relation~\cite{BJBP} of the trace 
to $\cal B$ as seen in their equation (1.10) therein and independently somewhat 
later~\cite{Mil97}.}. This relationship yields in the above 
equation (\ref{eq:epstemp}) simply $\theta^{\mu}_{\mu}=4{\cal B}$. 
This result gives an exact solution for the dilatation
current $D^{\mu}(x)$ of the form:
\begin{equation}   
D^{\mu}(x)~=~{\cal B}x^{\mu}.
\label{eq:bagdyxle}
\end{equation} 
\noindent
It is clear that this solution is a special case of the above general form
$x_{\nu}T^{\mu \nu}$ for the bag model. Similarly we derive the special
forms for the four conformal currents $K^{\alpha \mu}(x)$ using the same
general structure as for the dilatation current. Here there are two different
types of solutions: (a) $\alpha = \mu $ and (b) $\alpha \ne \mu$. In the first
case (a) using the propertime $\tau^2 = x_{\mu}x^{\mu}$ with the implicit 
summation over the indices $\mu$, we have for a single value of the index
chosen from $\alpha = \mu = 0$ for the temporal case
\begin{equation}   
K^{00}(x)~=~{\cal B}(2(x^{0})^2 - \tau^2),
\label{eq:bagconfdiagtemp}
\end{equation} 
\noindent
while for $\iota = 1,2,3$, where $\iota = \alpha = \mu \ne 0$ the spatial case
yields
\begin{equation}   
K^{\iota \iota}(x)~=~{\cal B}(2(x^{\iota})^2 + \tau^2).
\label{eq:bagconfdiagspac}
\end{equation} 
\noindent
For the case (b) we have $\iota \ne \kappa$ so that the result for 
$\iota,\kappa = 0,1,2,3$ is just
\begin{equation}   
K^{\iota \kappa}(x)~=~2{\cal B}x^{\iota}x^{\kappa}.
\label{eq:bagconfnondiag}
\end{equation} 
\noindent
Thus we see that in case (a) in the equations (\ref{eq:bagconfdiagtemp})
and we (\ref{eq:bagconfdiagspac}) have various combinations of 
quadratic terms depending upon which value of $\iota = 0$ or
$\iota = 1,2,3$. $K^{00}(x)$ is just a four dimensional shearing force.
Furthermore, we notice that it is truly positive definite so that for
both spacelike and timelike events in the bag it remains always positive. 
However, in case (b) in equation (\ref{eq:bagconfnondiag})
we find simply bilinear terms in two different space-time coordinates.
The simplicity of these solutions for $D^{\mu}(x)$ and $K^{\alpha \mu}(x)$ 
arises form the fact that the energy momentum tensor and the metric
tensor are both diagonal in the Minkowski metric for the bag model. 
All the conformal currents $K^{\alpha \mu}(x)$  act like internal 
shearing forces which all go against the preservation of the angular 
symmetries which includes the breaking of the scale symmetry associated
with $D^{\mu}(x)$. \\

\end{document}